\documentclass[a4paper,fleqn,usenatbib]{mnras}

\usepackage[T1]{fontenc}
\usepackage{ae,aecompl}

\usepackage{graphicx}	
\usepackage{amsmath}	
\usepackage{amssymb}	
\usepackage{subcaption}
\usepackage{cases}
\usepackage{multirow}
\defcitealias{demasi2018}{DM18}
\defcitealias{rosani2018}{R18}
\defcitealias{P04}{P04}
\defcitealias{labarbera2013}{LB13}

\title[IMF in ellipticals - clues from chemical abundances.]{Is the IMF in ellipticals bottom-heavy? Clues from their chemical abundances.}

\author[C. De Masi et al.]{
	C. De Masi,$^{1}$\thanks{E-mail: demasi@oats.inaf.it}
	F. Vincenzo,$^{2}$
	F. Matteucci,$^{1,3,4}$
	G. Rosani,$^{5,6}$
	F. La Barbera,$^{7}$
	\newauthor
	A. Pasquali$^{5}$
	E. Spitoni$^{1,8}$
	\\
	$^1$ Astronomy Department, University of Trieste, Via Tiepolo 11, I-34127 Trieste, Italy\\
	$^2$ Centre for Astrophysics Research, School of Physics, Astronomy and Mathematics, University of Hertfordshire,\\ College Lane, Hatfield, AL10 9AB, UK\\
	$^3$ I.N.A.F. Osservatorio Astronomico di Trieste, via G.B. Tiepolo 11, I-34131, Trieste, Italy\\
	$^4$ I.N.F.N. Sezione di Trieste, via Valerio 2, 34134 Trieste, Italy\\
	$^5$ Astronomisches Rechen-Istitut, Zentrum f\"ur Astronomie, Universit\"at Heidelberg, M\"onchhofstr. 12-14, D-69120 Heidelberg, Germany\\
	$^6$ Kapteyn Astronomical Institute, University of Groningen, P.O. Box 800, 9700AV Groningen, The Netherlands\\
	$^7$ INAF - Astronomical Observatory of Capodimonte, via Moiariello 16, 80131 Napoli, Italy\\
	$^8$ Stellar Astrophysics Centre, Department of Physics and Astronomy, Aarhus University, Ny Munkegade 120, DK-8000 Aarhus C, Denmark
}

\date{Accepted XXX. Received YYY; in original form ZZZ}

\pubyear{2015}

\begin{document}
\label{firstpage}
\pagerange{\pageref{firstpage}--\pageref{lastpage}}
\maketitle

\begin{abstract}
We tested the implementation of different IMFs in our model for the chemical evolution of ellipticals, with the aim of reproducing the observed relations of [Fe/H] and [Mg/Fe] abundances with galaxy mass in a sample of early-type galaxies selected from the SPIDER-SDSS catalog. Abundances in the catalog were derived from averaged spectra, obtained by stacking individual spectra according to central velocity dispersion, as a proxy of galaxy mass. We tested initial mass functions already used in a previous work, as well as two new models, based on low-mass tapered ("bimodal") IMFs, where the IMF becomes either (1) bottom-heavy in more massive galaxies, or (2) is time-dependent, switching from top-heavy to bottom-heavy in the course of galactic evolution. We found that observations could only be reproduced by models assuming either a constant, Salpeter IMF, or a time-dependent distribution, as other IMFs failed. We further tested the models by calculating their M/L ratios. We conclude that a constant, time-independent bottom-heavy IMF does not reproduce the data, especially the increase of the $[\alpha/Fe]$ ratio with galactic stellar mass, whereas a variable IMF, switching from top to bottom-heavy, can match observations. For the latter models, the IMF switch always occurs at the earliest possible considered time, i.e. $t_{\text{switch}}= 0.1$ Gyr.
\end{abstract}




\section{Introduction}\label{sec:model_intro}

The initial mass function (IMF) deeply affects the chemical evolution of a galaxy on many different levels, by determining the ratio between low and high mass stars. The former are known to produce the bulk of Fe in the galaxy via type Ia SNe over long time scales \citep{matteucci1986,matteucci2001}; additionally, even when not directly influencing the chemical abundances over a Hubble time, they still affect their evolution by locking away baryonic matter from the interstellar medium. On the opposite end of the mass range, massive stars are the main producers of $\alpha$ elements (O, Mg, Si, Ca), via processes characterized by much shorter timescales than for Fe-peak elements. The difference in production channels and timescales of the various chemical elements from stars in different mass ranges, when combined with the star formation history of a galaxy, leaves a characteristic mark on abundance ratios such as the $[\alpha/Fe]$, which in turn may allow the formation history itself to be reconstructed from observations \citep{matteucci1994,matteucci1998,matteucci2012}.\\
Other than the chemical evolution, many other properties of a galaxy are strictly related to the IMF. Low mass stars mainly contribute to build up the total present time stellar mass \citep{kennicutt1998}, while massive stars dominate the integrated light of galaxies \citep{conroyvandokkum2012b}, and determine the amount of energetic feedback produced after star formation episodes. Generally, different properties are determined by the slope of the IMF in different mass ranges. \cite{renzini2012} investigated the topic throughfully, and showed how the slope below $\approx 1\;M_{\odot}$ dominates the M/L ratio in local ellipticals, while its evolution is mainly influenced by the slope between $\approx1$ and $\approx1.4\;M_{\odot}$.\\
For these reasons, it does not come as a surprise that determining the exact shape of the IMF is one of the focal points of interest in the study of galaxies. Theoretically, a comprehensive physical picture explaining the origin and properties of the IMF does not exist yet; to this regard, \cite{silk1995} and \cite{krumholz2011} analyzed the effect of molecular flows and protostellar winds, \cite{larson1998,larson2005} tried to explain it in terms of the Jeans mass, while \cite{bonnell2007}, \cite{hopkins2013} and \cite{chabrier2014} explored the effect of gravitational fragmentation and of the thermal physics. Observationally, direct star counts in star forming regions and clusters of our Galaxy all seemed to point towards an invariant IMF, characterized as a Kroupa/Chabrier distribution, with a power-law for $m>1 M_{\odot}$, and a turn-off at lower masses \citep{scalo1986,kroupa2001,kroupa2002,bastian2010,kroupa2013}; this, in turn, generally led to the assumption of the universality of the IMF. A direct verification of this assumption in external galaxies, however, is well beyond our current observational capabilities, so that we are bound to employ indirect methods to obtain constraints on the IMF of galaxies with unresolved stellar populations. The main approach is that of observing gravity-sensitive features in the galaxy integrated spectra; to name a few, the presence of the NaI$\lambda\lambda8183, 8195 \AA$ doublet lines and of the Wing-Ford FeH band at 9900 \AA\, is an indicator of the presence of low-mass dwarfs, while the Ca triplet lines at $\lambda8498, 8542$, and $8662$ \AA\ are strong in giants and basically undetectable in dwarfs \citep{wing1969,faber1980,diaz1989}.\\
A number of works involving the observation of these features provided indications for the IMF becoming bottom-heavier than a Kroupa/Chabrier in massive early-type galaxies. \cite{cenarro2003} first proposed
a trend towards an excess of low-mass stars in massive galaxies, from a study of the CaT region. \cite{vandokkum2010,vandokkum2011} came to the same conclusion after analyzing a sample of eight massive ETGs in the Virgo and Coma clusters, and further confirmed it by using stellar population models accounting for variable element abundance ratios and using a full spectral fitting analysis on a set of 34 ETGs from the SAURON survey \citep{conroyvandokkum2012a,conroyvandokkum2012b}. \cite{ferreras2013}, \citet[hereafter LB13]{labarbera2013}, as well as \cite{spiniello2014} showed that a systematic trend is in place for the whole population of ETGs, with higher velocity dispersion (mass) galaxies having a bottom-heavier IMF (but see also \citealp{smith2013}, \citealp{smith2015} and \citealp{newman2017} for evidence of some massive ETGs with a "light" IMF normalization). A similar result was claimed by \cite{auger2010}, \cite{grillo2010}, \cite{treu2010}, \cite{barnabe2011}, \cite{cappellari2012} and \cite{spiniello2012} on the basis of kinematics and gravitational lensing studies, and by \cite{dutton2011,dutton2012,dutton2013} from scaling relations and models of light and dark-matter distribution in galaxies.\\
On the other hand, however, \cite{gunawardhana2011} observed a strong dependence of the IMF on star formation in a sample of low-to-moderate star-forming galaxies redshift galaxies from the GAMA survey, with the high mass slope of the initial mass function becoming flatter (hence providing a top-heavier IMF) in objects with higher formation activity, as it might be the case for the progenitors of more massive galaxies \citep{matteucci1998,matteucci2012}. Historically, galaxy formation models based on the hierarchical scenario failed in simultaneously reproducing two fundamental observational features of ellipticals, i.e. the increase of the $[\alpha/Fe]$ ratios with higher values of $\sigma$ (a proxy for mass) and the mass-metallicity relation \citep{pipino2008,okamoto2017}. Common solutions proposed to overcome this limit generally involved the introduction of AGN feedback and/or of variable IMFs, becoming top-heavier with mass.\\
In this sense, \cite{thomas1999} proposed two scenarios for the formation of giant ellipticals, either via fast $(\approx 1 Gyr)$ collapse of smaller entities or via merging of spiral galaxies similar to the Milky Way; in the latter case, the desired $[\alpha/Fe]$ overabundance could only be reproduced by assuming an IMF flatter than a Salpeter during the initial starburst triggered by the merging.\\
Similarly, a combination of IMFs top-heavier than a Salpeter one with other mechanisms was proposed by \cite{calura2009}, who assumed a star-formation-dependent IMF - with a slope switching from a Salpeter (x=1.35) to a slightly flatter value (x=1) for SFR >$100 M_{\odot}\,yr^{-1}$ - together with interaction-triggered starbursts and AGN feedback. \cite{arrigoni2010} used both a top-heavy IMF (with a slope x = 1.15) and a lower SNe Ia ratio. \cite{gargiulo2015} implemented SFR-dependent IMF together with a radio-mode AGN feedback quenching star formation.
\cite{fontanot2017,fontanot2018a,fontanot2018b} analyzed the implications of including the integrated galaxy-wide stellar initial mass function (IGIMF) in the semi-analytical model GAEA (GAlaxy Evolution and Assembly), and the effect of cosmic rays on its shape and evolution.\\
To conciliate the opposing indications as to whether the IMF in more massive ellipticals should be bottom or top-heavy, \cite{weidner2013} and \cite{ferreras2015} proposed a time dependent form of the IMF, switching from a top-heavier form during the initial burst of star formation to a bottom-heavier one at later times.\\
In \citet[hereafter DM18]{demasi2018}, we studied the chemical patterns observed in a sample of elliptical galaxies by adopting the chemical evolution model presented in \citet[hereafter P04]{P04}, describing the detailed time evolution of 21 different chemical elements. In that work, we generated the model galaxies by fine-tuning their initial parameters (star formation efficiency, infall time scale, effective radius and IMF) for different values of the mass, which yielded constraints on the formation and evolution of elliptical galaxies. Specifically, in accordance to the ``inverse wind scenario" \citep{matteucci1994}, we found that the best fitting models were those with higher star formation efficiency, larger effective radius and lower infall time scale in more massive galaxies. Moreover, at variance with what was concluded in \citetalias{P04}, we observed the necessity for a variation in the IMF as well, becoming top-heavier in more massive galaxies. As discussed in \citetalias{demasi2018}, we mainly ascribed this discrepancy - aside from the obvious consideration of using different data - to the operational definition of the quantities in play. The $[\alpha/Fe]$ ratios in the dataset we used in \citetalias{demasi2018} are related to the difference between the total metallicity $[Z/H]$ and the Fe abundance $[Fe/H]$, so that we derived a similar quantity from our model and compared it to the data. This quantity, although being consistent with observations, does not well represent the actual $[\alpha/Fe]$ ratio, since it also includes in the mixture of $\alpha$-elements other elements, such as C and N, which have a different behavior. In \citetalias{P04}, on the other hand, the comparison with observations was made by using the $[Mg/Fe]$ ratio directly predicted from the code, which is representative of the ``true'' $\alpha$-element behavior. As a matter of fact, in \citetalias{demasi2018}, when using the $[Mg/Fe]$ we obtain a better agreement with data (positive correlation with mass, although with a slightly flatter slope).\\
In this paper, we adopt a new dataset for the comparison, and we follow a different approach in generating the models, with the aim of better exploring the available parameter space. Instead of manually fine-tuning the parameters of the models, we assume a parameterization for the IMF, and for each choice of the latter we generate the models by varying all the initial parameters over a grid of values (see tables \ref{table:models_initialparameters_timeindependent} and \ref{table:models_initialparameters_timedependent}).\\
\begin{table*}
	\centering
	\caption{Possible values of the initial parameters used to generate the model galaxies in the time-independent cases (Models 01-04). For each choice of the IMF, we generated model galaxies using all the possible combinations of values reported in this table.}\label{table:models_initialparameters_timeindependent}
	\begin{tabular}[c]{lp{8cm}}
		Parameter & Value \\
		\hline
		Infall mass ($M_{\odot}$) & $5\times10^{9}$, $9\times10^{9}$, $1.62\times10^{10}$, $2.92\times10^{10}$, $5.25\times10^{10}$, $9.45\times10^{10}$, $1.70\times10^{11}$, $3.06\times10^{11}$, $5.51\times10^{11}$, $9.92\times10^{11}$, $1.79\times10^{12}$ \\
		Effective radius (kpc) & 1, 2, 3, 4, 5, 6, 7, 8, 9, 10 \\
		Star formation efficiency $\nu$ ($Gyr^{-1}$) & 5, 10, 15, 20, 25, 30, 35, 40, 45, 50 \\
		Infall time-scale $\tau$ ($Gyr$) & 0.2, 0.3, 0.4, 0.5
	\end{tabular}
\end{table*}
\begin{table*}
	\centering
	\caption{Possible values of the initial parameters used to generate the model galaxies in the time-dependent case (Model 05; see text). We generated model galaxies using all the possible combinations of values reported in this table.}\label{table:models_initialparameters_timedependent}
	\begin{tabular}[c]{lp{8cm}}
		Parameter & Value \\
		\hline
		Infall mass ($M_{\odot}$) & $5\times10^{9}$, $1\times10^{10}$, $2\times10^{10}$, $4\times10^{10}$, $8\times10^{10}$, $1.6\times10^{11}$, $3.2\times10^{11}$, $6.4\times10^{11}$, $1.3\times10^{12}$\\
		Effective radius (kpc) & 1, 3, 5 \\
		Star formation efficiency $\nu$ ($Gyr^{-1}$) & 5, 10, 30, 50, 70, 100 \\
		Infall time-scale $\tau$ ($Gyr$) & 0.2, 0.5\\
		$\mu_1$ & $0.4, 0.7, 1.0, 1.3$\\
		$\mu_2$ & $1.3, 1.6, 1.9, 2.3, 2.6, 2.9$\\
		$t_{switch}$ (Gyr) & 0.1, 0.3, 0.5, 0.8, 1.0
	\end{tabular}
\end{table*}
This paper is organized as follows.\\
In section \ref{sec:dataset}, we present the adopted dataset, in section \ref{sec:models} we describe our chemical evolution model, focusing on the comparison with the observed quantities, and describe the properties of the various adopted forms of the IMF.\\
In section \ref{sec:results}, we summarize the results of this work, indicating the IMFs which can provide the best fit to the dataset.\\
Finally, in section \ref{sec:MLratios} we present the analysis we performed on the calculation of the M/L ratios predicted by our best-fitting models, in an attempt to obtain further constraints.

\section{Dataset}\label{sec:dataset}

The dataset used in this work is a subsample of the catalogue of ETGs presented in \cite{labarbera2010}.\\
Details on the selection of the general dataset can be found in \citetalias{labarbera2013} and the final state of the dataset used in this work can be found in \citet[][hereafter R18]{rosani2018}. Briefly, we analyze stellar galaxy properties inferred from spectra stacked in central velocity dispersion from 20996 $(0.05<z<0.095)$ early-type galaxies, extracted from the 12th Data Release of the SDSS. The stacked spectra were collected to ensure a S/N ratio of the order of a few hundreds, needed to obtain constraints on the IMF from gravity-sensitive features \citep{conroyvandokkum2012a}.\\
The environment information for the galaxies in the dataset are derived from the catalog of \cite{wang2014}.
\begin{figure*}
	\includegraphics[width=.85\textwidth]{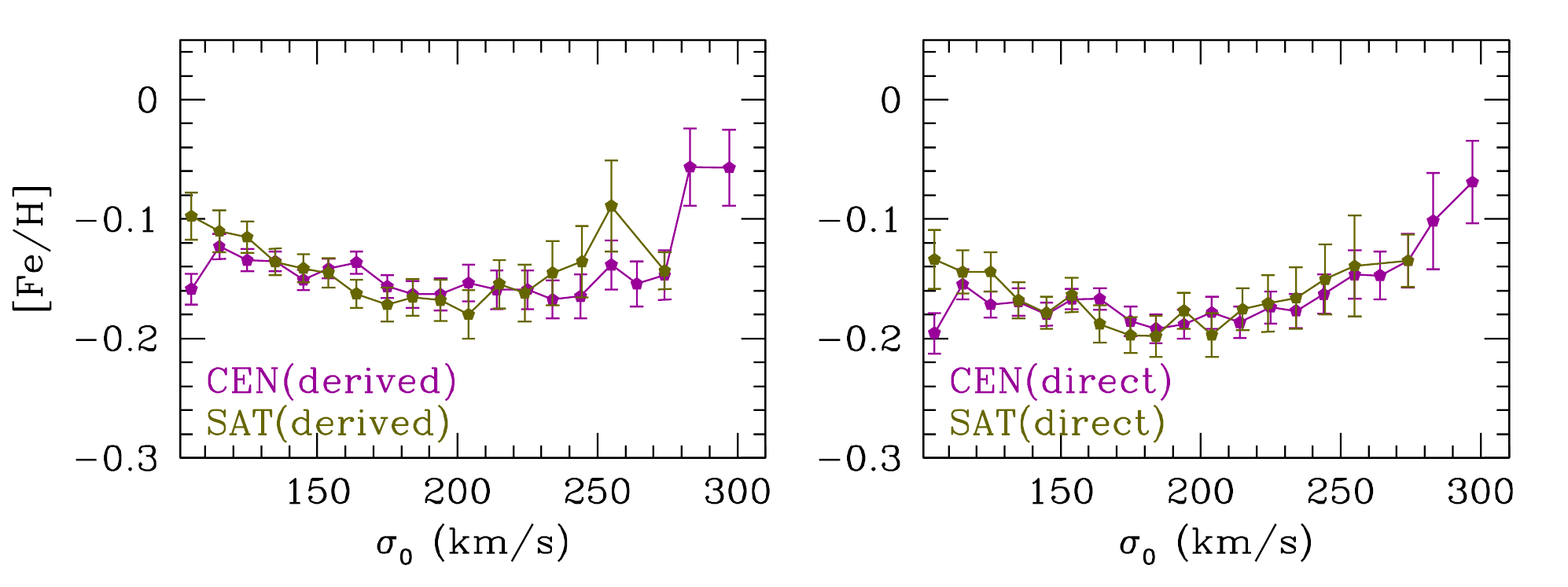}
	\caption{Comparison of $[Fe/H]$ values obtained using the ``direct'' method (fit of Fe5270 and Fe5335 iron lines)  and the ``derived'' method using the values presented in this paper}
	\label{fig:data_FeH_comparison}
\end{figure*}
As detailed in \citetalias{rosani2018}, stellar population properties and chemical abundances for various elements have been derived from the stacked spectra by fitting the equivalent widths of a set of line indices to the equivalent widths predicted by synthetic stellar population (SSP) models. The models used for the fitting are the EMILES SSPs of \cite{vazdekis2016}, with variable IMF slope, age, and total metallicity. Two approaches have been explored in the fitting by \citetalias{rosani2018}: i) the case in which only age, metallicity and IMF-sensitive indices were used; ii) the case in which, additionally to the ones of the previous case, indices sensitive to abundance pattern of different elements (among which $[Mg/Fe]$) were used. In this work, the values of IMF slope, age and total metallicity $[Z/H]$ used are those derived by \citetalias{rosani2018} for case i).\\
Since ETGs are found to be not solar-scaled in abundance pattern, but the EMILES models are, the abundances obtained in the fit for each stacked spectrum had to be corrected to reflect the $\alpha$-enhancement of ETGs.
The method used to derive [Mg/Fe] in \citetalias{rosani2018} is the same as in \citetalias{labarbera2013}. First, one estimates a proxy for $[Mg/Fe]$, defined as the difference between the metallicity derived from the $Mgb5177$ index and the metallicity derived using the $Fe_3$ index at fixed age (see \cite{trager1998} and \cite{kuntschner2000} respectively for index definition). Then, the proxy is converted into $[Mg/Fe]$. The conversion factor is established by comparing the proxy to $[Mg/Fe]$ estimates obtained with the ``direct method'', based on \cite{thomas2010} stellar population models with varying $[\alpha/Fe]$ (see Fig.6 of \citetalias{labarbera2013}). For the sample analyzed in \citetalias{labarbera2013} (and thus in \citetalias{rosani2018}), the conversion is very accurate, with an rms of $\approx0.025$ dex, i.e. well within the differences arising in a direct estimate of $[Mg/Fe]$ when adopting different stellar population models (see, e.g., Fig. 18 of \citealp{conroy2014}). The comparison of $[Mg/Fe]$ estimates based on the proxy and the direct method has been discussed in \citetalias{labarbera2013} and \cite{vazdekis2015}.
Finally, to obtain [Fe/H] for each of the stacked spectra, we invert the relation linking $[Mg/Fe]$, $[Fe/H]$ and total metallicity:
\begin{equation}
	[Z/H] = [Fe/H]+0.75 \times [Mg/Fe].
\end{equation}
The factor of 0.75 is the same as for the $\alpha$-enhanced MILES models of \cite{vazdekis2015}. For this set of models, theoretical $\alpha$-enhanced stellar spectra are produced by a uniform enhancement of $[X/Fe]=+0.4$, for elements O, Ne, Mg, Si, S, Ca and Ti, assuming the solar mixture from \cite{grevesse1998}. The above factor refers to these calculations (and is mostly driven by O and Mg). The factor varies for different studies in the literature, depending on the adopted partition table of enhanced vs. depressed elements in the theoretical calculations. For instance, in the case of \cite{thomas2003stellar} stellar population models, the conversion factor is as high as 0.94, most likely because the authors also include C, N, and Na, besides O, Mg, Si, Ca, Ti, in the $\alpha$ enhanced group. Since we estimate $[\alpha/Fe]$ from Mgb and Fe lines, our abundance estimates mostly reflect the behaviour of $[Mg/Fe]$ and other elements that (closely) track Mg. Hence, the scaling factor adopted in \cite{vazdekis2015} should suffice our purposes. As a test, we have also computed $[Fe/H]$ directly, as the metallicity estimate obtained when fitting only Fe line strengths. Figure \ref{fig:data_FeH_comparison} shows the fit obtained using the ``direct'' method (by fitting the mean value of equivalent width of the Fe5270 and Fe5335 iron lines) and the ``derived'' method using the values presented in this paper. The comparison shows good agreement between the ``direct'' values and the estimate of $[Fe/H]$ obtained with our adopted scaling factor.
\begin{figure*}
	\centering
	\includegraphics[width=.4\textwidth]{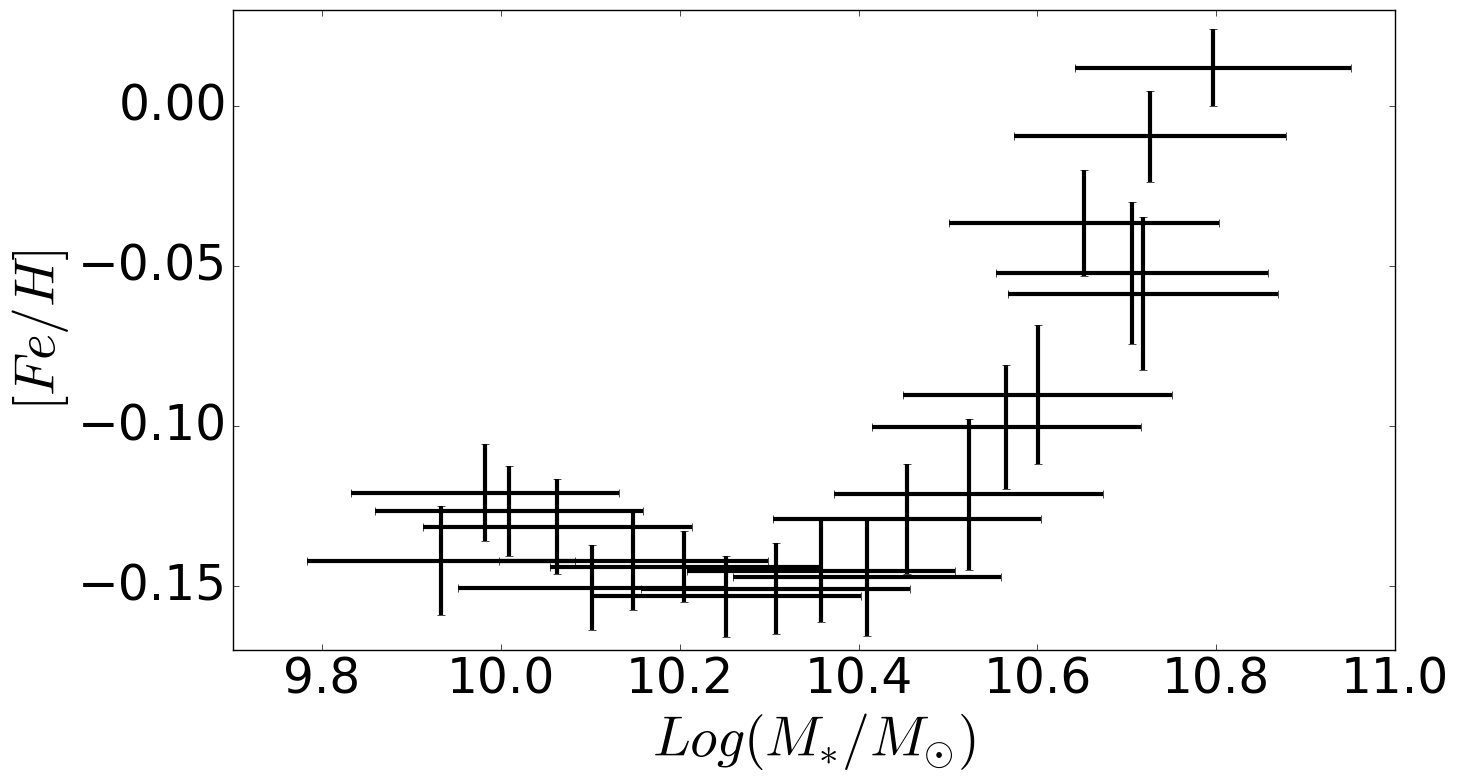}
	\includegraphics[width=.4\textwidth]{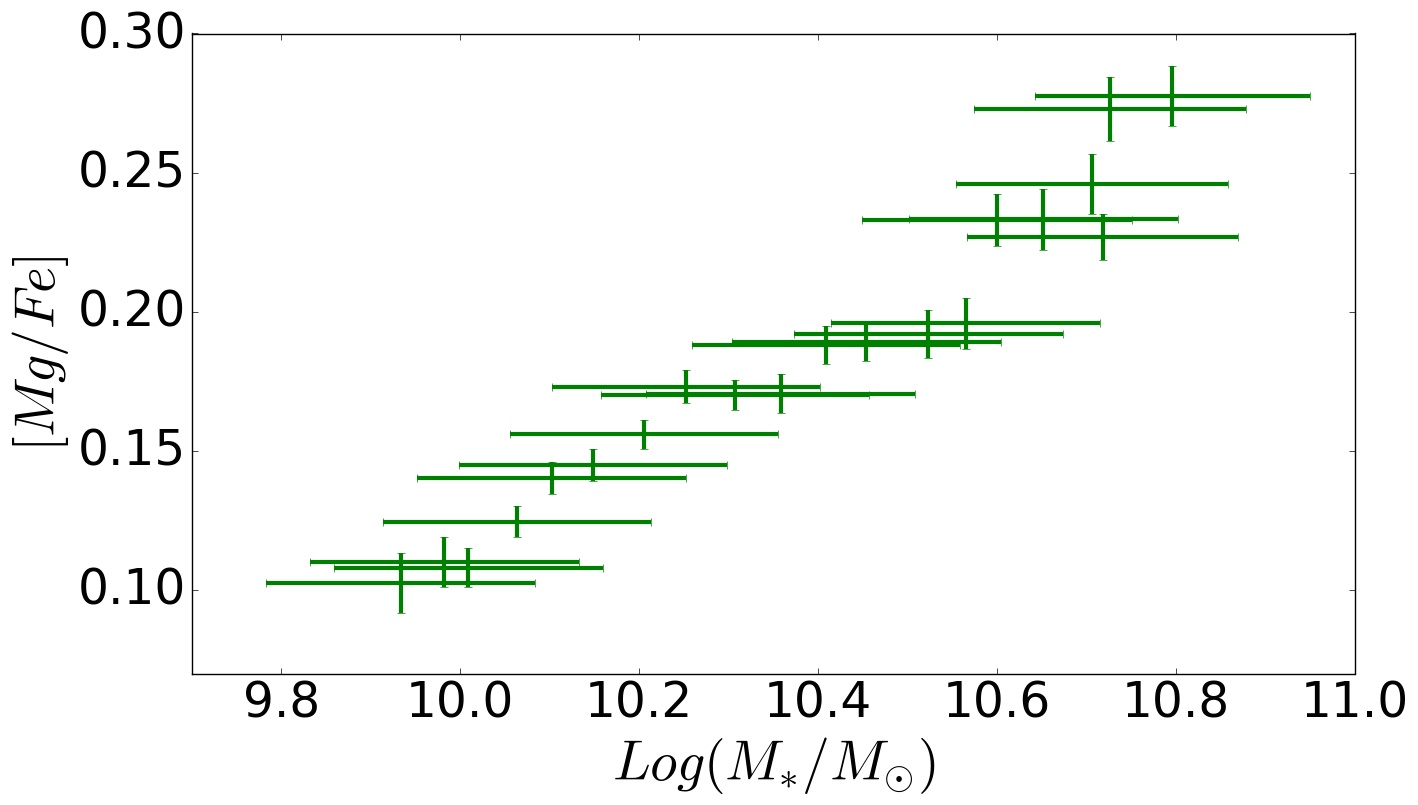}
	\caption{$[Fe/H]$ and $[Mg/Fe]$ ratios variation with stellar mass in the 20 mass bins provided in the dataset.}
	\label{data_overview01}
\end{figure*}
Both the [Fe/H] and the [Mg/Fe] abundances are compared to the analogous ratios as directly predicted by our chemical evolution code. Specifically, we use these values to test the mass-metallicity and $[Mg/Fe]$-mass relation predicted by our chemical evolution model.\\
In Figure \ref{data_overview01}, we show the variation of $[Fe/H]$ and $[Mg/Fe]$ as a function of galaxy mass in the SDSS stacked spectra, with their 1-$\sigma$ uncertainties. Since the stacking in \citetalias{rosani2018} is originally performed in central velocity dispersion $(\sigma_0)$ bins, we derived the stellar mass associated to a given stacked spectrum. Specifically, we took the stellar masses listed in the group catalog of \cite{wang2014}; as described by \cite{yang2007}, stellar masses are derived from the relation between stellar mass- to-light ratio and colour of \cite{bell2003}.\\
	\begin{figure*}
		\includegraphics[width=.45\textwidth]{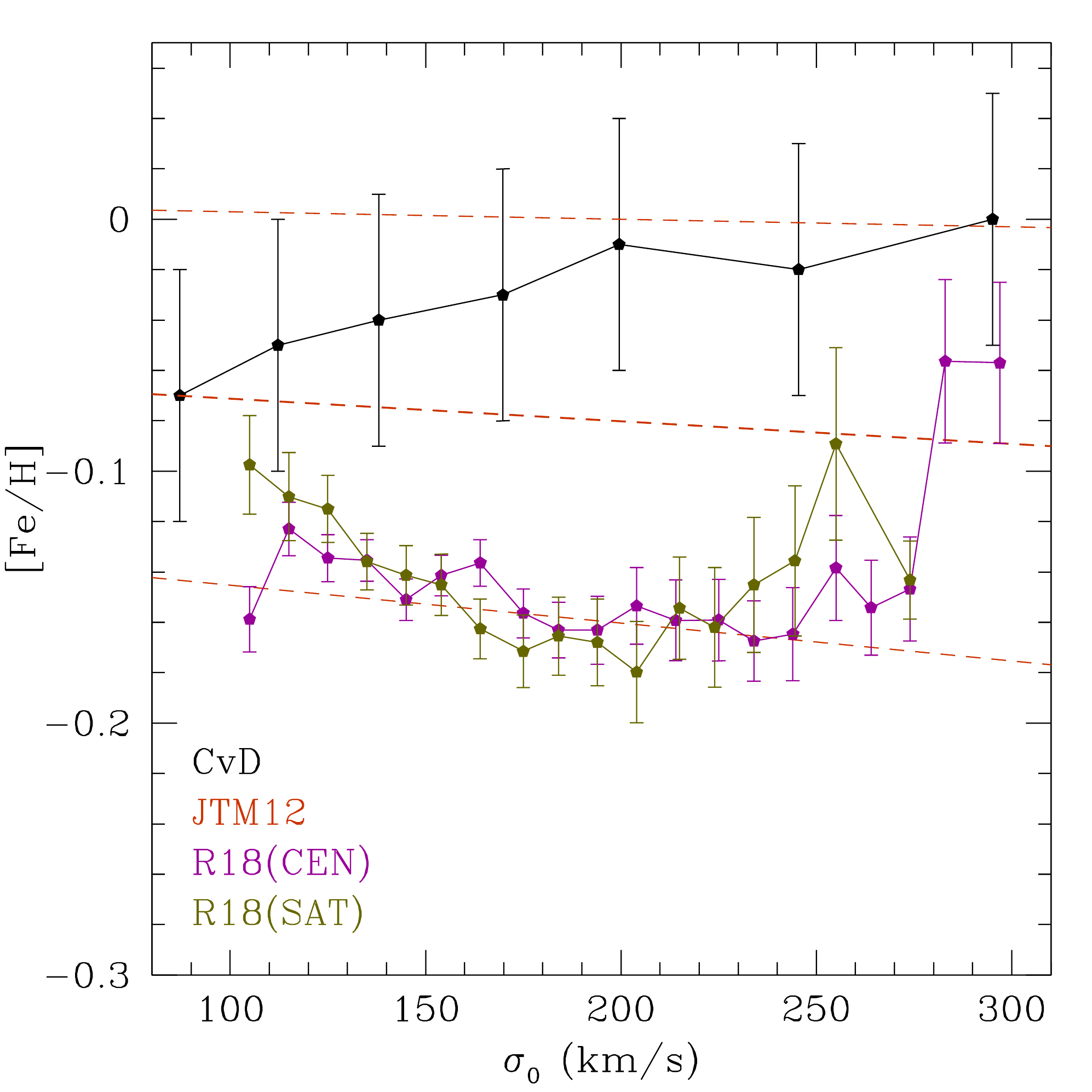}
		\caption{Comparison of the $[Fe/H]$-mass relation for the R18 CEN and SAT samples with the ones presented in \protect\cite{conroy2014}, in black, and \protect\cite{johansson2012}, in red.}
		\label{fig:data_FeH_relations}
	\end{figure*}
The resulting relation shows a slight turnover at $LogM = 10.3$, with a rise of iron abundance toward lower masses. Figure \ref{fig:data_FeH_relations} shows a comparison of the $[Fe/H]$-mass relation for the R18 CEN and SAT samples, with two literature relations by \cite{conroy2014}, in black, and \cite{johansson2012}, in red. While trends presented in literature do not show an upturn at low masses, one should also notice that our error bars on $[Fe/H]$ are far smaller than those from previous works. Therefore, the upturn might be just seen because of the very large number of spectra that we stack together in each sigma bin, and the corresponding very high S/N ratio (as discussed in \citetalias{rosani2018}) of our stacked spectra, compared to previous studies.\\
However, notice that for the sample of CENs (compared to SATs), the upturn is far less evident (almost absent at all). Still, our matching procedure does not provide significant differences between CENs and SATs, implying that the detailed shape of the $[Fe/H]$-mass relation at low mass is not affecting the main conclusions of our work.

\section{Models}\label{sec:models}
In this section, we present the implementation of our chemical evolution model. We start by giving a brief description of the model itself, of the calibrations needed to compare the results with the data, and we present the various forms of the IMF we tested in this work.

\subsection{Chemical evolution model}
A detailed description of the chemical evolution model adopted in this paper can be found in \citetalias{P04} and \citetalias{demasi2018}. Here, we briefly summarize its properties.\\
The model follows the detailed evolution with time of 21 different chemical elements in the various shells the galaxy is divided, by solving the equation of chemical evolution \citep[CEQ -][]{matteucci1986,matteucci1995} for each of the elements:
\begin{equation}\label{CEQ}
\begin{split}
\frac{dG_{i}(t)}{dt} = &-\psi(t)\,X_{i}(t) + \\ &+\int_{0.8M_{\odot}}^{3M_{\odot}}\psi(t-\tau_{m})\,Q_{mi}(t-\tau_m)\,\varphi(m)\,dm\, +\\
&+A\,\int_{3M_{\odot}}^{16M_{\odot}}dm\,\varphi(m)\times\\
&\times\left[\int_{\mu_{m}}^{0.5}f(\mu)\psi(t-\tau_{m_2})Q_{mi}(t-\tau_{m_2})d\mu \right]+\\
&+(1-A)\int_{3M_{\odot}}^{16M_{\odot}}\,\psi(t-\tau_{m})\,Q_{mi}(t-\tau_m)\,\varphi(m)\,dm+\\
&+\int_{16M_{\odot}}^{M_U}\,\psi(t-\tau_{m})\,Q_{mi}(t-\tau_m)\,\varphi(m)\,dm+\\
&+\left[ \frac{dG_i(t)}{dt} \right]_{infall}
\end{split}
\end{equation}
where each of the four integrals provides the quantity of the i-th chemical element restored to the ISM by dying stars of various masses. Single stars (both in the $0.8\,M_{\odot}-3\,M_{\odot}$ and the $3\,M_{\odot}$ - $16\,M_{\odot}$ mass ranges), binary systems generating type Ia SNe (with total mass $M_{B_m}$ in the $3\,M_{\odot}$-$16\,M_{\odot}$ range), and core collapse SNe ($m>16\,M_{\odot}$).\\
In the second integral, we made use of the Type Ia SNe rate for the single degenerate scenario \citep{whelan1973} as defined in \cite{greggio1983,matteucci1986,matteucci2001}:
\begin{equation}\label{SNI_rate}
R_{SNIa} = A\,\int_{M_{B_m}}^{M_{B_M}}dM_B\,\varphi(M_B)\,\int_{\mu_m}^{0.5}\,f(\mu)\,\psi(t-\tau_m)\,d\mu
\end{equation}
The mass fraction of the secondary star (the originally least massive one) with respect to the total mass of the binary system $\mu\equiv M_2/M_B$ is distributed according to:
\begin{equation}
f(\mu)=2^{\gamma+1}(\gamma + 1)\,\mu^{\gamma}
\end{equation}
with $\gamma=2$, and the free parameter A is constrained in order to reproduce the present-day observed rate of Type Ia SNe \citep{cappellaro1999}.\\
The core-collapse SNe rate is
\begin{equation}\label{SNII_rate}
\begin{split}
R_{cc} &= (1-A)\,\int_{8}^{16}dm\,\varphi(m)\,\psi(t-\tau_m) + \\
& + \int_{16}^{M_{WR}}\,dm\,\varphi(m)\,\psi(t-\tau_m) + \\
& + \int_{M_{WR}}^{M_{U}}\,dm\,\varphi(m)\,\psi(t-\tau_m) + \\
& + \alpha_{Ib/c}\,\int_{12}^{20}\,\varphi(m)\,\psi(t-\tau_m)
\end{split}
\end{equation}
where the first two integrals provide the Type II SNe rate, while the third and the fourth one express the Type Ib/c SN rate for single stars and binary systems, respectively. Again, $\alpha_{Ib/c}$ is a free parameter, representing the fraction of stars in the considered mass range which can actually produce Type Ib/c SNe, and its value is modified to reproduce the observed rate.\\
The quantity:
	\begin{equation*}
	X_i(t) \equiv \frac{M_{i}}{M_{gas}}
	\end{equation*}
is the abundance by mass of the i-th chemical species in the ISM, with the normalization
	\begin{equation*}
	\sum_{i=1}^{N}X_{i}=1
	\end{equation*}
while
	\begin{equation}
	G_{i}(t) = X_{i}(t)\, \rho_{gas}(t) 
	\end{equation}
is the ratio between the mass density of the element i at the time t and its initial value.\\
The star formation rate $\psi(t)$ is assumed to be described by a Kennicutt law \citep{kennicutt1998}, until the time at which the thermal energy, injected from stellar winds and SNe, overcomes the binding energy of the gas. At this point, a galactic wind starts, driving away the residual gas and quenching the star formation (\citealp{larson1974},\citetalias{P04}):
	\begin{equation}\label{eq_SF}
	\psi(t)=	
	\begin{cases}
	\displaystyle
	\nu\, \rho_{gas}(t) \qquad &\text{before GW}\\
	0\qquad &\text{after GW}\\
	\end{cases}
	\end{equation}
with a star formation efficiency $\nu$ getting higher in more massive galaxies \citep[``inverse wind model'' - ][]{matteucci1994,matteucci1998}.
In order to determine the thermal energy in the ISM and the time of the onset of the galactic wind, the code evaluates the contribution of both Type I and II SNe, assuming an average efficiency of energy release of the $\approx20\%$ between the two types \citep{cioffi1988,recchi2001,pipino2002}.\\
The assumed stellar yields are the same adopted in \citetalias{P04} and \citetalias{demasi2018}.

\subsection{Comparison between data and model output}\label{sec:data_model_comparison}

As detailed in \citetalias{demasi2018}, a comparison between the results of our chemical evolution model and data is in general only possible after taking an additional step. Specifically, chemical abundance estimates in ellipticals are mainly determined by the composition of stars dominating the visual light of the galaxy, whereas our code provides the evolution with time of the abundances in the ISM.\\
From the latter quantity, one has to perform an average, either on mass or luminosity. \cite{yoshii1987}, \cite{gibson1997a} and \cite{matteucci1998} showed that there is no significant difference for massive galaxies $(M>10^9\,M_{\odot})$ between light and mass weighted abundances. To this regard, although some of our models produce final stellar masses as low as $\approx10^8\;M_{\odot}$, the stellar mass in the data never gets lower than $\approx10^{9.8}\;M_{\odot}$ (see Fig. \ref{data_overview01}), so that when matching models and data, only models with stellar masses higher than this value have been retained.\\
For this reason, and to compare with our previous work (\citetalias{demasi2018}), when analyzing abundances for models matching the observed data, we always applied mass-weighted estimates, which are a natural outcome of the chemical evolution code, according to the relation:
	\begin{equation}\label{eq:mass_average}
		<X/H>_{mass} \equiv \frac{1}{M_0} \, \int_{0}^{M_0}Z(M)\,dM
	\end{equation}
where $M_0$ is the total mass of stars ever born contributing to light at the present time. However, we further tested the validity of this approach, by computing the light-averaged metallicities for a Salpeter IMF. The results, summarized in Appendix \ref{sec:mass_light_averages}, show that the light-averaged metallicities are slightly higher than the mass-weighted ones. The difference is almost constant, and always lower than 0.1 dex. It is worth noting that our conclusions cannot be significantly affected by this shift, since we are not interested in absolute abundances, but in abundance trends.\\
Using equation \ref{eq:mass_average} allows us to obtain abundance predictions that can be compared to the observed ones.

\subsection{Adopted IMFs}\label{sec:adopted_IMFs}

In this paper, we expand the investigation of the effects of different IMFs on the evolution of elliptical galaxies we previously carried out in \citetalias{demasi2018}, by testing the IMF parameterizations adopted in the previous paper, as well as some new IMF models.\\
Specifically, the adopted IMFs are:
\begin{itemize}
	\item \textbf{Model 01:}\\
	We obtained these galaxy models by using a fixed \citep{salpeter1955} IMF and by considering all possible combinations of values reported in Table \ref{table:models_initialparameters_timeindependent} for the initial parameters.
	
	\item \textbf{Model 02:}\\
	In \citetalias{demasi2018}, we applied the prescriptions of the ``inverse wind'' model \citep{matteucci1994,matteucci1998,matteucci2012}, where the star formation process is more efficient and shorter in more massive galaxies, to reproduce the higher $[\alpha/Fe]$ observed in more massive galaxies (``downsizing'' in star formation). This assumption, however, proved to be insufficient to reproduce the slope of the observed trends, so that we decided to test a variable IMF, switching to different  parameterizations in different mass ranges; specifically, the IMF variation which provided the best results was:
	\begin{itemize}
		\item \textbf{\cite{scalo1986}}: we used the approximate expression adopted in \cite{chiappini1997}:
		\begin{equation}
		\varphi(m)\,\propto
		\begin{cases}
		m^{-2.35}\qquad & 0.1 \leq m/M_{\odot} < 6\\
		m^{-2.7}\qquad & 6 \leq m/M_{\odot} \leq 100\\
		\end{cases}
		\end{equation}
		\item \textbf{\cite{salpeter1955}}, which is a simple power-law:
		\begin{equation}
		\varphi(m)\,\propto m^{-2.35}\qquad 0.1 \leq m/M_{\odot} < 100\\
		\end{equation}
		\item \textbf{\cite{chabrier2003}}:
		\begin{equation}
		\varphi(m)\,\propto
		\begin{cases}
		e^{ - \frac{(Log(m) - Log(0.079))^2}{2(0.69)^2}}     \qquad & 0.1 \leq m/M_{\odot} < 1\\
		m^{-2.2}\qquad & 1 \leq m/M_{\odot} \leq 100\\
		\end{cases}
		\end{equation}
	\end{itemize}
	Models 02 are produced by assuming the same IMF variation, as well as the parameters value reported in Table \ref{table:models_initialparameters_timeindependent}.

	\item \textbf{Model 03:}\\
	In these Models, we tested the effect of assuming an Integrated Galactic IMF \citep{recchi2009,vincenzo2014,weidner2010}.\\
	The IGIMF is obtained by combining the IMF describing the mass distribution of new-born stars within the star clusters - where star formation is assumed to take place - with the mass distribution of star clusters themselves (embedded cluster mass function, ECMF); assuming for the latter the form (with $\beta\approx 2$)
	\begin{equation}
	\xi_{ecl} \propto M_{ecl}^{\beta}
	\end{equation}
	the IGIMF is then defined as \citep{weidner2011,vincenzo2015}:
	\begin{equation}\label{eq:IGIMF_def}
	\begin{split}
	&\xi_{IGIMF}(m,t) \equiv\\ &\equiv\int_{M_{ecl}^{min}}^{M_{ecl}^{max}(\psi(t))}\,\varphi(m<m_{max}(M_{ecl}))\,\xi_{ecl}(M_{ecl})\,dM_{ecl}
	\end{split}
	\end{equation}
	with the mass normalization:
	\begin{equation}\label{eq:IGIMF_normalization}
	\int_{m_{min}}^{m_{max}}\,dm\,m\,\xi_{IGIMF}(m) = 1
	\end{equation}
	\begin{figure*}	
		\centering
		\includegraphics[width=.5\linewidth]{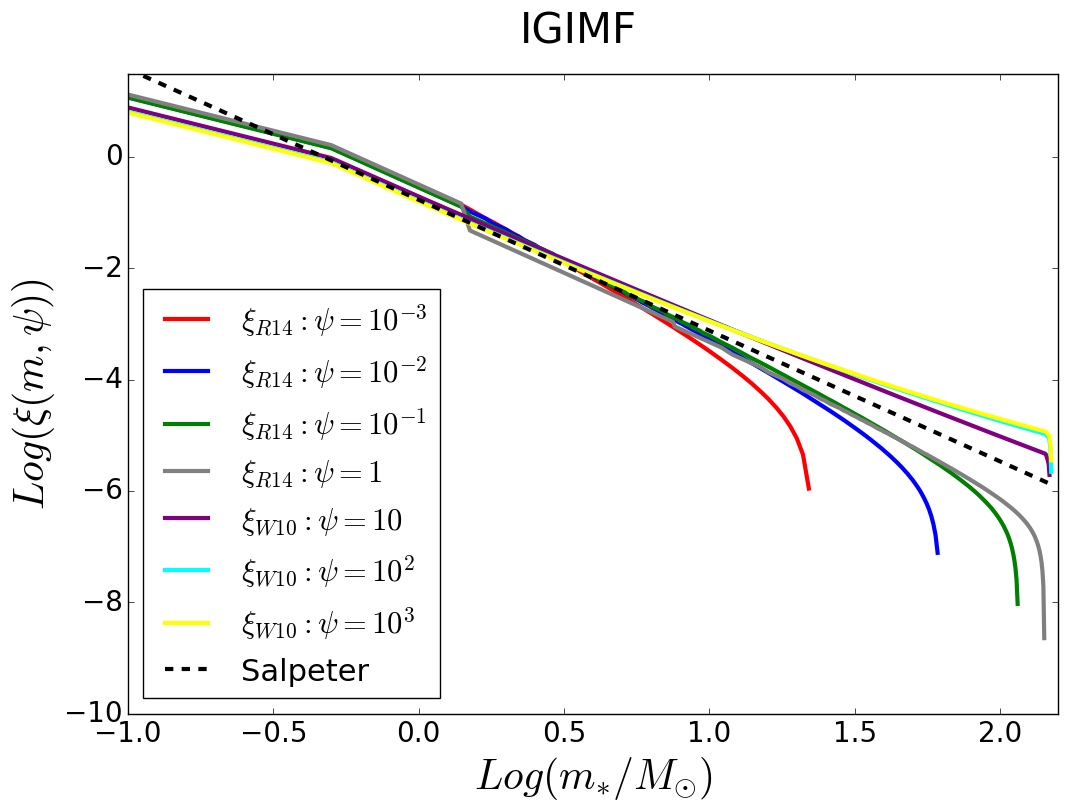}
		\caption{IGIMF for different SFRs (solid colours lines) and a canonical Salpeter IMF (black dashed line).}
		\label{fig:IGIMF_comparisons}
	\end{figure*}
	Briefly, for higher SFR values $M_{ecl}^{max}$, the maximum mass of the stellar clusters where star formation is taking place, increases, and hence the maximum mass of stars that can be formed within the cluster is larger as well; defined this way, the IGIMF becomes top-heavier as the SFR increases. This is shown in figure \ref{fig:IGIMF_comparisons}, where we compare the IGIMF for different star formation rates (SFRs), with the Salpeter IMF.\\

	\item \textbf{Model 04:}\\
	In these models, we tested the effect of adopting a low-mass tapered ("bimodal") IMF, as defined in \cite{vazdekis1997,vazdekis2003}.\\
	In this formulation, the IMF is defined as 
	\begin{equation}
	\xi(m)= \beta
	\begin{cases}
	m_1^{- \mu} \qquad 0.1 &< m/M_{\odot} < 0.2\\
	p(m) \qquad 0.2 &< m/M_{\odot} < 0.6\\
	m^{- \mu} \qquad 0.6 &< m/M_{\odot} < 100
	\end{cases}
	\end{equation}
	where $m_1=0.4$, and $p(m)$ is a third degree spline, i.e. 
	\begin{equation}
	p(m)=(A + B\,m + C\,m^2 + D\,m^3 )
	\end{equation}
	whose normalization constants are determined by solving the following boundary conditions:
	\begin{equation*}
	\begin{cases}
	p(0.2) = m_1^{-\mu}\\
	p'(0.2) = 0\\
	p(0.6) = 0.6^{-\mu}\\
	p'(0.6) = -\mu \; 0.6^{(-\mu-1)}\\
	\end{cases}
	\end{equation*}
	\begin{figure*}
		\includegraphics[width=.5\textwidth]{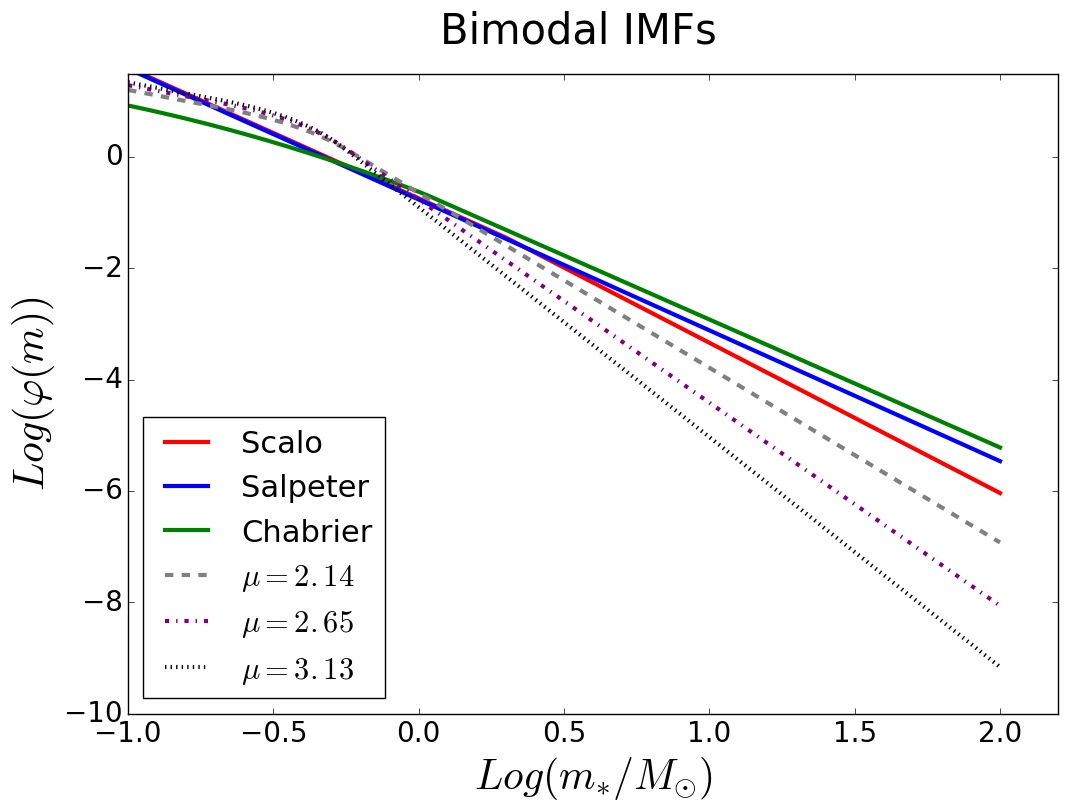}
		\caption{Comparison between bimodal IMF with varying slope $\mu$, and the IMFs used in our previous work (i.e. a Scalo, Salpeter, and Chabrier IMF;  see the inset panel).}
		\label{fig:comparison_canonical_bimodal_IMF}
	\end{figure*}	
	Notice that for $\mu=1.3$, the bimodal IMF closely matches a \cite{kroupa2001} distribution. For $\mu>1.3$, this IMF becomes more and more bottom-heavy, while for $\mu<1.3$ the IMF is top-heavy. We tested the effects of the bimodal IMF by assuming an increasing value for the slope $\mu$ (namely, a bottom heavier IMF) in more massive galaxies. Figure \ref{fig:comparison_canonical_bimodal_IMF} compares the bimodal IMFs with those adopted in our previous work (i.e. Models 01 and 02; see above).

	\item \textbf{Models 05}:\\
	In this final set of models, we tested a explicitly time dependent form for the bimodal IMF, as described in \cite{weidner2013} and \cite{ferreras2015}, by assuming that the slope value $\mu$ changes from an initial value $\mu_1$ to a final value $\mu_2$ after a time interval $t_{switch}$ (the IMF switches from top to bottom-heavy, so that by construction $\mu_2>\mu_1$). \\
	The Models are obtained by different combinations of $\mu1$, $\mu2$ and $t_{switch}$ values, summarized in table \ref{table:models_initialparameters_timedependent}.\\
\end{itemize}
In \citetalias{demasi2018}, we created the galaxy models manually, i.e. we selected a limited number ($\approx6-8$) of initial infall mass values, and we fine-tuned the other parameters of the code accordingly to reproduce the data.\\
\begin{figure*}
	\includegraphics[width=.45\textwidth]{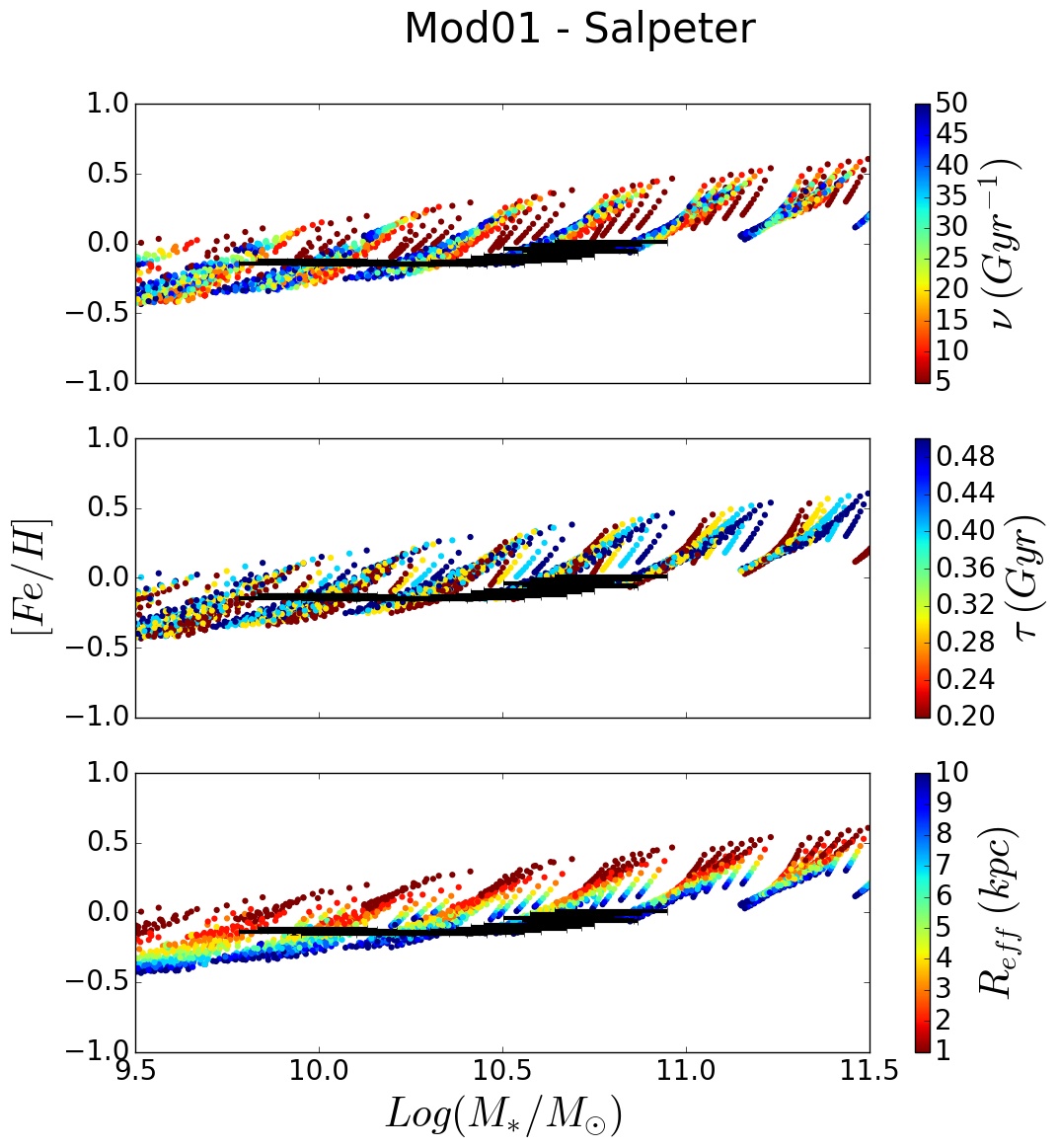}
	\includegraphics[width=.45\textwidth]{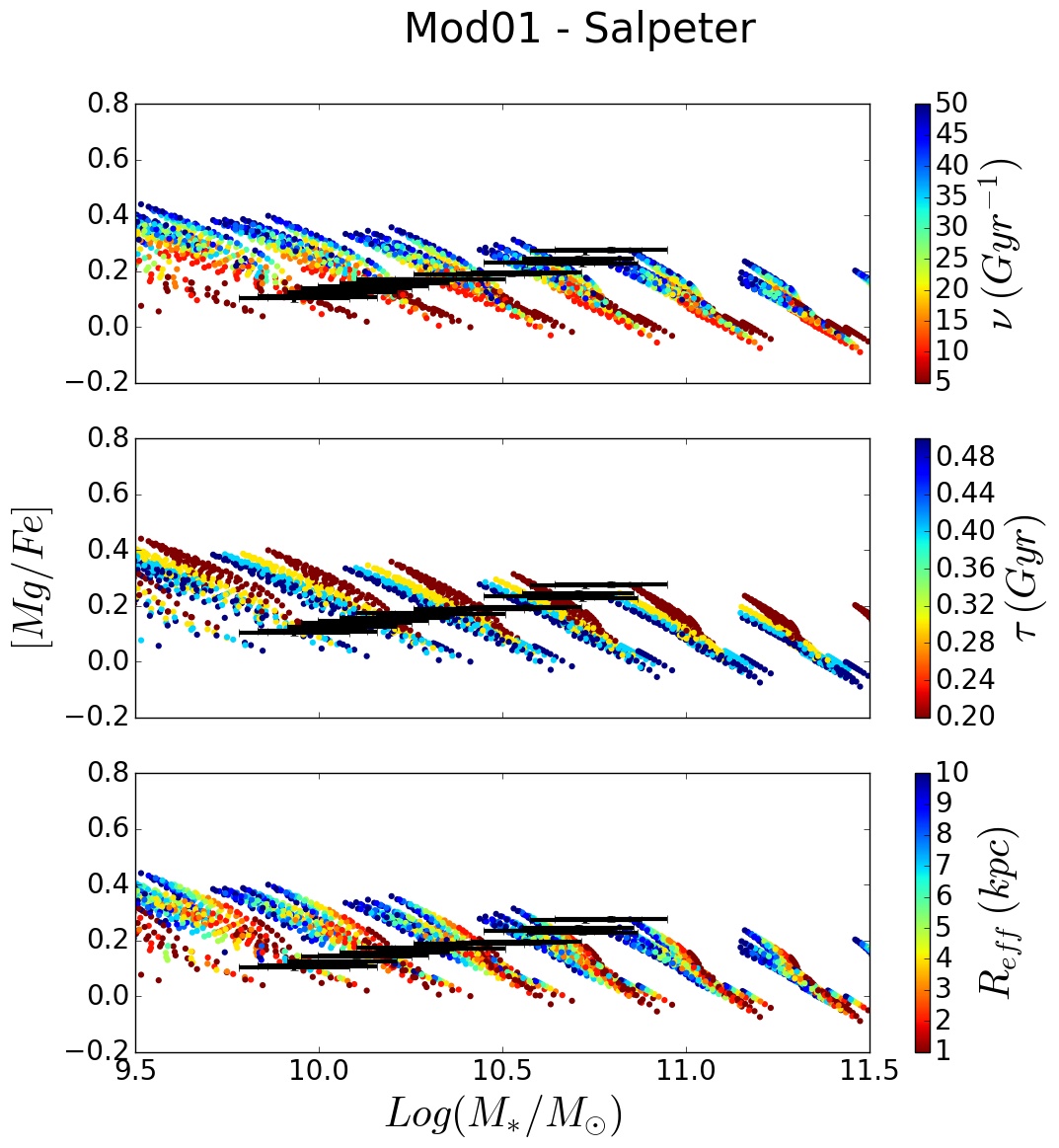}
	\caption{$[Fe/H]$ and $[Mg/Fe]$ ratios (left and right panel, respectively) for model galaxies, obtained by varying the model initial parameters over the grid of values reported in table \ref{table:models_initialparameters_timeindependent}, and assuming a fixed Salpeter IMF (Models 01). The plots show the variation of chemical abundances with total stellar mass, and are color coded to further show the dependance on the star formation efficiency $\nu$ (top panels), infall time-scale $\tau$ (central panels) and effective radius $R_{eff}$ (bottom panels). The black crosses represent the analogous quantities in observed data, with the corresponding error bars.}
	\label{fig:model01_properties_01}
\end{figure*}
\begin{figure*}
	\includegraphics[width=.45\textwidth]{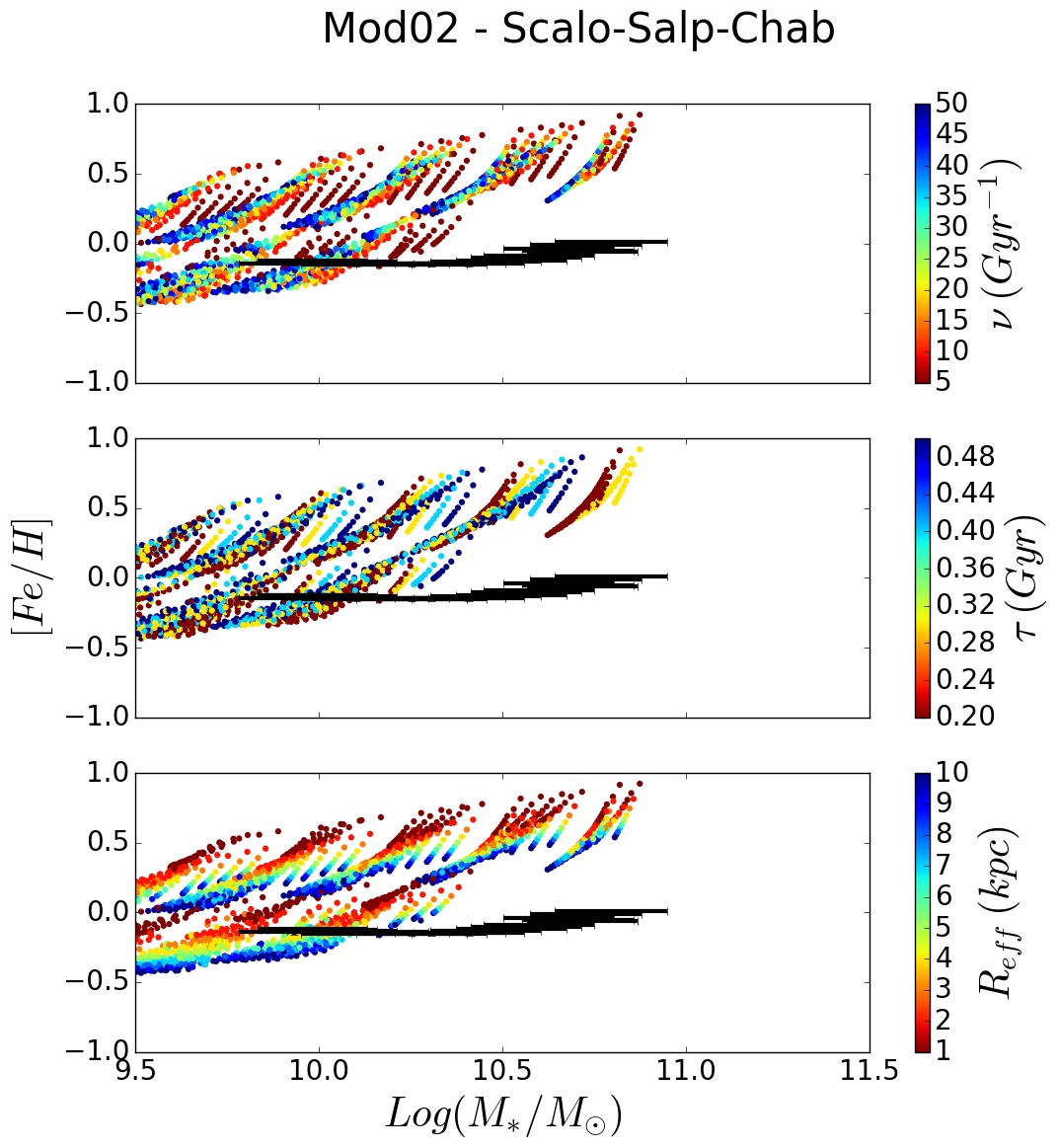}
	\includegraphics[width=.45\textwidth]{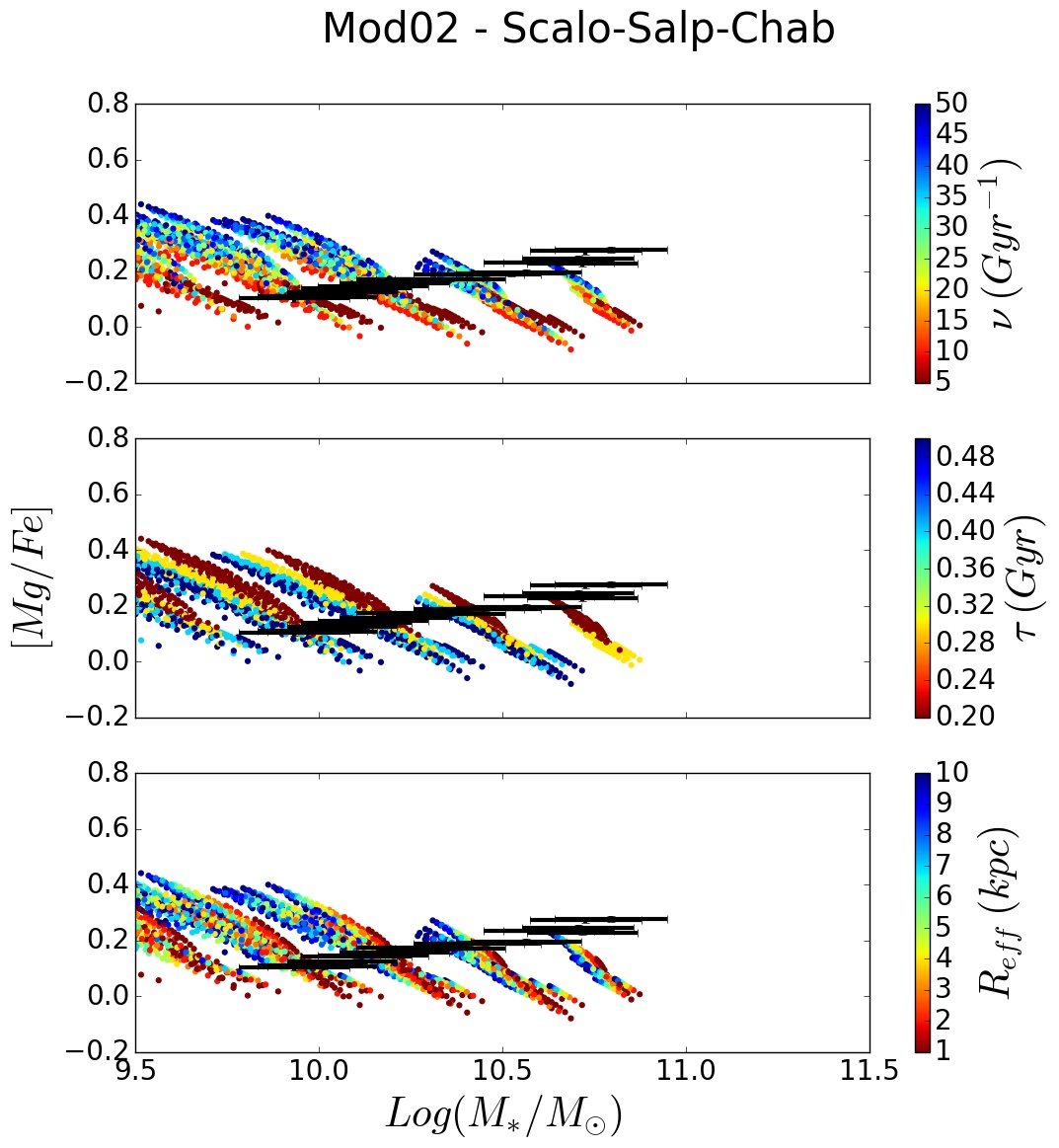}
	\caption{Same as figure \ref{fig:model01_properties_01}, but for Model 02, where we assume an IMF that varies with galaxy mass, becoming top-heavier in more massive galaxies (see sec. \ref{sec:adopted_IMFs}).}
	\label{fig:model02_properties_01}
\end{figure*}
\begin{figure*}
	\includegraphics[width=.45\textwidth]{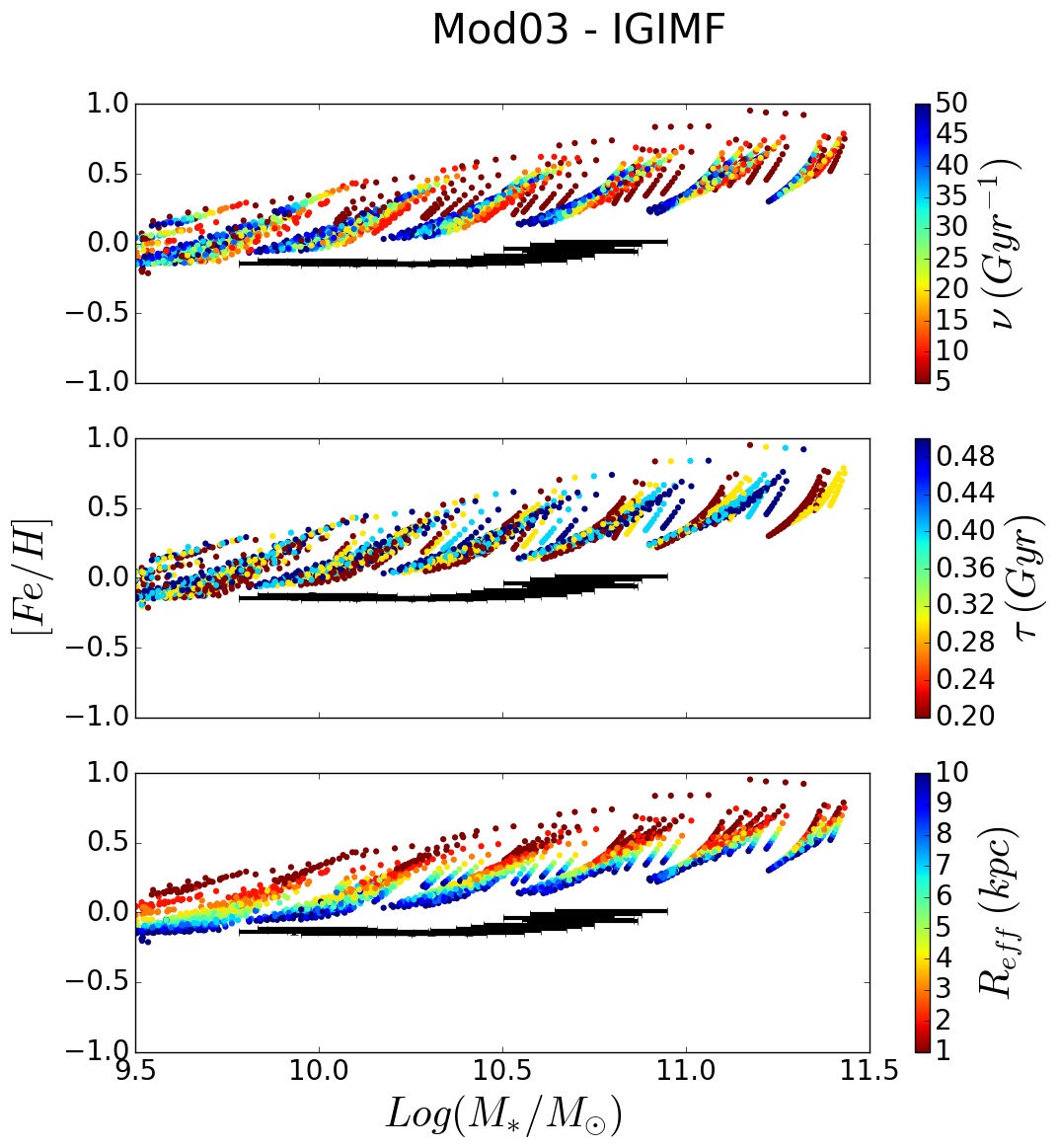}
	\includegraphics[width=.45\textwidth]{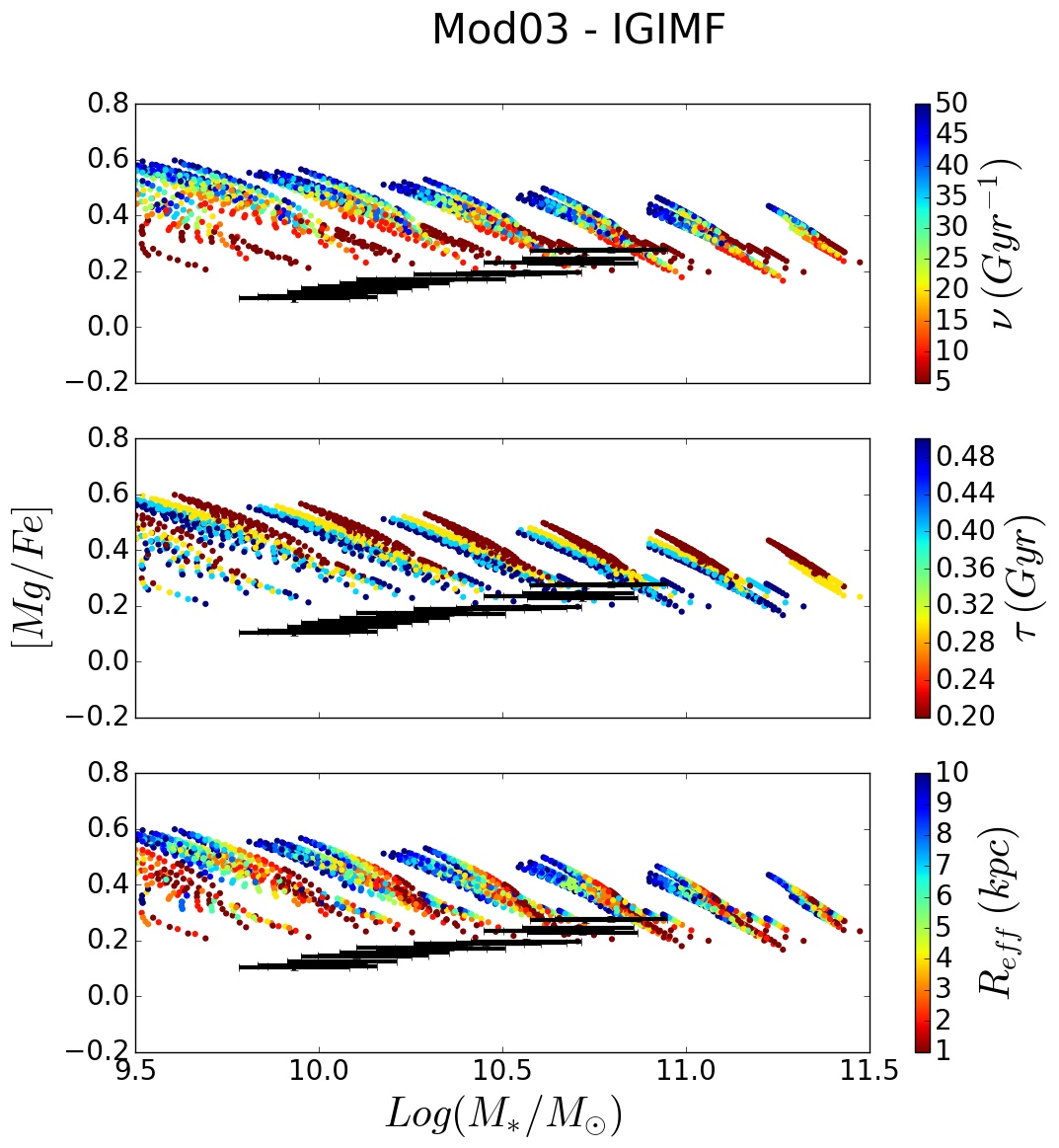}
	\caption{Same as figures \ref{fig:model01_properties_01} - \ref{fig:model02_properties_01}, for Models 03 assuming an IGIMF, becoming top-heavier for higher SFR values in more massive galaxies (see sec. \ref{sec:adopted_IMFs}).}
	\label{fig:model03_properties_01}
\end{figure*}
\begin{figure*}
	\includegraphics[width=.45\textwidth]{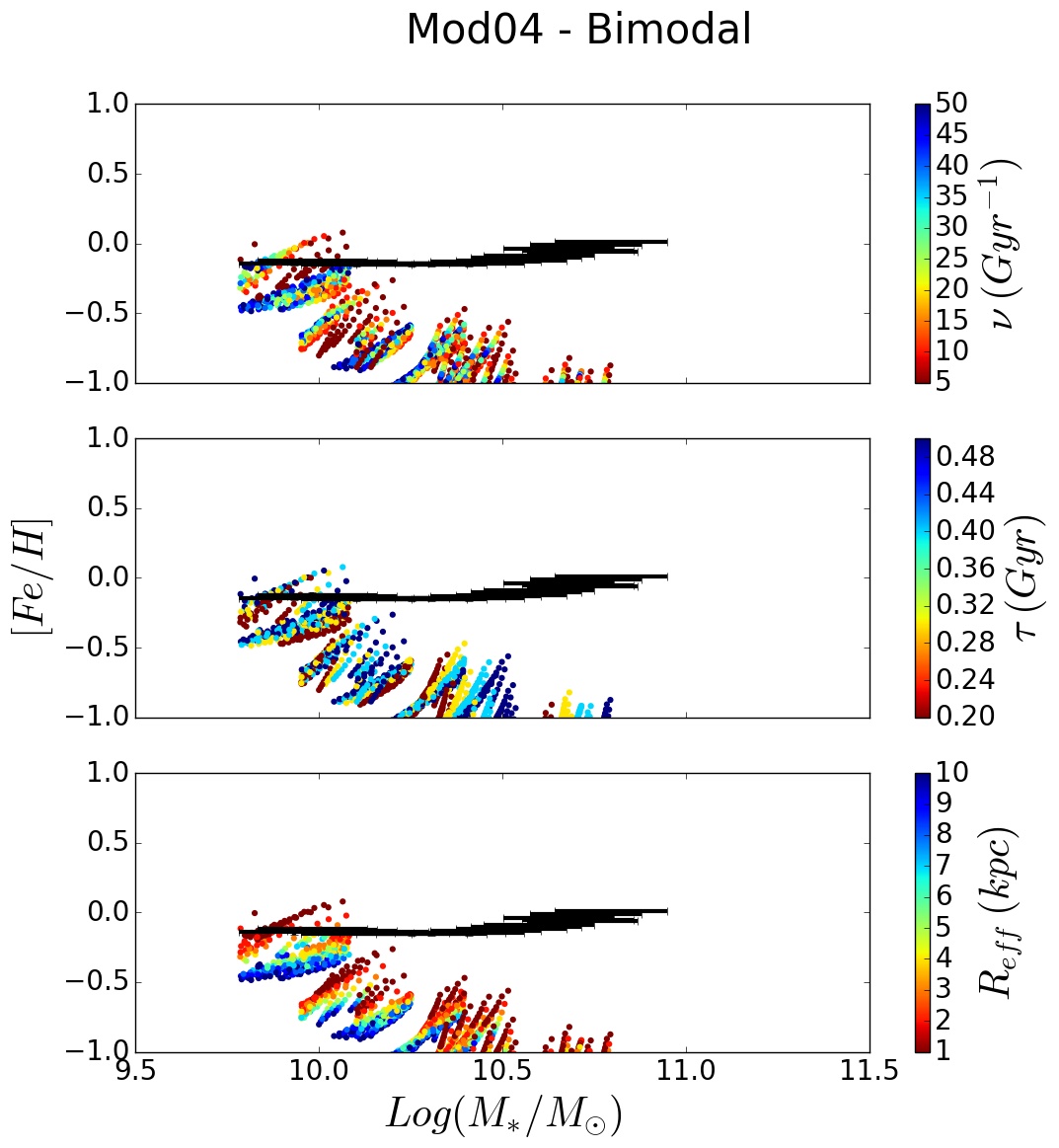}
	\includegraphics[width=.45\textwidth]{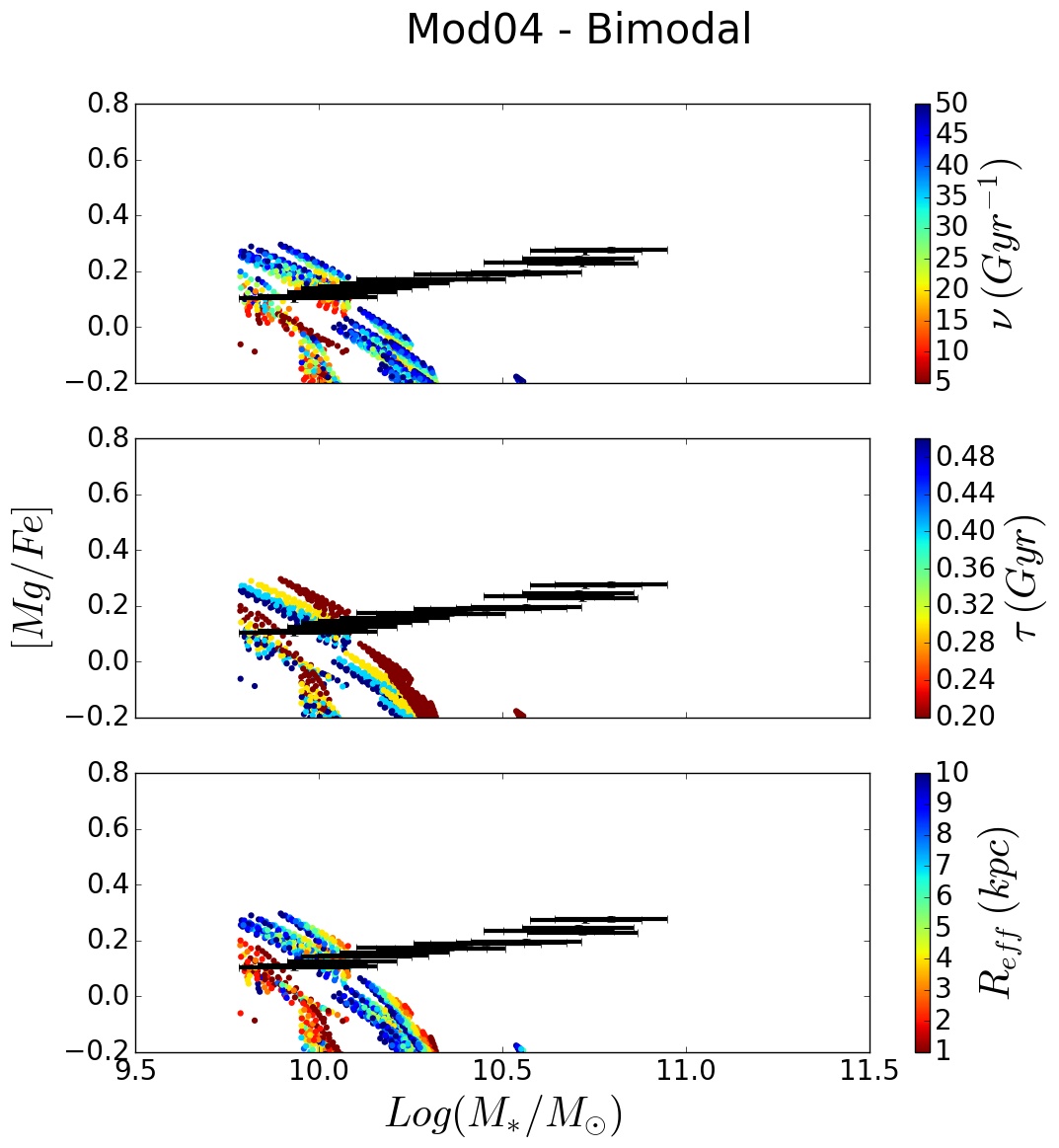}
	\caption{Same as figures \ref{fig:model01_properties_01} - \ref{fig:model03_properties_01} for Model 04, obtained by assuming a bimodal IMF, becoming bottom-heavier in more massive galaxies (see sec. \ref{sec:adopted_IMFs}).}
	\label{fig:model04_properties_01}
\end{figure*}
\begin{figure*}
	\includegraphics[width=.45\textwidth]{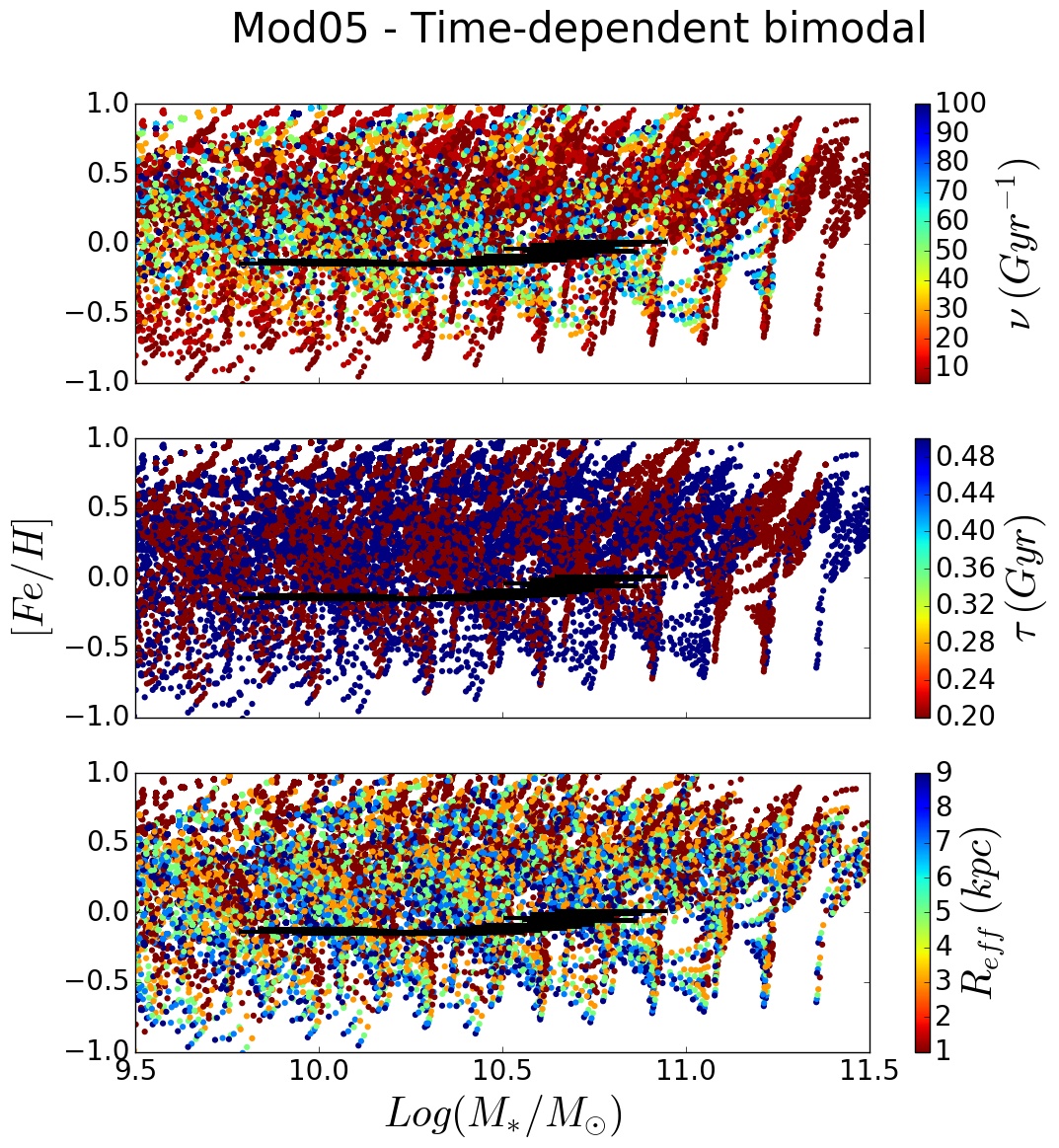}
	\includegraphics[width=.45\textwidth]{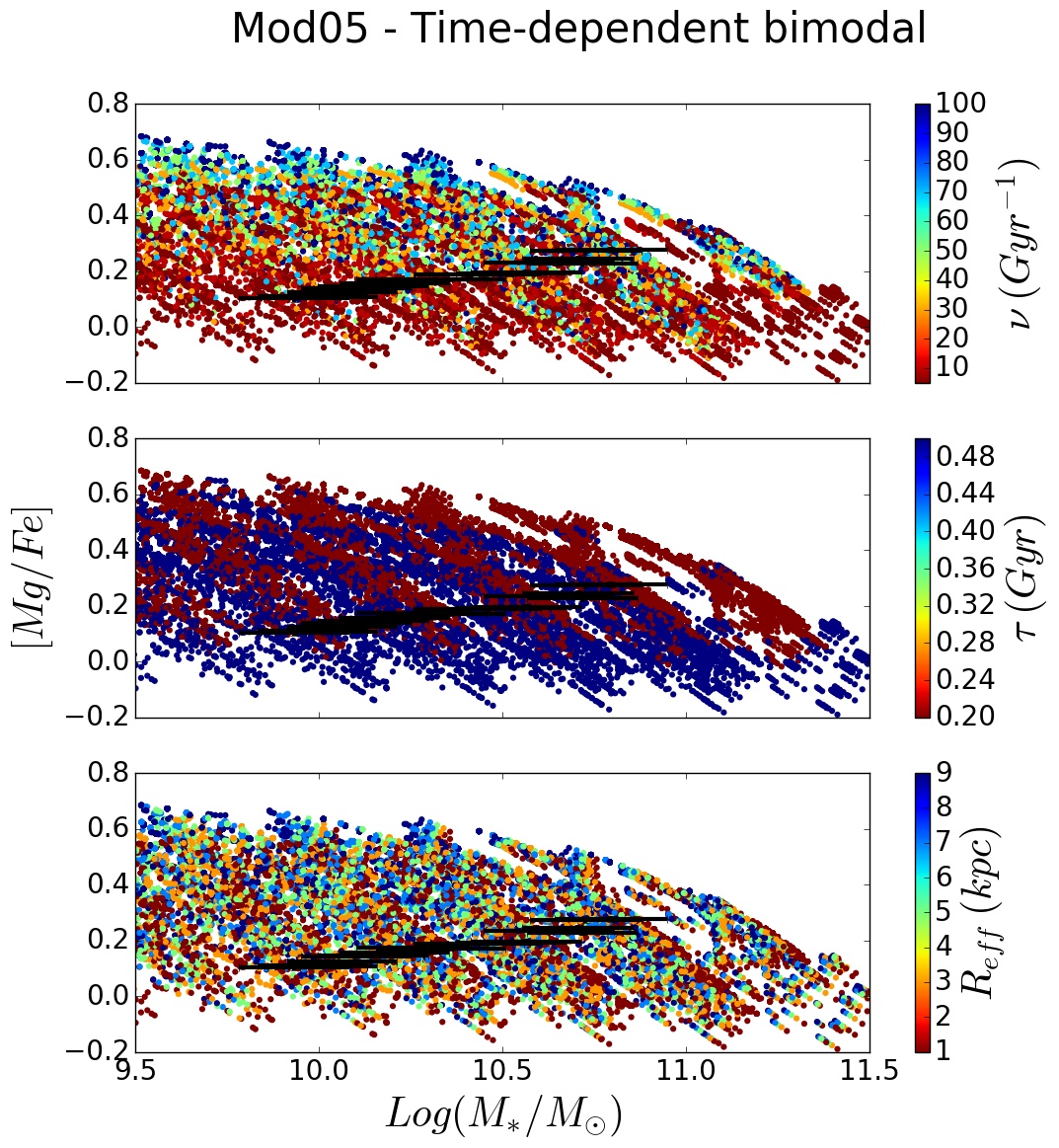}
	\caption{Same as figures \ref{fig:model01_properties_01} - \ref{fig:model04_properties_01}, this time for models with an explicitly time-dependent bimodal IMF, switching from a bottom-heavy (slope $\mu_1$) to a top-heavy (slope $\mu_2$) form after a time $t_{\text{switch}}$ (see sec. \ref{sec:adopted_IMFs}). Here, the color coding is analogous to the ones in previous pictures.}
	\label{fig:model05_properties_01}
\end{figure*}
\begin{figure*}
	\includegraphics[width=.45\textwidth]{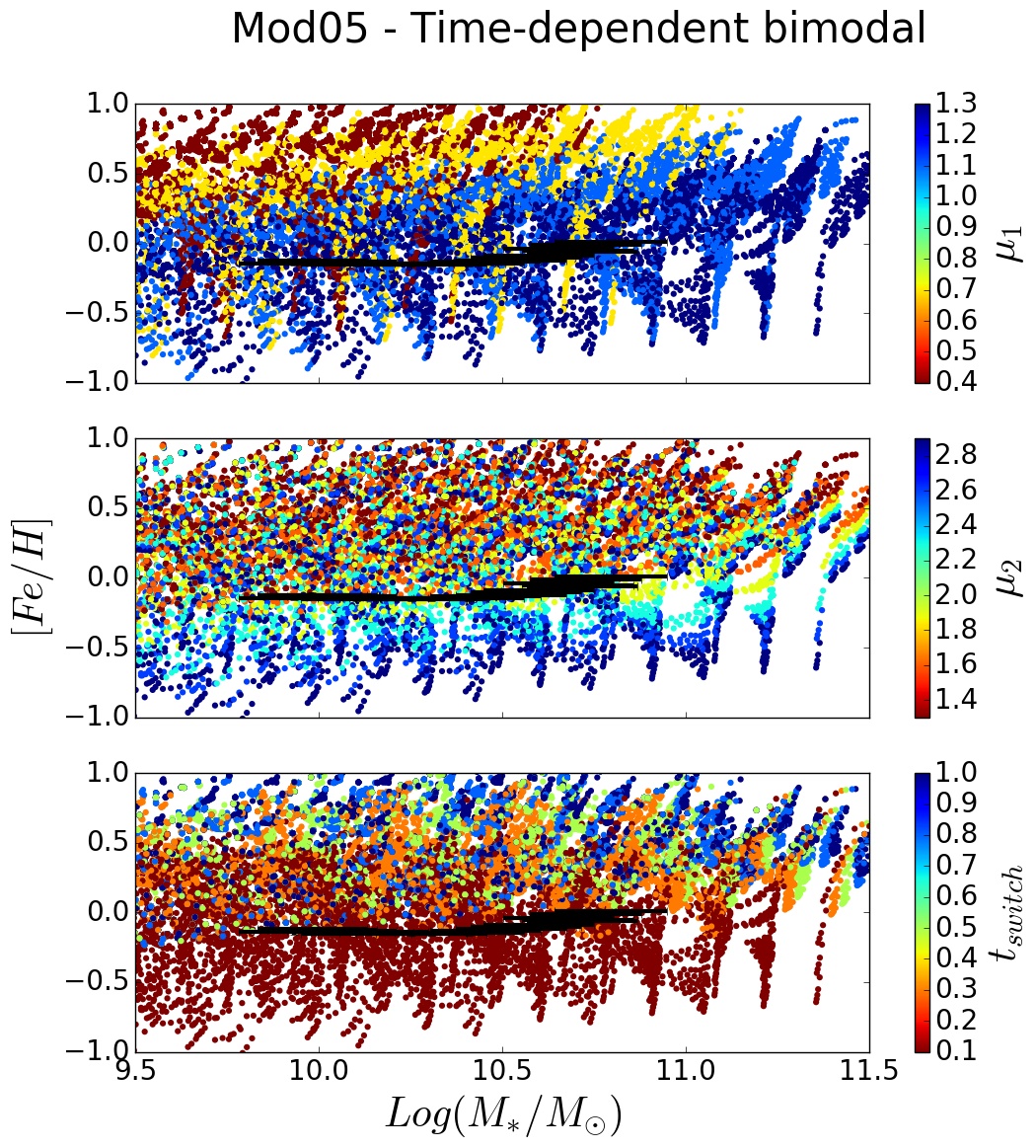}
	\includegraphics[width=.45\textwidth]{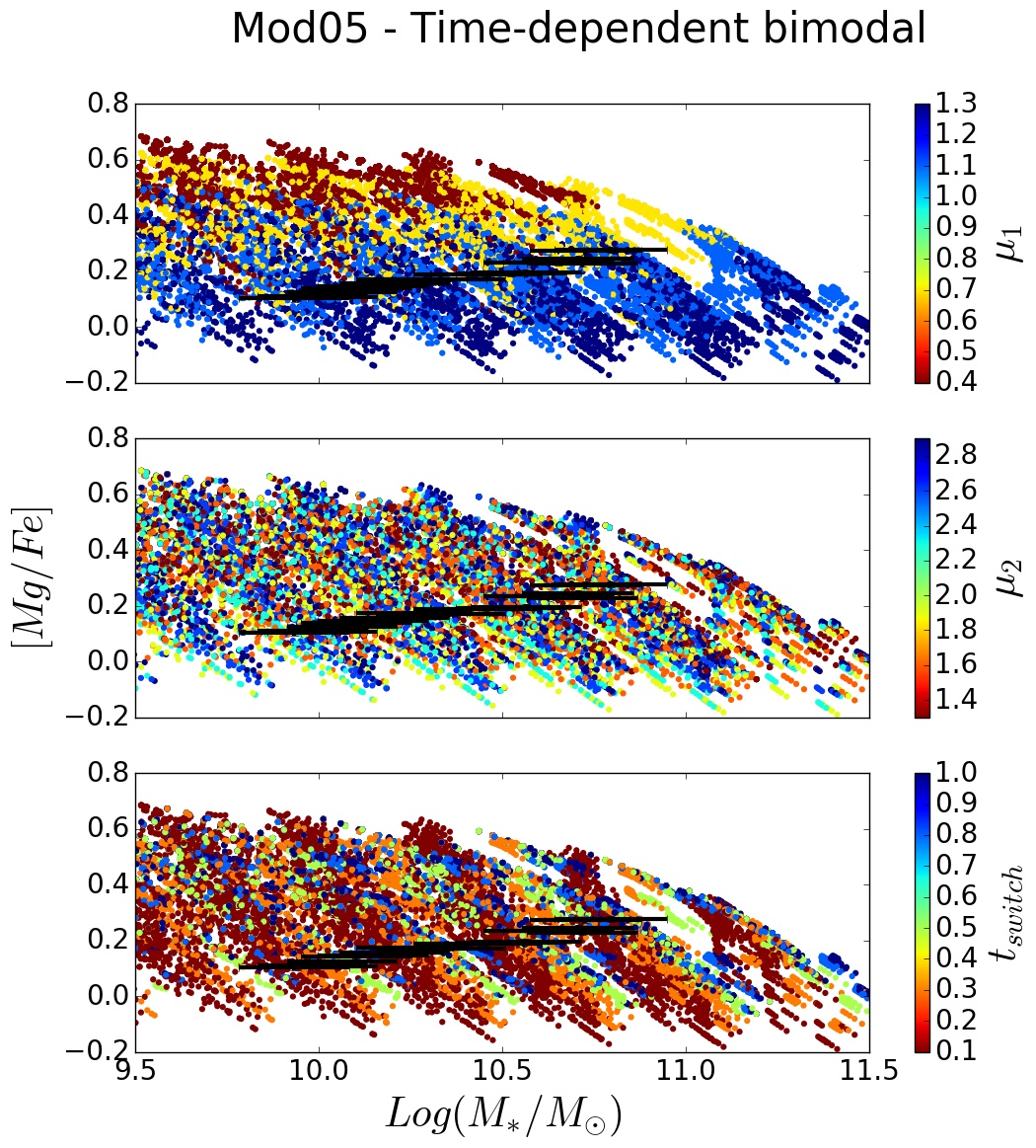}
	\caption{Same as figures \ref{fig:model01_properties_01} - \ref{fig:model05_properties_01}, for models with an explicitly time-dependent bimodal IMF, switching from a bottom-heavy to a top-heavy form after a time $t_{\text{switch}}$. In this case, plots are color coded to show the dependency on the slope values $\mu_1$ and $\mu_2$ (before and after the switch, respectively), and on the value of $t_{\text{switch}}$}
	\label{fig:model05_properties_02}
\end{figure*}
In this work, we extended this approach, with the aim of fully exploring the model parameter space. Once we assumed one of the IMF parameterizations described above, we generated the models by varying all the initial parameters over a grid of values, and considering all their possible combinations.
The results of this procedure are shown in Figures \ref{fig:model01_properties_01} to \ref{fig:model05_properties_02}, where we plot the variation of the $[Fe/H]$ and $[Mg/Fe]$ abundance ratios with stellar mass, as calculated by the chemical evolution code for all  model galaxies. For each ratio, we provide various versions of the same plot, color-coded to show the dependency on the other parameters.\\
Specifically, for Models 01 to 04, we show the dependency on:
\begin{itemize}
	\item the star formation efficiency $\nu$, defined as in equation \ref{eq_SF} as the proportionality constant between the SFR and the gas density;
	\item the infall time scale $\tau$, describing the time-scale of the initial infall, according to: 
	\begin{equation*}
	\displaystyle \left( \frac{dG_i(t)}{dt} \right)_{infall} = X_{i,infall}\,C\,e^{-t/\tau}
	\end{equation*}
	where the term on the LHS of the equation is the abundance variation of the i-th element in the gas due to infall alone, $X_{i,infall}$ is the abundance of said element in the infalling gas and C is a normalization constant;
	\item the effective radius $R_{eff}$ achieved after the collapse is over.
\end{itemize}
For Model 05, we produced additional plots (Fig \ref{fig:model05_properties_02}), showing the dependency of the $[Fe/H]$-mass and $[Mg/Fe]$-mass relations on three additional parameters:
\begin{itemize}
	\item $\mu_1$ and $\mu_2$, i.e. the value of the bimodal IMF slope before and after the switch, respectively;
	\item $t_{\text{switch}}$, i.e. the time when the switch from the bottom to the top-heavy form of the bimodal IMF occurs;
\end{itemize}
It is generally evident from figures \ref{fig:model01_properties_01} - \ref{fig:model05_properties_01} that the $[Mg/Fe]$ ratio in galaxies of the same stellar mass are higher in models with increasing $\nu$, where the larger thermal energy injected by stellar winds and SNe into the ISM leads to an earlier onset of a galactic wind, which drives the gas away from the galaxy and quenches star formation.\\
The effect of decreasing the infall time scale $\tau$, which is similar to increasing $\nu$ and $R_{eff}$, appears to be less significant.\\
Similarly, in Fig. \ref{fig:model05_properties_02}, it is noticeable how galaxies with higher values of $\mu_1$ (i.e., galaxies whose IMF was bottom-heavier before the switch) present lower $[Mg/Fe]$ for a given mass, accordingly to theoretical expectations (no significant trend with $\mu_2$ and $t_{\text{switch}}$ is noticeable from the plot).\\
Again, the presented grid of models have been produced by simply considering all the possible combinations of the values for the initial input parameters of the code. As a result, some of these parameter configurations end up being physically less plausible, and some degeneracies in the model galaxies are present. In spite of this, here we chose to present the whole grids without any selection on the model galaxies, in order to highlight the response of our model to the change of its parameters. A complete discussion will be presented in later sections, only considering models actually matching the data.

\section{Results}\label{sec:results}

In this section, we compare predictions from different models with observations.\\
For every IMF, we selected Models matching the observed mass-$[Fe/H]$ and mass-$[Mg/Fe]$ relations within the observational errors.
\begin{figure*}
	\includegraphics[width=.8\textwidth]{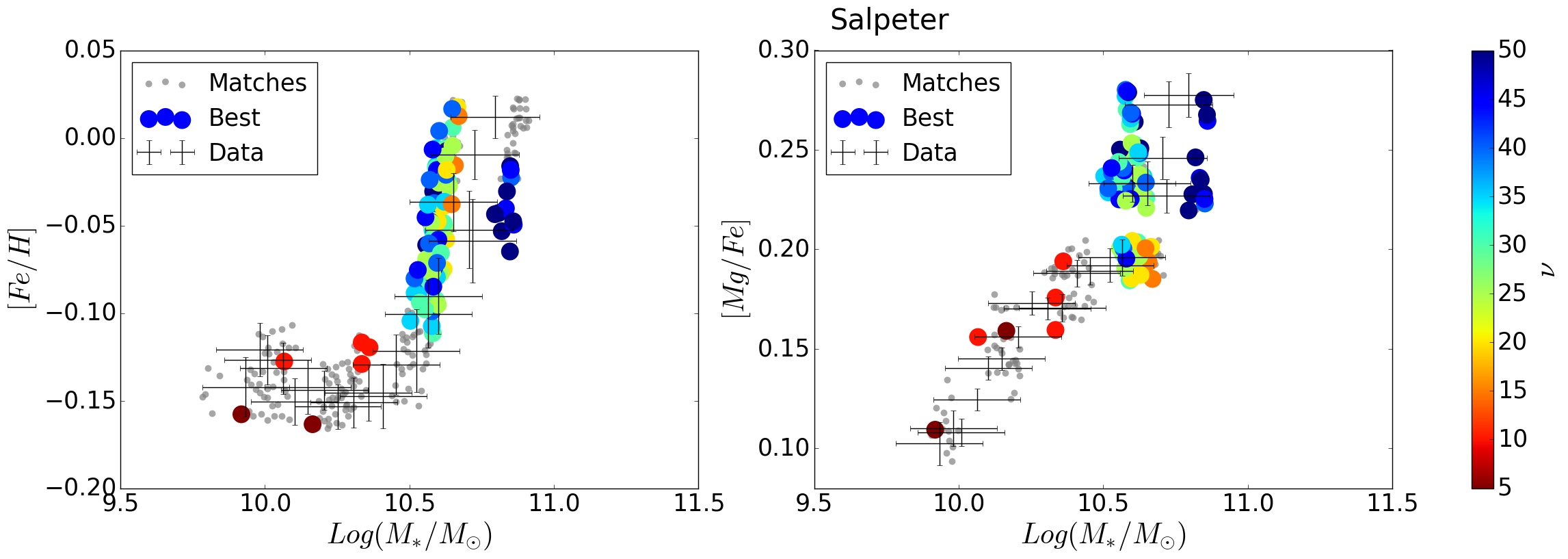}
	\includegraphics[width=.8\textwidth]{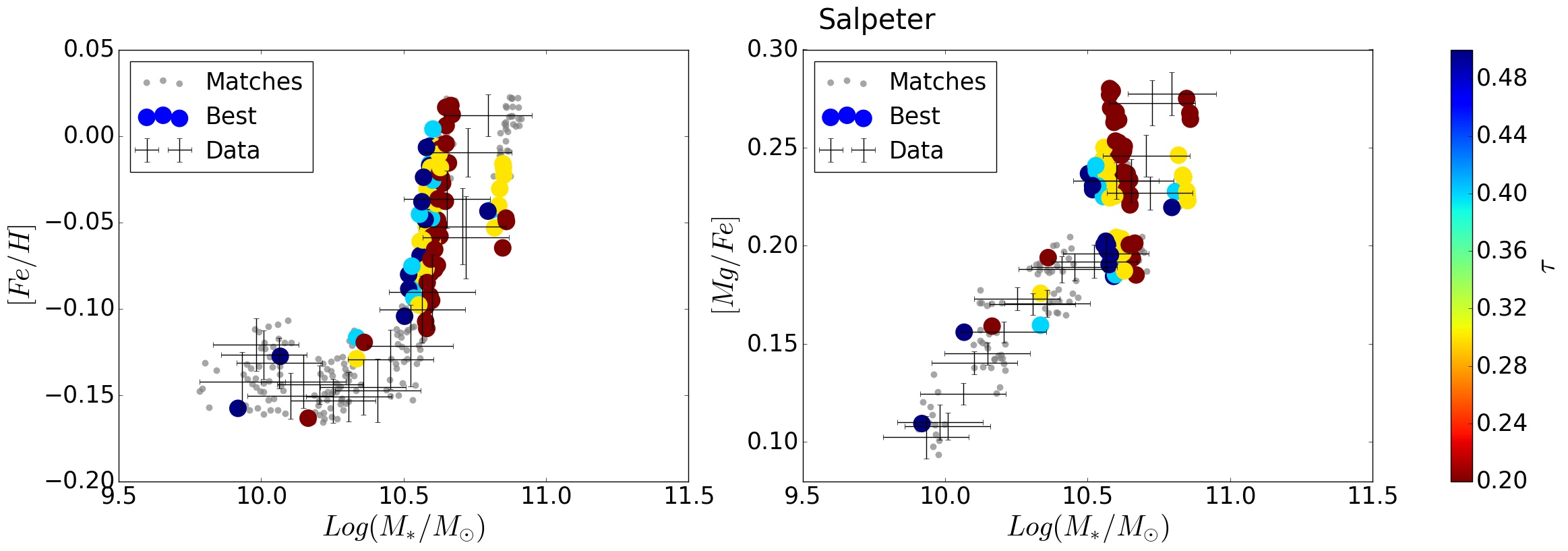}
	\includegraphics[width=.8\textwidth]{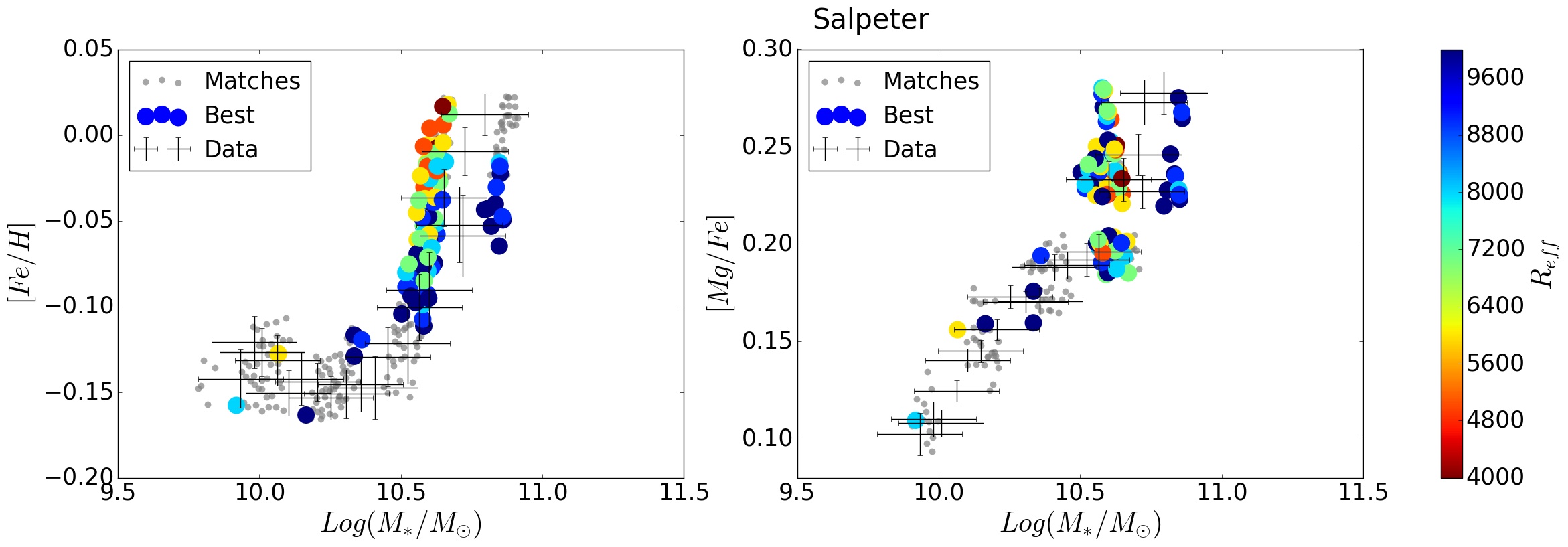}
	\caption{Comparison between data and Models01 for the $[Fe/H]$ and $[Mg/Fe]$ abundance ratios. Models matching the $[Fe/H]$-mass and the $[Mg/Fe]$-mass relations simultaneously are color-coded according to their star formation efficiency $\nu$ (top row), infall time-scale $\tau$ (middle row) and effective radius $R_{eff}$ (bottom row), while the ones matching only one of the two relations are shown with fading, smaller markers.}
	\label{fig:match_01}
\end{figure*}
\begin{figure*}
	\includegraphics[width=.8\textwidth]{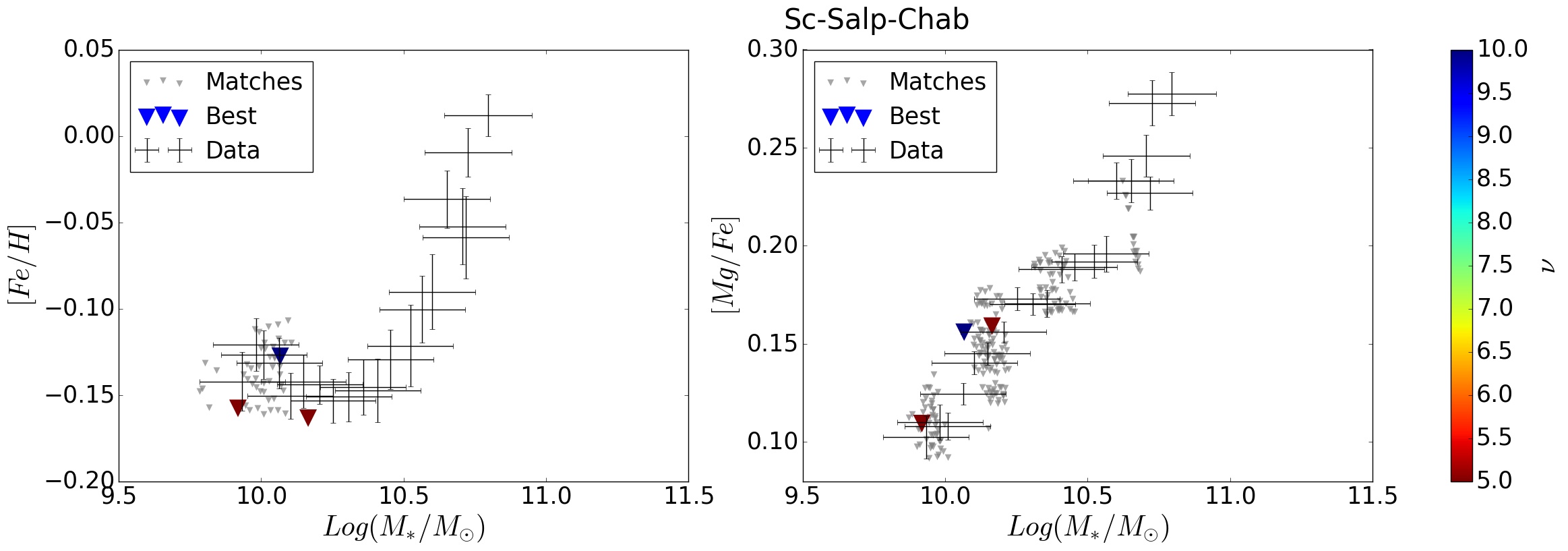}
	\includegraphics[width=.8\textwidth]{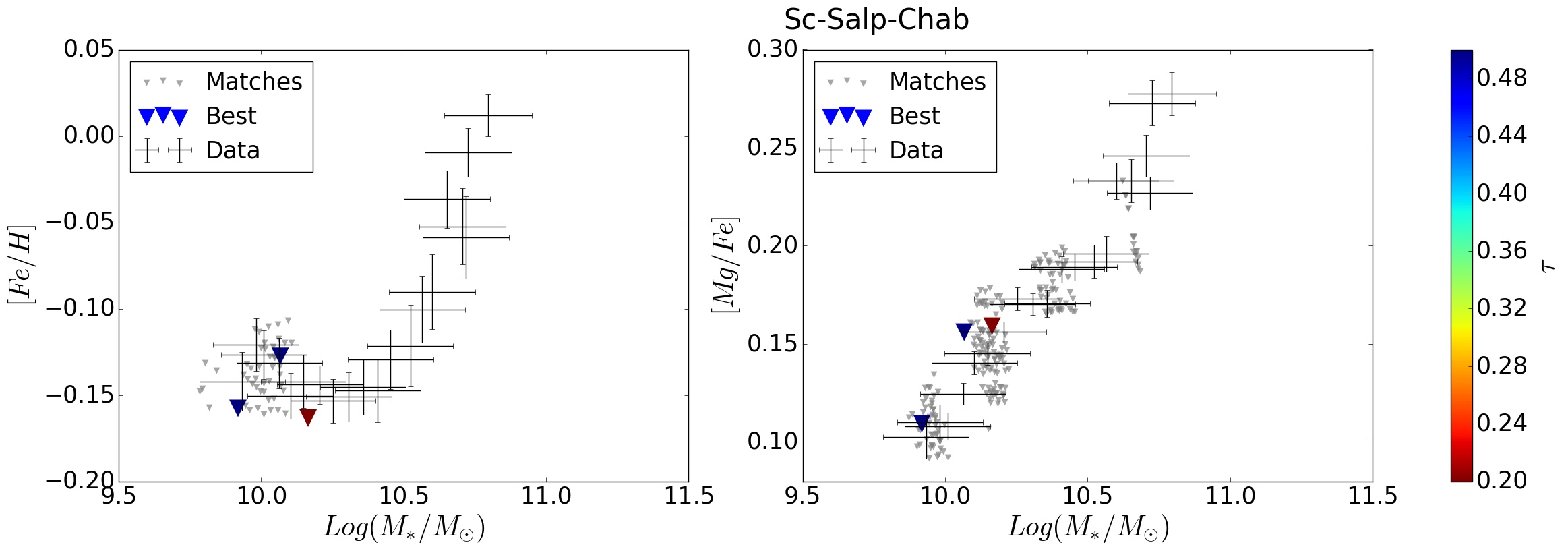}
	\includegraphics[width=.8\textwidth]{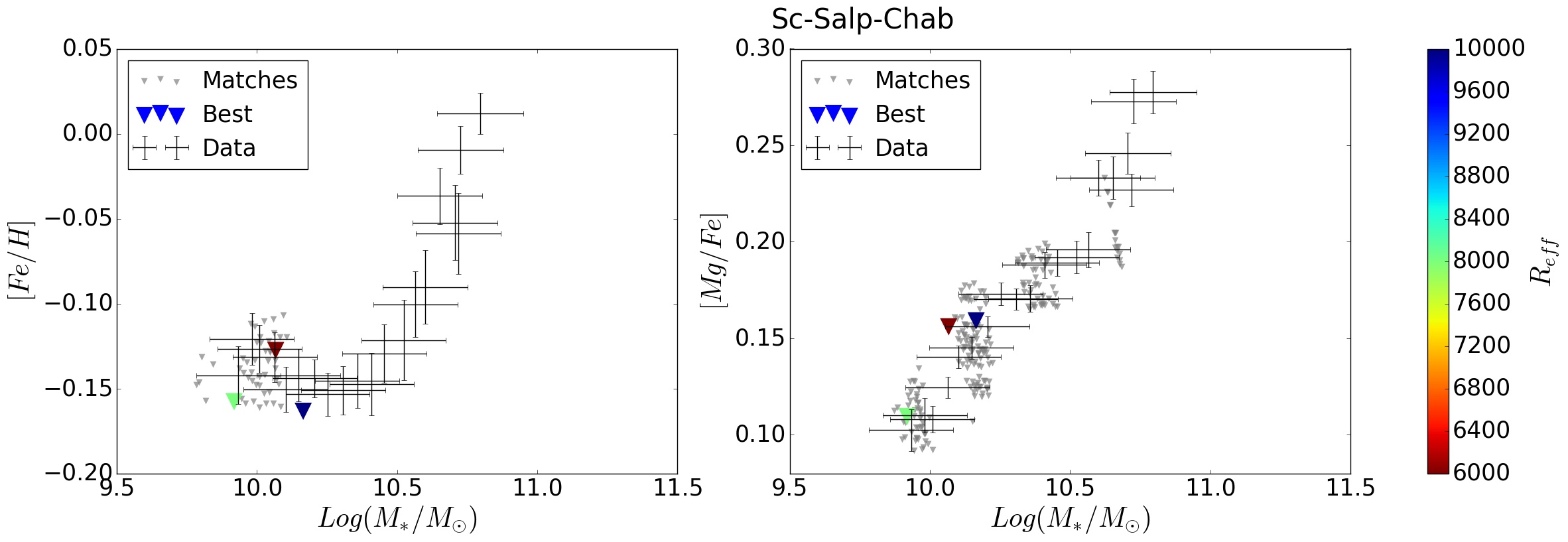}
	\caption{As Figure \ref{fig:match_01}, but for Model02}
	\label{fig:match_02}
\end{figure*}
\begin{figure*}
	\includegraphics[width=.8\textwidth]{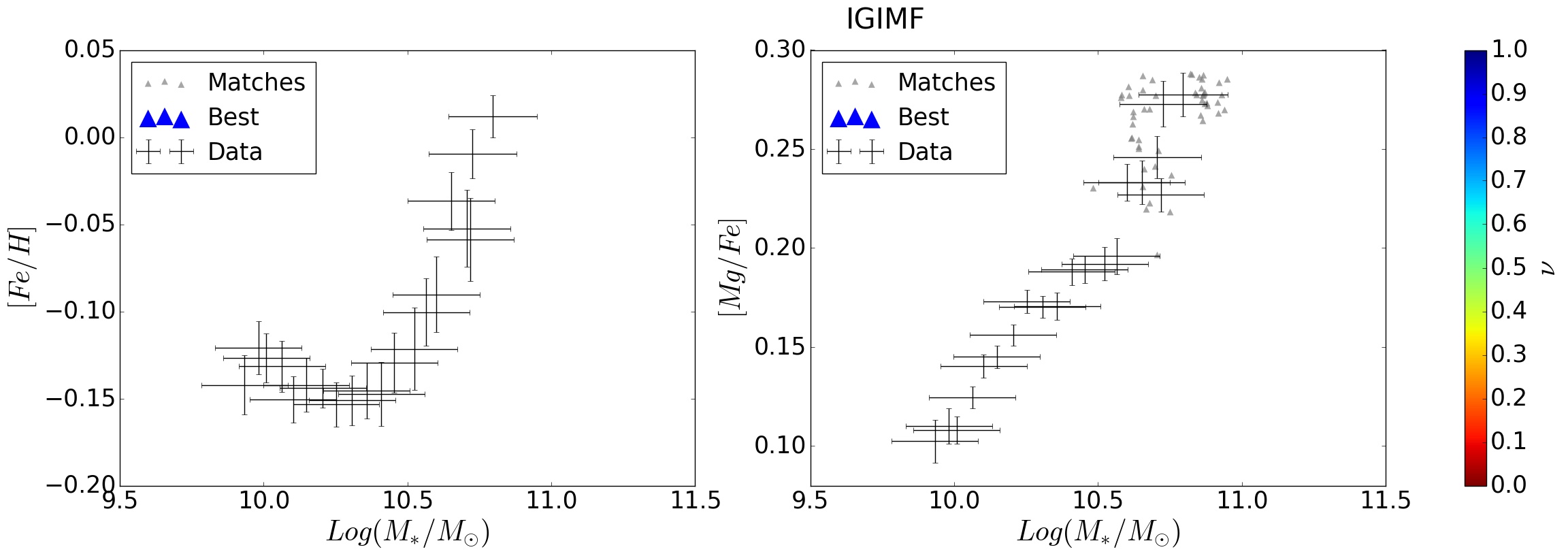}
	\includegraphics[width=.8\textwidth]{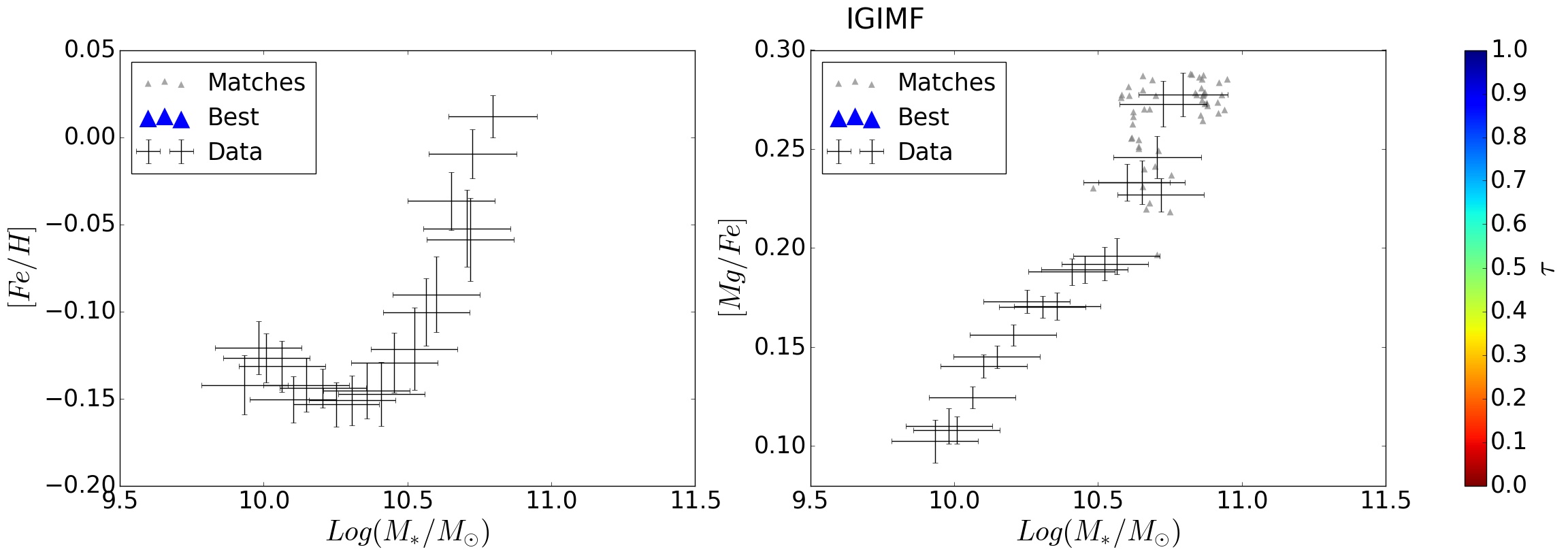}
	\includegraphics[width=.8\textwidth]{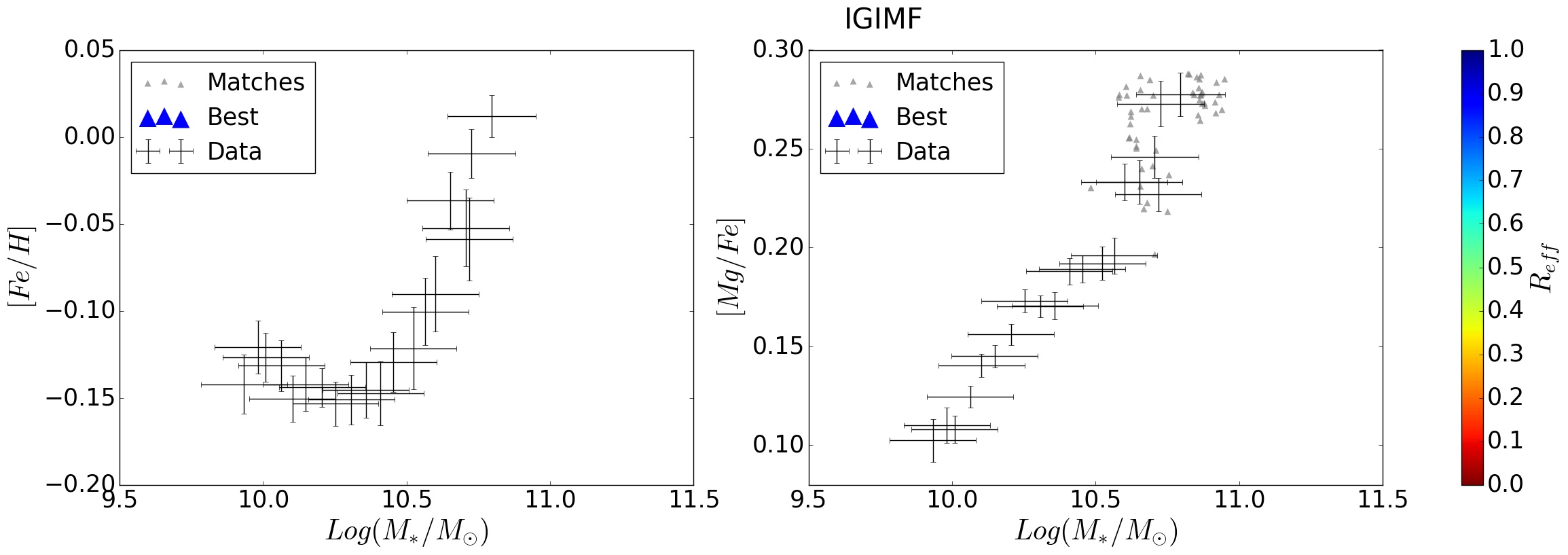}
	\caption{As Figures \ref{fig:match_01} and \ref{fig:match_02}, but for Model02}
	\label{fig:match_03}
\end{figure*}
\begin{figure*}
	\includegraphics[width=.8\textwidth]{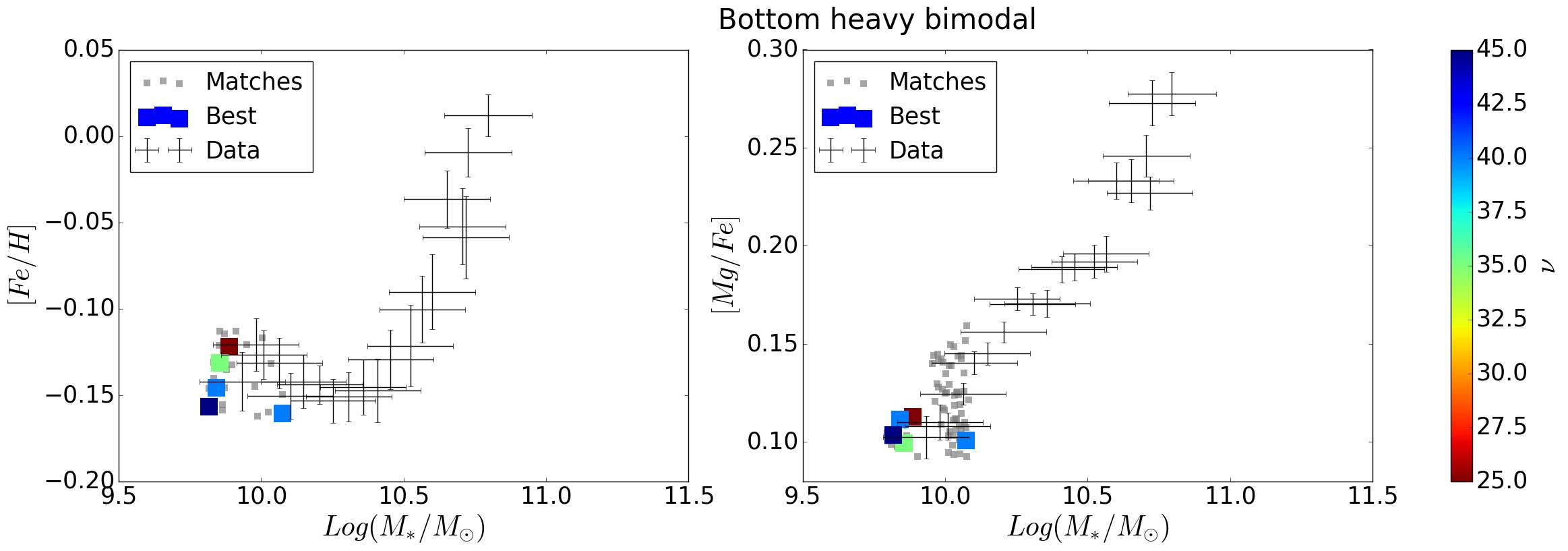}
	\includegraphics[width=.8\textwidth]{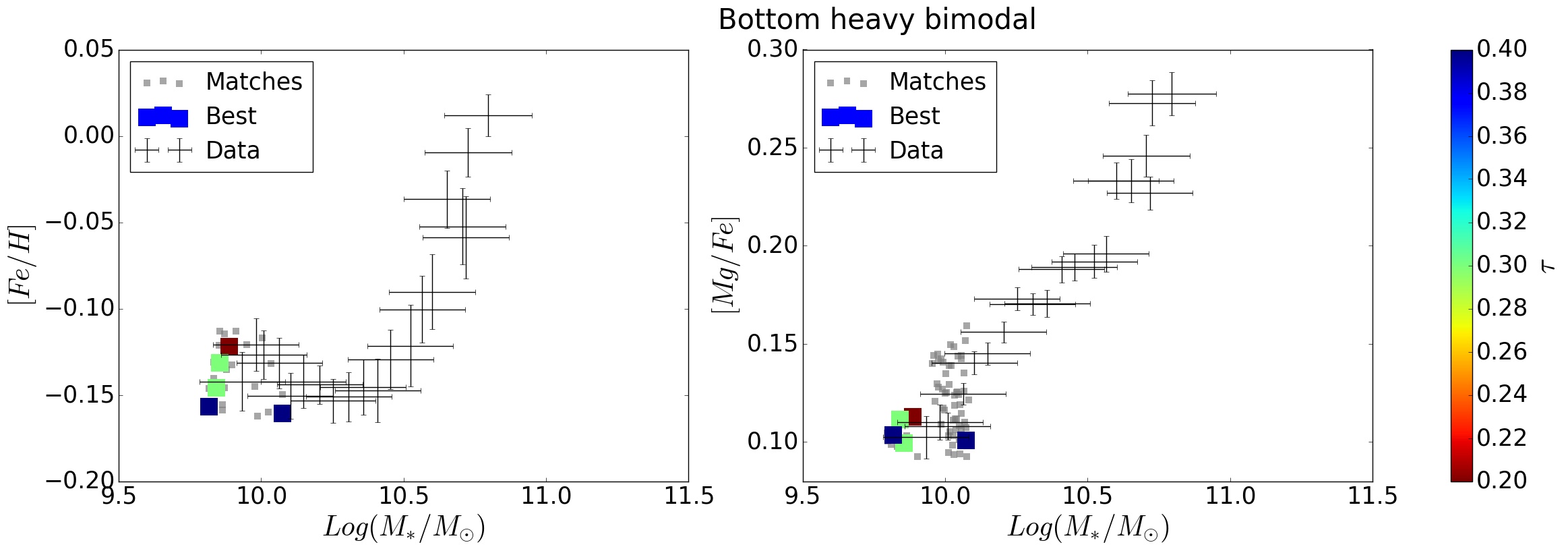}
	\includegraphics[width=.8\textwidth]{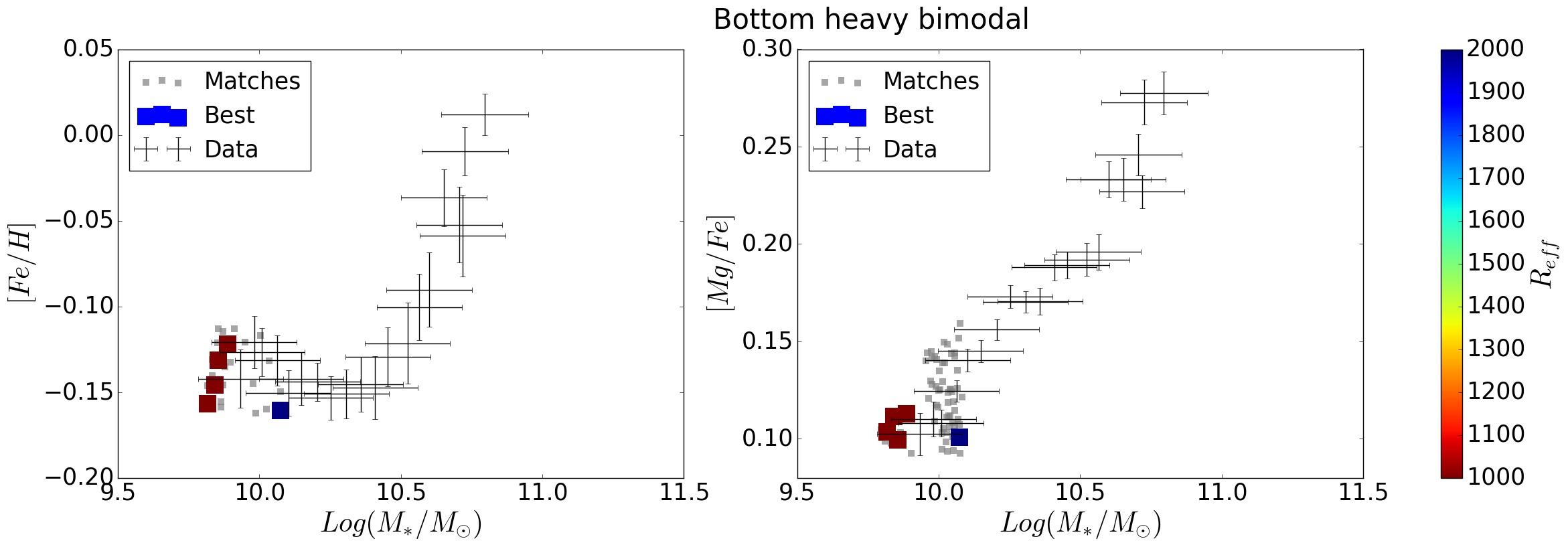}
	\caption{As Figures \ref{fig:match_01} - \ref{fig:match_03}, but for Model04}
	\label{fig:match_04}
\end{figure*}
\begin{figure*}
	\includegraphics[width=.8\textwidth]{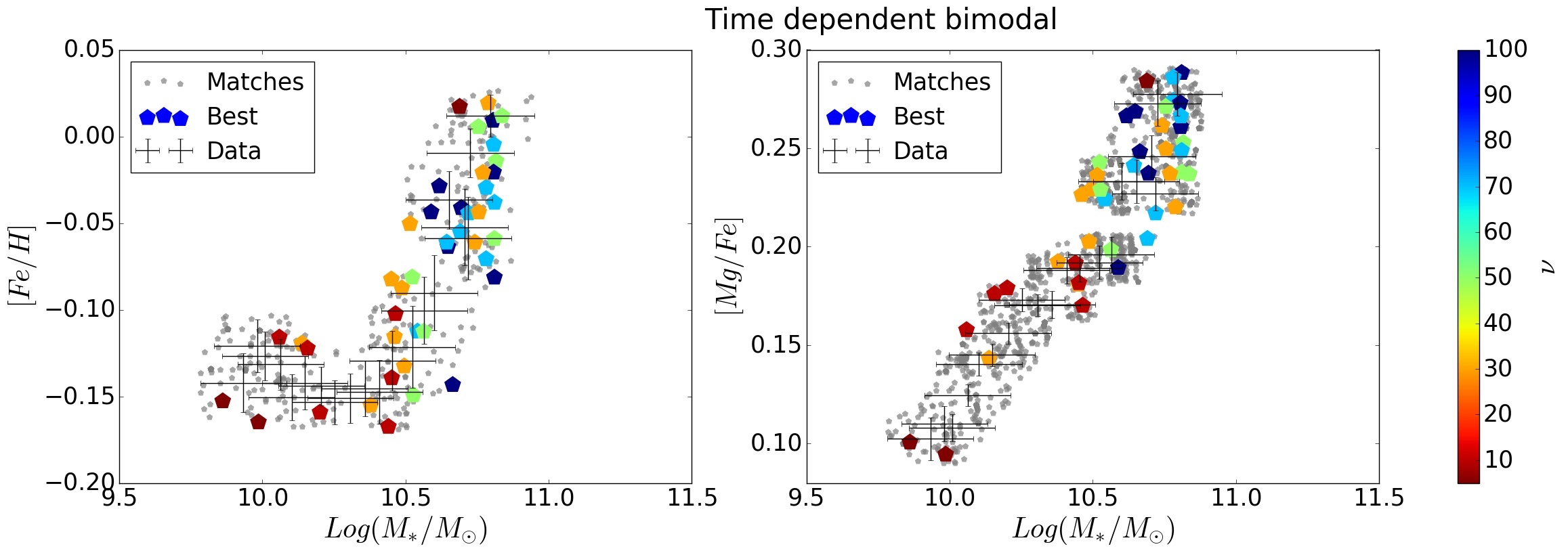}
	\includegraphics[width=.8\textwidth]{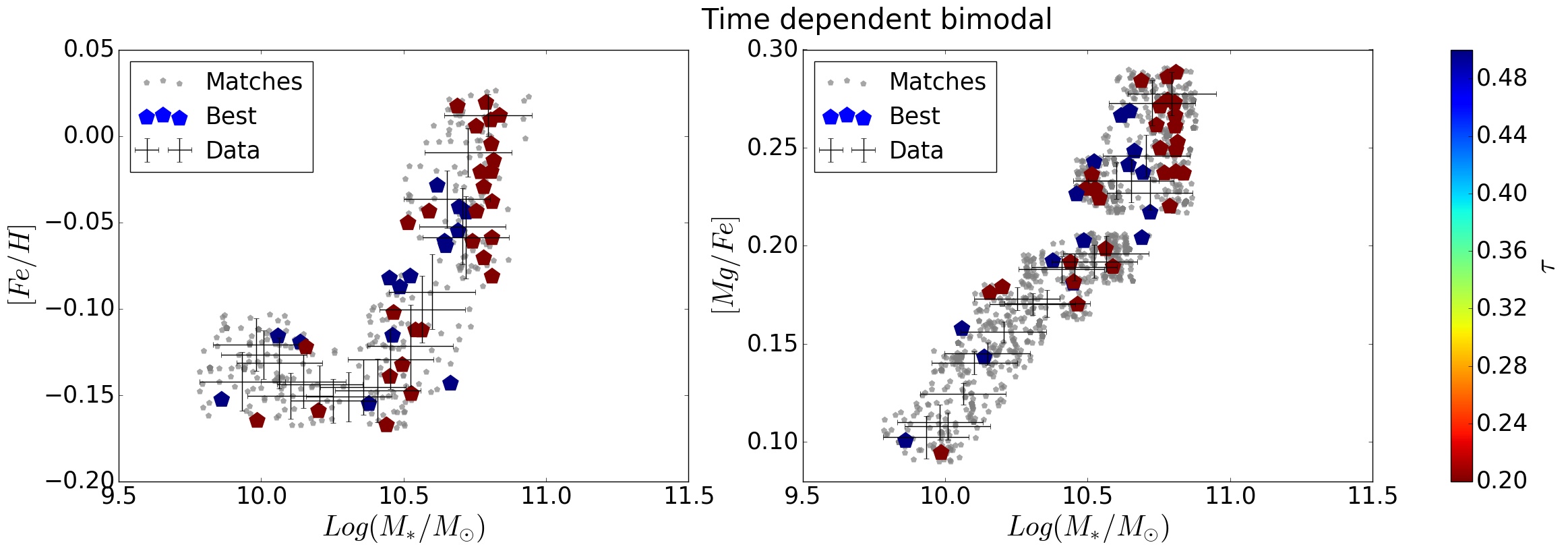}
	\includegraphics[width=.8\textwidth]{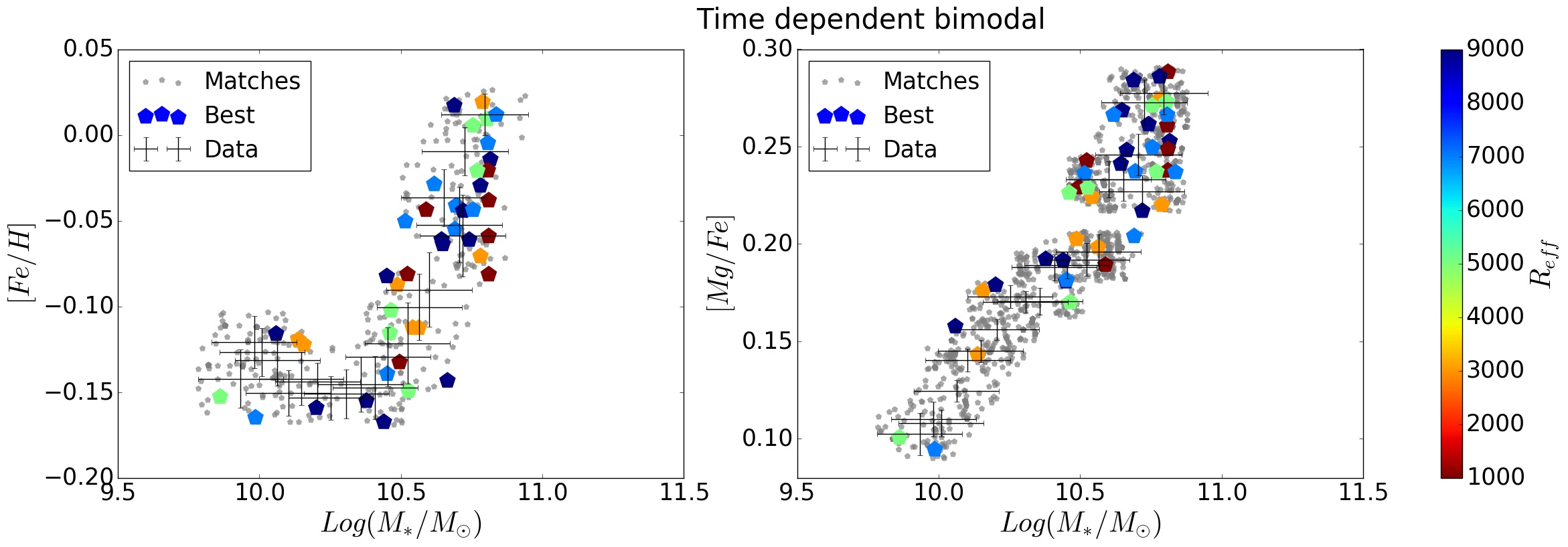}
	\caption{As Figures \ref{fig:match_01} - \ref{fig:match_04}, but for Model05}
	\label{fig:match_05a}
\end{figure*}
\begin{figure*}
	\includegraphics[width=.8\textwidth]{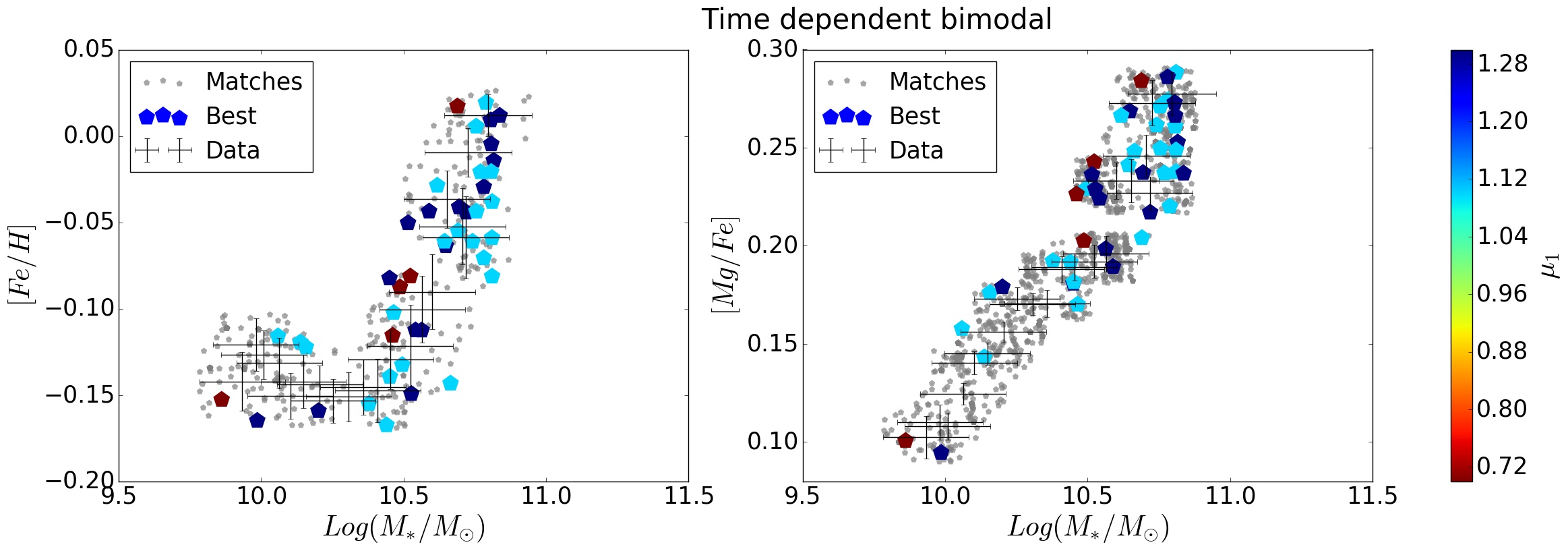}
	\includegraphics[width=.8\textwidth]{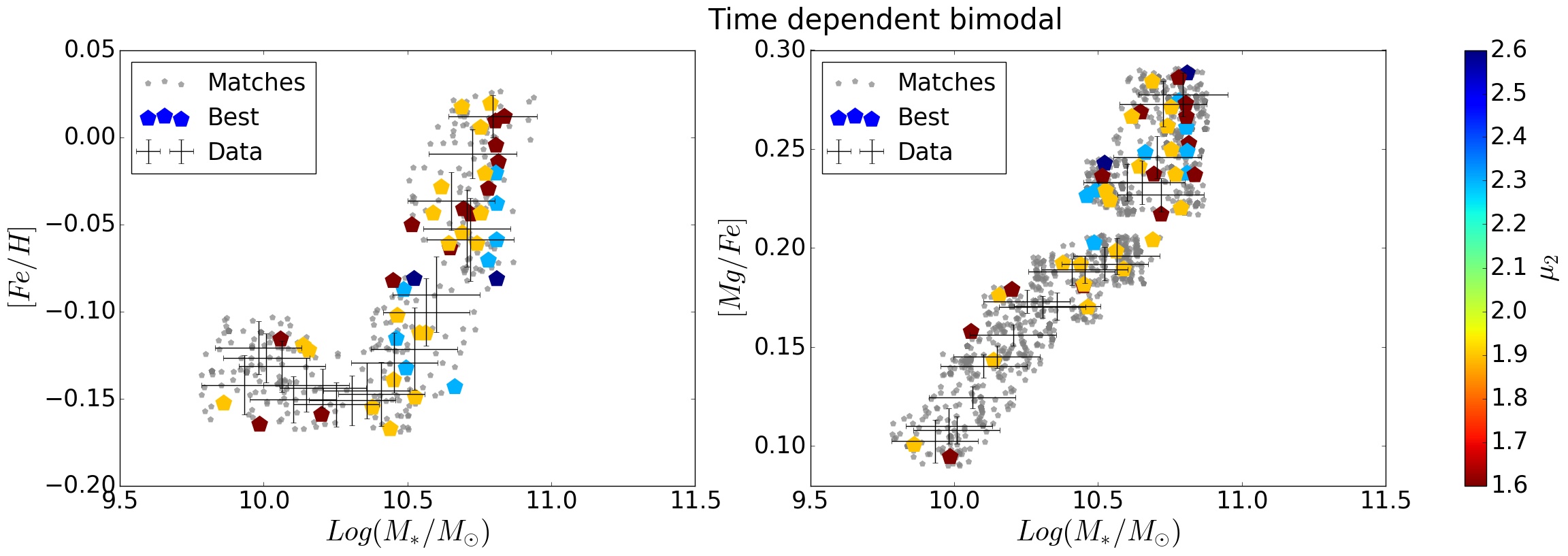}
	\caption{Comparison between data and Models05 for the $[Fe/H]$ and $[Mg/Fe]$ abundance ratios. Models matching the $[Fe/H]$-mass and the $[Mg/Fe]$-mass relations simultaneously are color-coded according to the slope of the bimodal IMF before ($\mu_1$, top row) and after ($\mu_2$, bottom row) the switch (all these models switch IMF at $t_{\text{switch}}=0.1$ Gyr). Models matching only one of the two relations are shown with fading, smaller markers.}
	\label{fig:match_05b}
\end{figure*}
The results of the matching procedure are summarized in figures \ref{fig:match_01} to \ref{fig:match_05a} and in table \ref{table:model_matches}. In figures \ref{fig:match_01} - \ref{fig:match_05a}, galaxy models matching one of the analyzed relations (either the $[Fe/H]$-mass or the $[Mg/Fe]$-mass relation) are shown as gray points, while we highlight and color-code, based on their star formation efficiency $\nu$, the models matching the two relations simultaneously.\\
For each IMF, table \ref{table:model_matches} reports the number of model galaxies matching the data, for three different mass ranges and in total, for the $[Fe/H]$ ratio (columns 2-5), the $[Mg/Fe]$ ratio (columns 6-9) and for both these quantities simultaneously (columns 10-13).\\
While all the suggested IMFs - aside from the IGIMF, which provide the worst results - produce model galaxies matching the abundance ratios of the data in the lower mass bin, the number of matches decreases dramatically at higher masses, especially for the $[Fe/H]$ ratio. This happens for all the Models, except for the ones with a Salpeter (Models 01) or time-dependent bimodal IMFs (Models 05). Moreover, these two sets of Models are the only ones producing a significant number of matches for both the abundance ratios simultaneously.\\
For this reason, we select from Models 01 and 05 the ones matching both $[Fe/H]$ and $[Mg/Fe]$, and analyze their properties.\\
In both the classes of Models, we confirm the results of \citetalias{P04}, with the best matching models presenting a trend of increasing star formation efficiency at higher masses (see fig. \ref{fig:best_models_params}).\\
In Model 05, we observe that, despite the wide range of possible values for $t_{switch}$ (the time at which the slope changes from the initial value $\mu_1$ to the present day value $\mu_2$), all model galaxies reproducing the two abundance ratios simultaneously switch slope at the same time; specifically, at $t_{switch}=0.1$ Gyr, the lowest value (for reference, \citealt{weidner2013} found the optimal time for the switch to be $t_{switch}\ge0.3$ Gyr). So, if the switch has to occur, it has to be in the early stages of the chemical evolution in order to reproduce the data.\\
Taking this into account, in Figure \ref{fig:mass_created_before_switch} (left panel), we plot the distribution of the galactic wind onset time for the 20 best-matching Model05 galaxies; in all of them, the wind starts (and so star formation is quenched) some time after the slope switch in the IMF. The percentage of stellar mass created before the switch (so, under a top-heavy IMF regime) is always smaller than 50\% (see Fig. \ref{fig:mass_created_before_switch}, right panel).
\begin{figure*}
	\centering
	\begin{subfigure}{.4\textwidth}
		\includegraphics[width=1\linewidth]{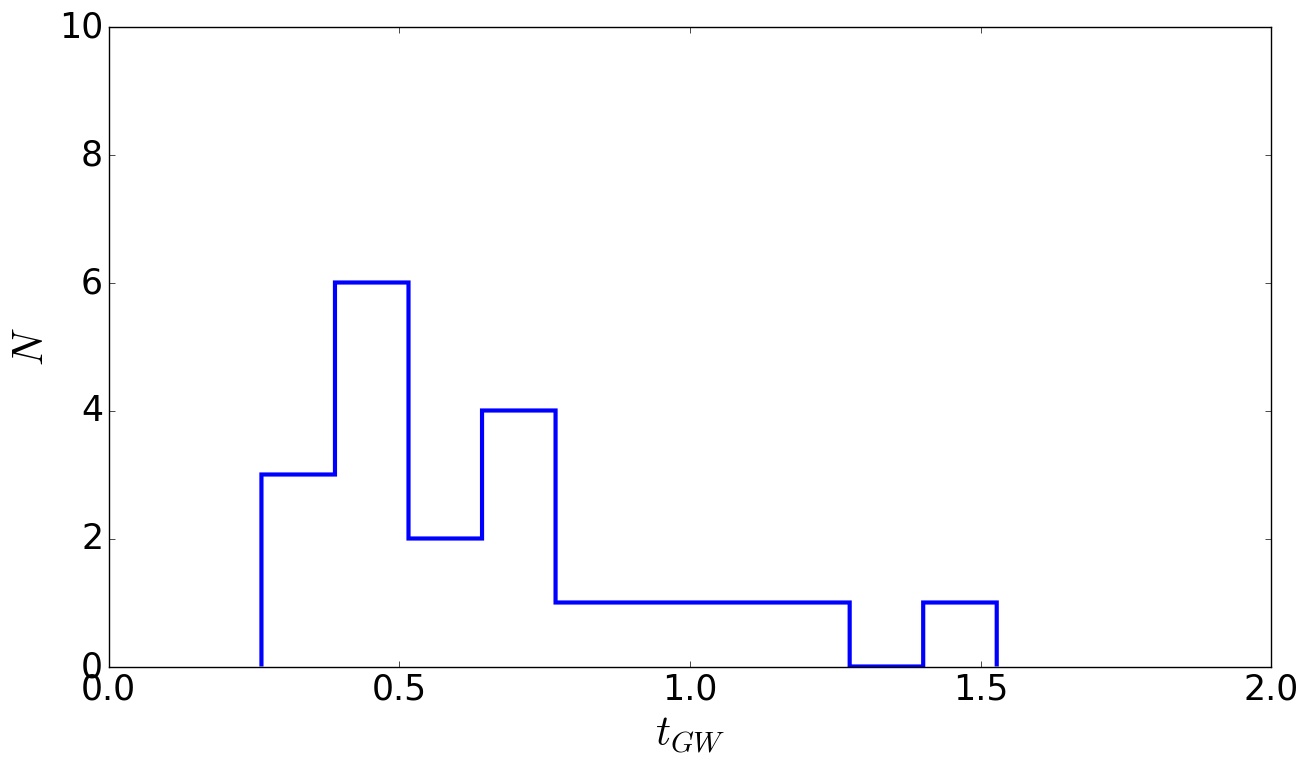}
	\end{subfigure}
	\begin{subfigure}{.4\textwidth}
		\includegraphics[width=1\linewidth]{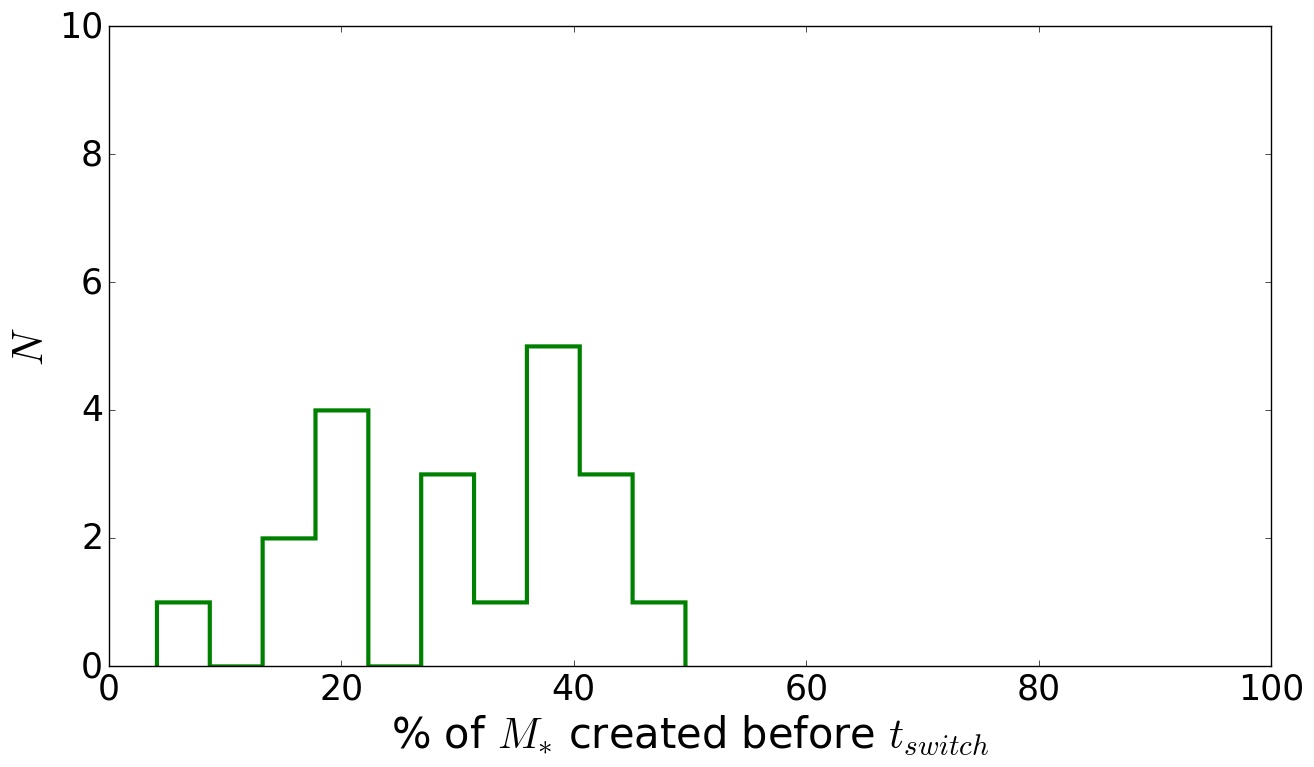}
	\end{subfigure}
	\caption{Left panel: Distribution of the onset time of the galactic wind for the Model05 galaxies matching the observed  $[Fe/H]$-mass and the $[Mg/Fe]$-mass relations simultaneously. Right panel: Distribution of the percentage of stellar mass created before the IMF switching from bottom to top-heavy in the same galaxies. .}\label{fig:mass_created_before_switch}
\end{figure*}
A weak positive trend with mass can be observed in the slope value before and after the switch (see fig. \ref{fig:best_models_mu1_mu2}). At low masses, we have mostly models with both $\mu_1$ and $\mu_2$ in the range from 1 to 2, i.e. not so different from the Kroupa-like slope (1.3). At higher masses, the slope before the switch $\mu_1$ becomes as low as $0.5$ (top-heavier), while the slope after the switch $\mu_2$ gets as high as 2.6 (bottom-heavier).
\begin{figure*}
	\centering
	\begin{subfigure}{.4\textwidth}
		\includegraphics[width=1\linewidth]{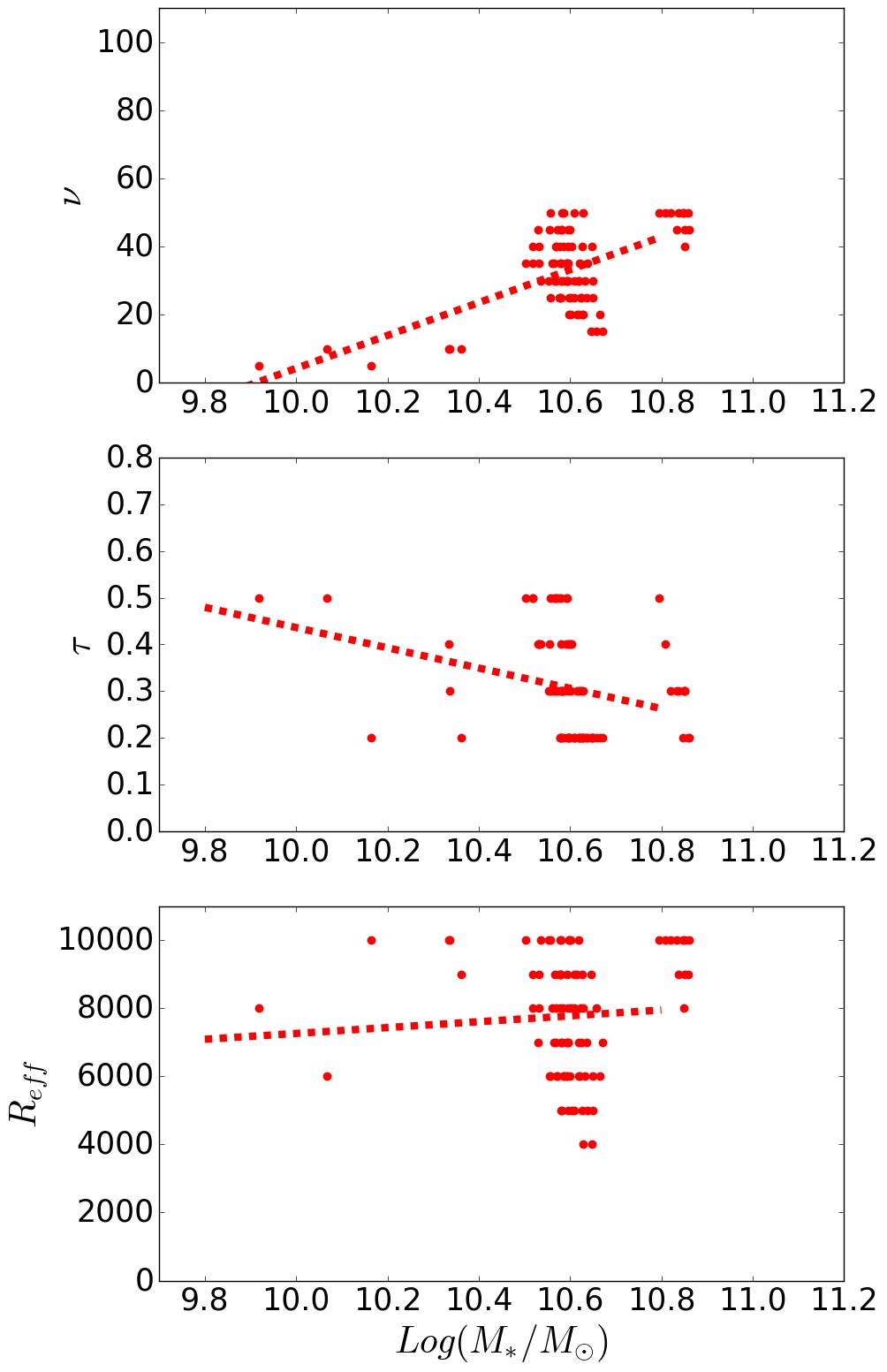}
	\end{subfigure}
	\begin{subfigure}{.4\textwidth}
		\includegraphics[width=1\linewidth]{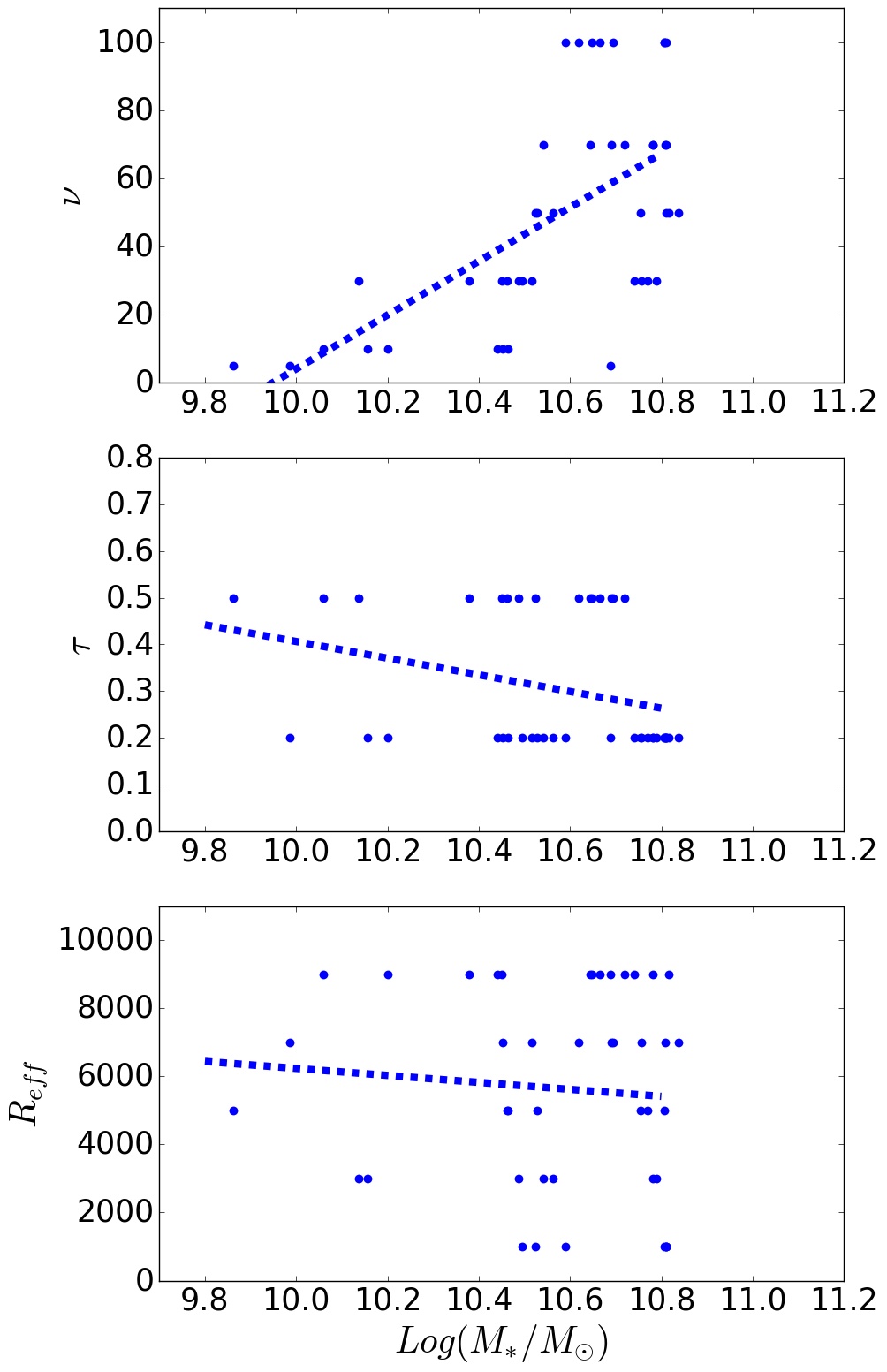}
	\end{subfigure}
	\caption{Variation with mass of the star formation efficiency (top row), infall time scale (middle row) and effective radius (bottom row) for the models matching the two abundance ratios observed in data  simultaneously for a Salpeter (Model01, left) or a time-dependent bimodal (Model05, right) IMFs. Dotted lines show linear fits to the models.}\label{fig:best_models_params}
\end{figure*}

\begin{figure*}
	\centering
	\includegraphics[width=.5\textwidth]{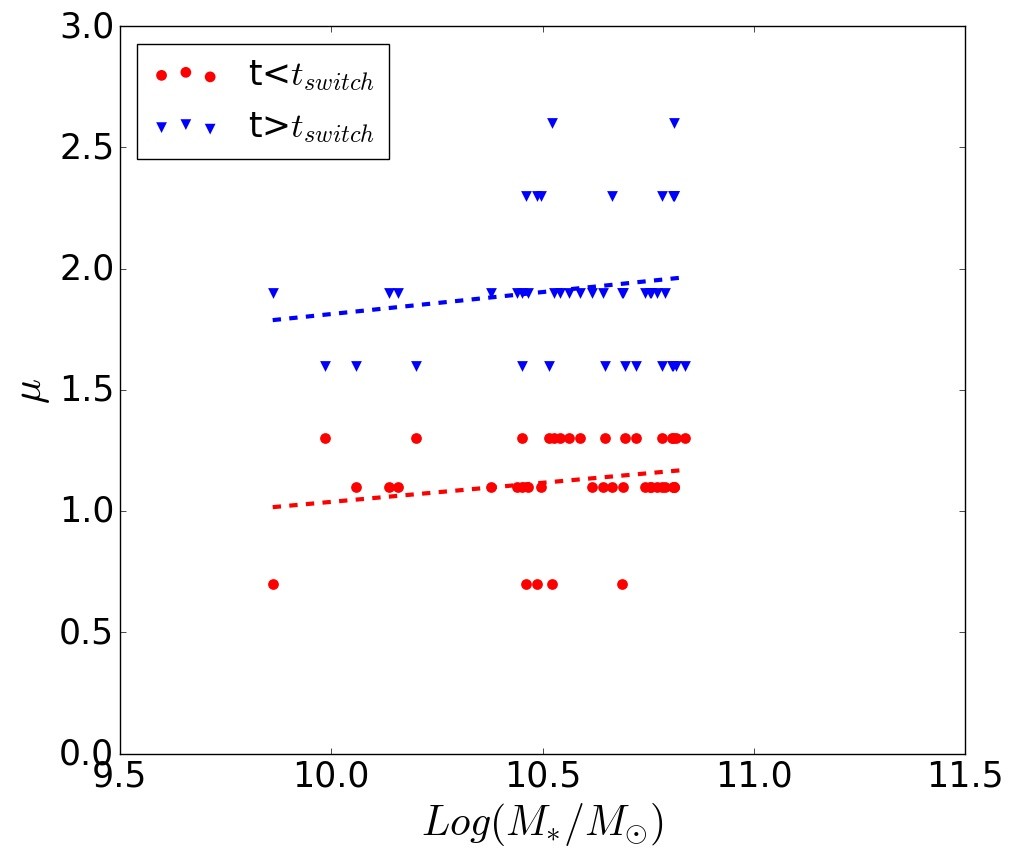}
	\caption{Variation with mass of the IMF slope before and after the switch for the time-dependent bimodal models matching the observed $[Fe/H]$-mass and the $[Mg/Fe]$-mass relations simultaneously.}
	\label{fig:best_models_mu1_mu2}
\end{figure*}

\begin{table*}
	\centering
	\caption{Number of models matching the data, either for $[Fe/H]$ or $[Mg/Fe]$ separately, or for both of them simultaneously. In each case, the number of matches are reported in three different mass ranges referred to as "low", "middle", and "high", respectively, and for all masses ("tot").}\label{table:model_matches}
	\begin{tabular}[c]{|l|c|c|c|c|c|c|c|c|c|c|c|r|}
		\hline
		\multirow{2}{*}{Model} & \multicolumn{4}{c}{$[Fe/H]$}  & \multicolumn{4}{|c|}{$[Mg/Fe]$}  & \multicolumn{4}{|c|}{both} \\
		\cline{2-13}
		& low & middle & high & tot & low & middle & high & tot & low & middle & high & tot \\
		\hline
		01 &  73 & 188 &  44 & 305 &  44 &  137 &  12 &  193 &   3 & 72 & 11 & 86 \\
		02 &  50 &   0 &   0 &  50 & 155 &   71 &   0 &  226 &   3 &  0 &  0 &  3 \\
		03 &   0 &   0 &   0 &   0 &   0 &   27 &  25 &   52 &   0 &  0 &  0 &  0 \\
		04 &  24 &   0 &   0 &  24 &  66 &    0 &   0 &   66 &   5 &  0 &  0 &  5 \\
		05 & 218 & 181 &  54 & 453 & 372 & 1004 & 437 & 1813 &   6 & 23 & 14 & 43 \\
		\hline
	\end{tabular}
\end{table*}

\section{M/L ratios}\label{sec:MLratios}
We find two classes of models providing a good match to the observed stacked spectra:
\begin{enumerate}
	\item Models with a constant, \cite{salpeter1955} IMF;
	\item Models with the bimodal IMF by \cite{vazdekis1997,vazdekis2003}, with a slope switching from an initial $\mu_1$ value (top-heavy) to a different one (bottom heavy) after a time $t_{\text{switch}}=0.1\text{Gyr}$.
\end{enumerate}
As stated in the introduction of the paper, changing the IMF can have strong consequences on the properties of a galaxy, especially on its M/L ratio. In order to verify the plausibility of these two best-fitting sets of models, we investigated the expected M/L ratios by combining luminosities derived from the population synthesis code by \cite{vincenzo2016} with the stellar masses provided by our chemical evolution code. Our predicted $(M/L)_B$ are in the range $11.4-13.5$ for Model01 (Salpeter IMF), and in the range $7.8-12.9$ for Model05 (time-dependent form of the bimodal IMF) depending on the total stellar mass. These ratios turn out to be slightly higher than modern M/L ratios estimates. For comparison, \cite{labarbera2016} showed the stellar r-band (M/L) variation with $\sigma$ for a local sample of elliptical galaxies extracted from the $ATLAS^{3D}$ survey \citep{cappellari2013b}. These ratios were computed from the SDSS r-band luminosities, and were converted into the analogous for the B band by using EMILES SSP models \citep{vazdekis2015,vazdekis2016}; the resulting conversion factor varies between 1.45 and 1.7, according to the mass of the galaxy and not depending on the IMF. After such conversion, their estimated $(M/L)_B$ ratios are in the range (4.9-12.9).\\
The match with our results is actually very good for massive galaxies, but our M/L are larger than the observed ones in less massive objects; in other words, since the trend of M/L with M is related to the tilt of the fundamental plane (FP), our models would imply a shallower tilt than observations suggest.\\
These differences could be mainly ascribed to the use of different prescriptions (i.e., different evolutionary tracks, assumed metallicity).\\
However, it should be stressed that changing the IMF can lead to M/L ratios even order of magnitudes in disagreement with observations \citep{padovani1993}. Since the discrepancies we observe do not go to this extent, and since our main focus is to reproduce the chemical properties of the observed galaxies, specifically the mass-metallicity and the $[Mg/Fe]$-mass relations, we considered the comparison between our models and the dataset satisfying, leaving further investigation to future works.

\section{Environment dependence}\label{sec:environment}

\citetalias{rosani2018} analyzed the environmental dependence of the IMF -mass relation for the SPIDER sample, investigating the impact of hierarchy (central/satellite) and of the mass of the dark matter host halo where galaxies reside.\\
They concluded that while age, $[Z/H]$ and $[Mg/Fe]$ do show a dependence on environment, the IMF slope is not influenced either by hierarchy or by host halo mass, showing a constant trend of increasing (bottom-heavier IMF) in more massive galaxies.
\begin{table*}
	\centering
	\caption{Number of models matching the data, either for $[Fe/H]$ or $[Mg/Fe]$ separately, or for both of them simultaneously. We show the number matches between models and galaxies, for the whole dataset (columns 1-3), for central galaxies only (4-6) and satellites (7-9), respectively.}\label{table:central_satellites_matches}
	\begin{tabular}[c]{|l|c|c|c|c|c|c|c|c|r|}
		\hline
		\multirow{2}{*}{Model} & \multicolumn{3}{c}{All}  & \multicolumn{3}{|c|}{Central}  & \multicolumn{3}{|c|}{Satellites} \\
		\cline{2-10}
		& $[Fe/H]$ & $[Mg/Fe]$ & both & $[Fe/H]$ & $[Mg/Fe]$ & both & $[Fe/H]$ & $[Mg/Fe]$ & both \\
		\hline
		01 & 305 & 193 &  86 & 309 & 193 &   76 & 342 &  204 &  93 \\
		02 &  50 & 226 &   3 &  62 & 194 &    1 &  67 &  284 &   9 \\
		03 &   0 &  52 &   0 &   0 &  61 &    0 &   0 &   14 &   0 \\
		04 &  24 &  66 &   5 &  30 &  43 &    8 &  23 &   82 &   7 \\
		05 & 453 &1813 &  43 & 373 &1713 &   38 & 639 & 1526 &  40\\
		\hline
	\end{tabular}
\end{table*}
We re-applied all of our tests by repeating the matching procedure between models and observed galaxies, which were separated into centrals and satellites. This test gives no particular indication of a dependence of the results on hierarchy, as shown in table \ref{table:central_satellites_matches}, where we report - similarly to table \ref{table:model_matches} - the number of matching models for galaxies with different hierarchy.
In the Appendix, we show plots analogous to figures \ref{fig:match_01} and \ref{fig:match_02} of section \ref{sec:results}, showing models matching the galaxies in the observed central/satellite subsets.

\section{Summary and conclusions}

The IMF is a crucial parameter in establishing the properties of a galaxy. In particular, from the point of view of the chemical properties, the variation of the ratio between low mass and massive stars induced by different IMFs has a very significant effect on the $[\alpha/Fe]$ abundance ratios.\\
In this work, we extend our previous investigation of the effects of IMF on the chemical evolution of elliptical galaxies, by testing the same implementations of the initial mass function used in \cite{demasi2018}, plus two forms of the bimodal IMF by \cite{vazdekis1997,vazdekis2003}. Specifically, a form becoming bottom-heavier in more massive galaxies, as suggested in \cite{ferreras2013}, \citetalias{labarbera2013}, and an explicitly time-dependent form of the latter, switching from top-heavy to bottom-heavy after a time $t_{switch}$ \citep{ferreras2015,weidner2013}. We implement these new IMFs into our chemical evolution code for ellipticals, and test its predictions against a dataset of early-type galaxies extracted from the SPIDER sample \citep{labarbera2010}. For each IMF form, we generate models by varying all parameters of the code over a grid of values, and select the best ones matching the observed $[Fe/H]$-mass and the $[Mg/Fe]$-mass relations within the errors on masses and chemical ratios.\\
All IMF choices provide models matching the data at least in some limited mass bins. However, when it comes to fitting the two observed relations simultaneously over the whole considered mass range, all models fail, aside from Models01 and 05.
These matching models confirm that ``downsizing'' in star formation is required to match the data, meaning that more massive galaxies are characterized by a more efficient, shorter period of star formation (\citealp{matteucci1994,matteucci1998};\citetalias{P04}.)\\
In the first case, we obtain a crucial difference with the previous work, in that we no longer find evidence to advocate for a change in the IMF. This discrepancy can be accounted to a few main reasons:
\begin{itemize}
	\item Simply, the use of a different dataset, characterized by a different slope of the mass-metallicity and $[\alpha/Fe]$ relations;
	\item In \cite{demasi2018}, the main indication for the need of a IMF variation came from the analysis of the spectral indices $Mg_2$ and $<Fe>$, which we derived from the average abundances of the stellar population by applying the calibration relations by \cite{tantalo1998}. Since different calibrations generally yield different results \citep{P04}, this procedure is always plagued by uncertainties, so that results based on such a comparison should be taken with a grain of salt.
	\item One of the main problems one has to deal with when comparing models with data lies in the operational definition of the various considered quantities. Our model directly provides us with the abundances of single chemical elements, whereas the abundance ratios reported for the objects in the catalog are often resulting  from other quantities (see section \ref{sec:dataset} for the definitions of $[Fe/H]$ and $[Mg/Fe]$ adopted in this case). We always tried to be consistent in the comparison, and derived similar quantities from our models by adopting the same definitions as in the observed data. However, this means that the comparison of a given abundance ratio performed on two different catalogs may lead to a discrepancy in the resulting trends.
\end{itemize}
In spite of this difference, we confirm the main result that, as far as the chemical properties of ellipticals are concerned, scenarios involving IMFs which are bottom-heavier through the whole evolution of more massive galaxies should be discarded, since they invariably lead to drastic underestimation of the values of $[\alpha/Fe]$ ratios.\\
Regarding this point, the second successful scenario we describe, i.e. a time-dependent bimodal IMF, allows us to reconcile the indications obtained from chemical abundances (i.e., higher $[\alpha/Fe]$ ratios in massive galaxies) with the results derived from the spectra of stellar population of ellipticals, favoring bottom-heavier IMFs. This IMF is top-heavy in the first period of the chemical evolution of galaxies, thus accounting for the characteristic $[\alpha/Fe]$ trends with mass, and then switch to a different, bottom-heavy form. We stress how, in spite of the various possible values we tested, only models switching IMF at a early time, namely $t_{\text{switch}}=0.1\;Gyr$, were able to fit the data. The bottom-heavy phase would account for observations, as  IMF-sensitive features in the integrated spectra of ETGs at $z\approx0$ are dominated by stars still alive at the present time, i.e. less massive stars, whereas the more massive ones, born during the initial top-heavier phase, do not contribute to the spectra since they died a long time ago. We test different possible values for the switching time, but said switching is always found to take place at the same time, specifically at the earliest possible one, i.e. $t_{\text{switch}}\approx0.1\text{Gyr}$. This, again, is in accordance with observations, since ellipticals are old objects, and consequently the IMF constraints we observe at the present time are related to old stellar populations.\\
Whereas the bimodal IMFs in lower mass galaxies mostly present slopes similar to a canonical Kroupa IMF ($\gamma\approx1.3$), more massive ones span a wider range of values, ranging from $\mu_1=0.5$ (top-heavier) before the switch to $\mu_2=2.6$ (bottom-heavier) after.\\
We decided to investigate the M/L ratios of these two sets of models, to obtain another, independent constraint for the IMF, by combining our masses with luminosities from the population synthesis model of \cite{vincenzo2016}. Our M/L ratio estimates are generally higher than recent observations \citep{cappellari2013b,labarbera2016}, particularly at lower masses, whereas we obtain a good agreement for more massive galaxies. The discrepancies we find are not large enough to provide indication able to discard these models, and we reserve a further analysis of the topic for future works.\\
Finally, we repeat all the tests with a different version of the dataset, where the mass stacking was performed by separating central and satellites galaxies; the obtained results, however, do not show any significant difference, thus reinforcing the idea that the IMF is an intrinsic galaxy property, and is not affected by other ``external'' effects (such as the environment, where galaxies reside).

\section*{Acknowledgements}
CDM and FM acknowledge research funds from the University of Trieste (FRA2016). FLB acknowledges support from grant AYA2016-77237-C3-1-P from the Spanish Ministry of Economy and Competitiveness (MINECO). FV acknowledges funding from the United Kingdom Science and Technology Facility Council through grant ST/M000958/1. This research has made use of the University of Hertfordshire's high-performance computing facility. We also thank the anonymous referee for their useful comments and suggestions, which improved the content and clarity of the paper.

\newpage



\bibliographystyle{apalike}
\bibliography{bibliography.bib} 

\begin{thebibliography}{}
\makeatletter
\relax
\def\mn@urlcharsother{\let\do\@makeother \do\$\do\&\do\#\do\^\do\_\do\%\do\~}
\def\mn@doi{\begingroup\mn@urlcharsother \@ifnextchar [ {\mn@doi@}
  {\mn@doi@[]}}
\def\mn@doi@[#1]#2{\def\@tempa{#1}\ifx\@tempa\@empty \href
  {http://dx.doi.org/#2} {doi:#2}\else \href {http://dx.doi.org/#2} {#1}\fi
  \endgroup}
\def\mn@eprint#1#2{\mn@eprint@#1:#2::\@nil}
\def\mn@eprint@arXiv#1{\href {http://arxiv.org/abs/#1} {{\tt arXiv:#1}}}
\def\mn@eprint@dblp#1{\href {http://dblp.uni-trier.de/rec/bibtex/#1.xml}
  {dblp:#1}}
\def\mn@eprint@#1:#2:#3:#4\@nil{\def\@tempa {#1}\def\@tempb {#2}\def\@tempc
  {#3}\ifx \@tempc \@empty \let \@tempc \@tempb \let \@tempb \@tempa \fi \ifx
  \@tempb \@empty \def\@tempb {arXiv}\fi \@ifundefined
  {mn@eprint@\@tempb}{\@tempb:\@tempc}{\expandafter \expandafter \csname
  mn@eprint@\@tempb\endcsname \expandafter{\@tempc}}}

\bibitem[\protect\citeauthoryear{{Arrigoni}, {Trager}, {Somerville}  \&
  {Gibson}}{{Arrigoni} et~al.}{2010}]{arrigoni2010}
{Arrigoni} M.,  {Trager} S.~C.,  {Somerville} R.~S.,   {Gibson} B.~K.,  2010,
  \mn@doi [\mnras] {10.1111/j.1365-2966.2009.15924.x}, \href
  {http://adsabs.harvard.edu/abs/2010MNRAS.402..173A} {402, 173}

\bibitem[\protect\citeauthoryear{{Auger}, {Treu}, {Gavazzi}, {Bolton},
  {Koopmans}  \& {Marshall}}{{Auger} et~al.}{2010}]{auger2010}
{Auger} M.~W.,  {Treu} T.,  {Gavazzi} R.,  {Bolton} A.~S.,  {Koopmans}
  L.~V.~E.,   {Marshall} P.~J.,  2010, \mn@doi [\apjl]
  {10.1088/2041-8205/721/2/L163}, \href
  {http://adsabs.harvard.edu/abs/2010ApJ...721L.163A} {721, L163}

\bibitem[\protect\citeauthoryear{Barnab{\`e}, Czoske, Koopmans, Treu  \&
  Bolton}{Barnab{\`e} et~al.}{2011}]{barnabe2011}
Barnab{\`e} M.,  Czoske O.,  Koopmans L.~V.,  Treu T.,   Bolton A.~S.,  2011,
  Monthly Notices of the Royal Astronomical Society, 415, 2215

\bibitem[\protect\citeauthoryear{{Bastian}, {Covey}  \& {Meyer}}{{Bastian}
  et~al.}{2010}]{bastian2010}
{Bastian} N.,  {Covey} K.~R.,   {Meyer} M.~R.,  2010, \mn@doi [\araa]
  {10.1146/annurev-astro-082708-101642}, \href
  {http://adsabs.harvard.edu/abs/2010ARA%26A..48..339B} {48, 339}

\bibitem[\protect\citeauthoryear{{Bell}, {McIntosh}, {Katz}  \&
  {Weinberg}}{{Bell} et~al.}{2003}]{bell2003}
{Bell} E.~F.,  {McIntosh} D.~H.,  {Katz} N.,   {Weinberg} M.~D.,  2003, \mn@doi
  [\apjs] {10.1086/378847}, \href
  {http://adsabs.harvard.edu/abs/2003ApJS..149..289B} {149, 289}

\bibitem[\protect\citeauthoryear{{Bonnell}, {Larson}  \& {Zinnecker}}{{Bonnell}
  et~al.}{2007}]{bonnell2007}
{Bonnell} I.~A.,  {Larson} R.~B.,   {Zinnecker} H.,  2007, Protostars and
  Planets V, \href {http://adsabs.harvard.edu/abs/2007prpl.conf..149B} {pp
  149--164}

\bibitem[\protect\citeauthoryear{{Calura} \& {Menci}}{{Calura} \&
  {Menci}}{2009}]{calura2009}
{Calura} F.,  {Menci} N.,  2009, \mn@doi [\mnras]
  {10.1111/j.1365-2966.2009.15440.x}, \href
  {http://adsabs.harvard.edu/abs/2009MNRAS.400.1347C} {400, 1347}

\bibitem[\protect\citeauthoryear{{Cappellari} et~al.,}{{Cappellari}
  et~al.}{2012}]{cappellari2012}
{Cappellari} M.,  et~al., 2012, \mn@doi [\nat] {10.1038/nature10972}, \href
  {http://adsabs.harvard.edu/abs/2012Natur.484..485C} {484, 485}

\bibitem[\protect\citeauthoryear{{Cappellari} et~al.,}{{Cappellari}
  et~al.}{2013}]{cappellari2013b}
{Cappellari} M.,  et~al., 2013, \mn@doi [\mnras] {10.1093/mnras/stt562}, \href
  {http://adsabs.harvard.edu/abs/2013MNRAS.432.1709C} {432, 1709}

\bibitem[\protect\citeauthoryear{{Cappellaro}, {Evans}  \&
  {Turatto}}{{Cappellaro} et~al.}{1999}]{cappellaro1999}
{Cappellaro} E.,  {Evans} R.,   {Turatto} M.,  1999, \aap, \href
  {http://adsabs.harvard.edu/abs/1999A%26A...351..459C} {351, 459}

\bibitem[\protect\citeauthoryear{Cenarro, Gorgas, Vazdekis, Cardiel  \&
  Peletier}{Cenarro et~al.}{2003}]{cenarro2003}
Cenarro A.,  Gorgas J.,  Vazdekis A.,  Cardiel N.,   Peletier R.,  2003,
  Monthly Notices of the Royal Astronomical Society, 339, L12

\bibitem[\protect\citeauthoryear{{Chabrier}}{{Chabrier}}{2003}]{chabrier2003}
{Chabrier} G.,  2003, \mn@doi [\pasp] {10.1086/376392}, \href
  {http://adsabs.harvard.edu/abs/2003PASP..115..763C} {115, 763}

\bibitem[\protect\citeauthoryear{Chabrier, Hennebelle  \& Charlot}{Chabrier
  et~al.}{2014}]{chabrier2014}
Chabrier G.,  Hennebelle P.,   Charlot S.,  2014, The Astrophysical Journal,
  796, 75

\bibitem[\protect\citeauthoryear{{Chiappini}, {Matteucci}  \&
  {Padoan}}{{Chiappini} et~al.}{1997}]{chiappini1997}
{Chiappini} C.,  {Matteucci} F.,   {Padoan} P.,  1997, in {Valls-Gabaud} D.,
  {Hendry} M.~A.,  {Molaro} P.,   {Chamcham} K.,  eds,  Astronomical Society of
  the Pacific Conference Series Vol. 126, From Quantum Fluctuations to
  Cosmological Structures. p.~545

\bibitem[\protect\citeauthoryear{{Cioffi}, {McKee}  \& {Bertschinger}}{{Cioffi}
  et~al.}{1988}]{cioffi1988}
{Cioffi} D.~F.,  {McKee} C.~F.,   {Bertschinger} E.,  1988, \mn@doi [\apj]
  {10.1086/166834}, \href {http://adsabs.harvard.edu/abs/1988ApJ...334..252C}
  {334, 252}

\bibitem[\protect\citeauthoryear{{Conroy} \& {van Dokkum}}{{Conroy} \& {van
  Dokkum}}{2012a}]{conroyvandokkum2012a}
{Conroy} C.,  {van Dokkum} P.,  2012a, \mn@doi [\apj]
  {10.1088/0004-637X/747/1/69}, \href
  {http://adsabs.harvard.edu/abs/2012ApJ...747...69C} {747, 69}

\bibitem[\protect\citeauthoryear{{Conroy} \& {van Dokkum}}{{Conroy} \& {van
  Dokkum}}{2012b}]{conroyvandokkum2012b}
{Conroy} C.,  {van Dokkum} P.~G.,  2012b, \mn@doi [\apj]
  {10.1088/0004-637X/760/1/71}, \href
  {http://adsabs.harvard.edu/abs/2012ApJ...760...71C} {760, 71}

\bibitem[\protect\citeauthoryear{{Conroy}, {Graves}  \& {van Dokkum}}{{Conroy}
  et~al.}{2014}]{conroy2014}
{Conroy} C.,  {Graves} G.~J.,   {van Dokkum} P.~G.,  2014, \mn@doi [\apj]
  {10.1088/0004-637X/780/1/33}, \href
  {http://adsabs.harvard.edu/abs/2014ApJ...780...33C} {780, 33}

\bibitem[\protect\citeauthoryear{{De Masi}, {Matteucci}  \& {Vincenzo}}{{De
  Masi} et~al.}{2018}]{demasi2018}
{De Masi} C.,  {Matteucci} F.,   {Vincenzo} F.,  2018, \mn@doi [\mnras]
  {10.1093/mnras/stx3044}, \href
  {http://adsabs.harvard.edu/abs/2018MNRAS.474.5259D} {474, 5259}

\bibitem[\protect\citeauthoryear{{Diaz}, {Terlevich}  \& {Terlevich}}{{Diaz}
  et~al.}{1989}]{diaz1989}
{Diaz} A.~I.,  {Terlevich} E.,   {Terlevich} R.,  1989, \mn@doi [\mnras]
  {10.1093/mnras/239.2.325}, \href
  {http://adsabs.harvard.edu/abs/1989MNRAS.239..325D} {239, 325}

\bibitem[\protect\citeauthoryear{{Dutton} et~al.,}{{Dutton}
  et~al.}{2011}]{dutton2011}
{Dutton} A.~A.,  et~al., 2011, \mn@doi [\mnras]
  {10.1111/j.1365-2966.2011.19038.x}, \href
  {http://adsabs.harvard.edu/abs/2011MNRAS.416..322D} {416, 322}

\bibitem[\protect\citeauthoryear{{Dutton}, {Mendel}  \& {Simard}}{{Dutton}
  et~al.}{2012}]{dutton2012}
{Dutton} A.~A.,  {Mendel} J.~T.,   {Simard} L.,  2012, \mn@doi [\mnras]
  {10.1111/j.1745-3933.2012.01230.x}, \href
  {http://adsabs.harvard.edu/abs/2012MNRAS.422L..33D} {422, L33}

\bibitem[\protect\citeauthoryear{{Dutton}, {Macci{\`o}}, {Mendel}  \&
  {Simard}}{{Dutton} et~al.}{2013}]{dutton2013}
{Dutton} A.~A.,  {Macci{\`o}} A.~V.,  {Mendel} J.~T.,   {Simard} L.,  2013,
  \mn@doi [\mnras] {10.1093/mnras/stt608}, \href
  {http://adsabs.harvard.edu/abs/2013MNRAS.432.2496D} {432, 2496}

\bibitem[\protect\citeauthoryear{{Faber} \& {French}}{{Faber} \&
  {French}}{1980}]{faber1980}
{Faber} S.~M.,  {French} H.~B.,  1980, \mn@doi [\apj] {10.1086/157644}, \href
  {http://adsabs.harvard.edu/abs/1980ApJ...235..405F} {235, 405}

\bibitem[\protect\citeauthoryear{{Ferreras}, {La Barbera}, {de la Rosa},
  {Vazdekis}, {de Carvalho}, {Falc{\'o}n-Barroso}  \&
  {Ricciardelli}}{{Ferreras} et~al.}{2013}]{ferreras2013}
{Ferreras} I.,  {La Barbera} F.,  {de la Rosa} I.~G.,  {Vazdekis} A.,  {de
  Carvalho} R.~R.,  {Falc{\'o}n-Barroso} J.,   {Ricciardelli} E.,  2013,
  \mn@doi [\mnras] {10.1093/mnrasl/sls014}, \href
  {http://adsabs.harvard.edu/abs/2013MNRAS.429L..15F} {429, L15}

\bibitem[\protect\citeauthoryear{{Ferreras}, {Weidner}, {Vazdekis}  \& {La
  Barbera}}{{Ferreras} et~al.}{2015}]{ferreras2015}
{Ferreras} I.,  {Weidner} C.,  {Vazdekis} A.,   {La Barbera} F.,  2015, \mn@doi
  [\mnras] {10.1093/mnrasl/slv003}, \href
  {http://adsabs.harvard.edu/abs/2015MNRAS.448L..82F} {448, L82}

\bibitem[\protect\citeauthoryear{{Fontanot}, {De Lucia}, {Hirschmann},
  {Bruzual}, {Charlot}  \& {Zibetti}}{{Fontanot} et~al.}{2017}]{fontanot2017}
{Fontanot} F.,  {De Lucia} G.,  {Hirschmann} M.,  {Bruzual} G.,  {Charlot} S.,
   {Zibetti} S.,  2017, \mn@doi [\mnras] {10.1093/mnras/stw2612}, \href
  {http://adsabs.harvard.edu/abs/2017MNRAS.464.3812F} {464, 3812}

\bibitem[\protect\citeauthoryear{{Fontanot}, {Barbera}, {De Lucia}, {Pasquali}
  \& {Vazdekis}}{{Fontanot} et~al.}{2018a}]{fontanot2018b}
{Fontanot} F.,  {Barbera} F.~L.,  {De Lucia} G.,  {Pasquali} A.,   {Vazdekis}
  A.,  2018a, \mn@doi [\mnras] {10.1093/mnras/sty1768}, \href
  {http://adsabs.harvard.edu/abs/2018MNRAS.tmp.1691F} {}

\bibitem[\protect\citeauthoryear{{Fontanot}, {De Lucia}, {Xie}, {Hirschmann},
  {Bruzual}  \& {Charlot}}{{Fontanot} et~al.}{2018b}]{fontanot2018a}
{Fontanot} F.,  {De Lucia} G.,  {Xie} L.,  {Hirschmann} M.,  {Bruzual} G.,
  {Charlot} S.,  2018b, \mn@doi [\mnras] {10.1093/mnras/stx3323}, \href
  {http://adsabs.harvard.edu/abs/2018MNRAS.475.2467F} {475, 2467}

\bibitem[\protect\citeauthoryear{{Gargiulo} et~al.,}{{Gargiulo}
  et~al.}{2015}]{gargiulo2015}
{Gargiulo} I.~D.,  et~al., 2015, \mn@doi [\mnras] {10.1093/mnras/stu2272},
  \href {http://adsabs.harvard.edu/abs/2015MNRAS.446.3820G} {446, 3820}

\bibitem[\protect\citeauthoryear{{Gibson}}{{Gibson}}{1997}]{gibson1997a}
{Gibson} B.~K.,  1997, \mn@doi [\mnras] {10.1093/mnras/290.3.471}, \href
  {http://adsabs.harvard.edu/abs/1997MNRAS.290..471G} {290, 471}

\bibitem[\protect\citeauthoryear{{Greggio} \& {Renzini}}{{Greggio} \&
  {Renzini}}{1983}]{greggio1983}
{Greggio} L.,  {Renzini} A.,  1983, \aap, \href
  {http://adsabs.harvard.edu/abs/1983A%26A...118..217G} {118, 217}

\bibitem[\protect\citeauthoryear{{Grevesse} \& {Sauval}}{{Grevesse} \&
  {Sauval}}{1998}]{grevesse1998}
{Grevesse} N.,  {Sauval} A.~J.,  1998, \mn@doi [\ssr]
  {10.1023/A:1005161325181}, \href
  {https://ui.adsabs.harvard.edu/#abs/1998SSRv...85..161G} {85, 161}

\bibitem[\protect\citeauthoryear{{Grillo} \& {Gobat}}{{Grillo} \&
  {Gobat}}{2010}]{grillo2010}
{Grillo} C.,  {Gobat} R.,  2010, \mn@doi [\mnras]
  {10.1111/j.1745-3933.2009.00803.x}, \href
  {http://adsabs.harvard.edu/abs/2010MNRAS.402L..67G} {402, L67}

\bibitem[\protect\citeauthoryear{{Gunawardhana} et~al.,}{{Gunawardhana}
  et~al.}{2011}]{gunawardhana2011}
{Gunawardhana} M.~L.~P.,  et~al., 2011, \mn@doi [\mnras]
  {10.1111/j.1365-2966.2011.18800.x}, \href
  {http://adsabs.harvard.edu/abs/2011MNRAS.415.1647G} {415, 1647}

\bibitem[\protect\citeauthoryear{Hopkins}{Hopkins}{2013}]{hopkins2013}
Hopkins P.~F.,  2013, Monthly Notices of the Royal Astronomical Society, 433,
  170

\bibitem[\protect\citeauthoryear{{Johansson}, {Thomas}  \&
  {Maraston}}{{Johansson} et~al.}{2012}]{johansson2012}
{Johansson} J.,  {Thomas} D.,   {Maraston} C.,  2012, \mn@doi [\mnras]
  {10.1111/j.1365-2966.2011.20316.x}, \href
  {http://adsabs.harvard.edu/abs/2012MNRAS.421.1908J} {421, 1908}

\bibitem[\protect\citeauthoryear{{Kennicutt}}{{Kennicutt}}{1998}]{kennicutt1998}
{Kennicutt} Jr. R.~C.,  1998, \mn@doi [\apj] {10.1086/305588}, \href
  {http://adsabs.harvard.edu/abs/1998ApJ...498..541K} {498, 541}

\bibitem[\protect\citeauthoryear{{Kroupa}}{{Kroupa}}{2001}]{kroupa2001}
{Kroupa} P.,  2001, \mn@doi [\mnras] {10.1046/j.1365-8711.2001.04022.x}, \href
  {http://adsabs.harvard.edu/abs/2001MNRAS.322..231K} {322, 231}

\bibitem[\protect\citeauthoryear{{Kroupa}}{{Kroupa}}{2002}]{kroupa2002}
{Kroupa} P.,  2002, \mn@doi [Science] {10.1126/science.1067524}, \href
  {http://adsabs.harvard.edu/abs/2002Sci...295...82K} {295, 82}

\bibitem[\protect\citeauthoryear{{Kroupa}, {Weidner}, {Pflamm-Altenburg},
  {Thies}, {Dabringhausen}, {Marks}  \& {Maschberger}}{{Kroupa}
  et~al.}{2013}]{kroupa2013}
{Kroupa} P.,  {Weidner} C.,  {Pflamm-Altenburg} J.,  {Thies} I.,
  {Dabringhausen} J.,  {Marks} M.,   {Maschberger} T.,  2013, {The Stellar and
  Sub-Stellar Initial Mass Function of Simple and Composite Populations}.
p.~115, \mn@doi{10.1007/978-94-007-5612-0_4}

\bibitem[\protect\citeauthoryear{{Krumholz}}{{Krumholz}}{2011}]{krumholz2011}
{Krumholz} M.~R.,  2011, \mn@doi [\apj] {10.1088/0004-637X/743/2/110}, \href
  {http://adsabs.harvard.edu/abs/2011ApJ...743..110K} {743, 110}

\bibitem[\protect\citeauthoryear{{Kuntschner}}{{Kuntschner}}{2000}]{kuntschner2000}
{Kuntschner} H.,  2000, \mn@doi [\mnras] {10.1046/j.1365-8711.2000.03377.x},
  \href {http://adsabs.harvard.edu/abs/2000MNRAS.315..184K} {315, 184}

\bibitem[\protect\citeauthoryear{{La Barbera}, {de Carvalho}, {de La Rosa},
  {Lopes}, {Kohl-Moreira}  \& {Capelato}}{{La Barbera}
  et~al.}{2010}]{labarbera2010}
{La Barbera} F.,  {de Carvalho} R.~R.,  {de La Rosa} I.~G.,  {Lopes} P.~A.~A.,
  {Kohl-Moreira} J.~L.,   {Capelato} H.~V.,  2010, \mn@doi [\mnras]
  {10.1111/j.1365-2966.2010.16850.x}, \href
  {http://adsabs.harvard.edu/abs/2010MNRAS.408.1313L} {408, 1313}

\bibitem[\protect\citeauthoryear{{La Barbera}, {Ferreras}, {Vazdekis}, {de la
  Rosa}, {de Carvalho}, {Trevisan}, {Falc{\'o}n-Barroso}  \&
  {Ricciardelli}}{{La Barbera} et~al.}{2013}]{labarbera2013}
{La Barbera} F.,  {Ferreras} I.,  {Vazdekis} A.,  {de la Rosa} I.~G.,  {de
  Carvalho} R.~R.,  {Trevisan} M.,  {Falc{\'o}n-Barroso} J.,   {Ricciardelli}
  E.,  2013, \mn@doi [\mnras] {10.1093/mnras/stt943}, \href
  {http://adsabs.harvard.edu/abs/2013MNRAS.433.3017L} {433, 3017}

\bibitem[\protect\citeauthoryear{{La Barbera}, {Vazdekis}, {Ferreras},
  {Pasquali}, {Cappellari}, {Mart{\'{\i}}n-Navarro}, {Sch{\"o}nebeck}  \&
  {Falc{\'o}n-Barroso}}{{La Barbera} et~al.}{2016}]{labarbera2016}
{La Barbera} F.,  {Vazdekis} A.,  {Ferreras} I.,  {Pasquali} A.,  {Cappellari}
  M.,  {Mart{\'{\i}}n-Navarro} I.,  {Sch{\"o}nebeck} F.,   {Falc{\'o}n-Barroso}
  J.,  2016, \mn@doi [\mnras] {10.1093/mnras/stv2996}, \href
  {http://adsabs.harvard.edu/abs/2016MNRAS.457.1468L} {457, 1468}

\bibitem[\protect\citeauthoryear{{Larson}}{{Larson}}{1974}]{larson1974}
{Larson} R.~B.,  1974, \mn@doi [\mnras] {10.1093/mnras/166.3.585}, \href
  {http://adsabs.harvard.edu/abs/1974MNRAS.166..585L} {166, 585}

\bibitem[\protect\citeauthoryear{{Larson}}{{Larson}}{1998}]{larson1998}
{Larson} R.~B.,  1998, \mn@doi [\mnras] {10.1046/j.1365-8711.1998.02045.x},
  \href {http://adsabs.harvard.edu/abs/1998MNRAS.301..569L} {301, 569}

\bibitem[\protect\citeauthoryear{{Larson}}{{Larson}}{2005}]{larson2005}
{Larson} R.~B.,  2005, \mn@doi [\mnras] {10.1111/j.1365-2966.2005.08881.x},
  \href {http://adsabs.harvard.edu/abs/2005MNRAS.359..211L} {359, 211}

\bibitem[\protect\citeauthoryear{{Matteucci}}{{Matteucci}}{1994}]{matteucci1994}
{Matteucci} F.,  1994, \aap, \href
  {http://adsabs.harvard.edu/abs/1994A%26A...288...57M} {288, 57}

\bibitem[\protect\citeauthoryear{{Matteucci}}{{Matteucci}}{2012}]{matteucci2012}
{Matteucci} F.,  2012, {Chemical Evolution of Galaxies},
  \mn@doi{10.1007/978-3-642-22491-1.
}

\bibitem[\protect\citeauthoryear{{Matteucci} \& {Gibson}}{{Matteucci} \&
  {Gibson}}{1995}]{matteucci1995}
{Matteucci} F.,  {Gibson} B.~K.,  1995, \aap, \href
  {http://adsabs.harvard.edu/abs/1995A%26A...304...11M} {304, 11}

\bibitem[\protect\citeauthoryear{{Matteucci} \& {Greggio}}{{Matteucci} \&
  {Greggio}}{1986}]{matteucci1986}
{Matteucci} F.,  {Greggio} L.,  1986, \aap, \href
  {http://adsabs.harvard.edu/abs/1986A%26A...154..279M} {154, 279}

\bibitem[\protect\citeauthoryear{{Matteucci} \& {Recchi}}{{Matteucci} \&
  {Recchi}}{2001}]{matteucci2001}
{Matteucci} F.,  {Recchi} S.,  2001, \mn@doi [\apj] {10.1086/322472}, \href
  {http://adsabs.harvard.edu/abs/2001ApJ...558..351M} {558, 351}

\bibitem[\protect\citeauthoryear{{Matteucci}, {Ponzone}  \&
  {Gibson}}{{Matteucci} et~al.}{1998}]{matteucci1998}
{Matteucci} F.,  {Ponzone} R.,   {Gibson} B.~K.,  1998, \aap, \href
  {http://adsabs.harvard.edu/abs/1998A%26A...335..855M} {335, 855}

\bibitem[\protect\citeauthoryear{{Newman}, {Smith}, {Conroy}, {Villaume}  \&
  {van Dokkum}}{{Newman} et~al.}{2017}]{newman2017}
{Newman} A.~B.,  {Smith} R.~J.,  {Conroy} C.,  {Villaume} A.,   {van Dokkum}
  P.,  2017, \mn@doi [\apj] {10.3847/1538-4357/aa816d}, \href
  {http://adsabs.harvard.edu/abs/2017ApJ...845..157N} {845, 157}

\bibitem[\protect\citeauthoryear{{Okamoto}, {Nagashima}, {Lacey}  \&
  {Frenk}}{{Okamoto} et~al.}{2017}]{okamoto2017}
{Okamoto} T.,  {Nagashima} M.,  {Lacey} C.~G.,   {Frenk} C.~S.,  2017, \mn@doi
  [\mnras] {10.1093/mnras/stw2729}, \href
  {http://adsabs.harvard.edu/abs/2017MNRAS.464.4866O} {464, 4866}

\bibitem[\protect\citeauthoryear{{Padovani} \& {Matteucci}}{{Padovani} \&
  {Matteucci}}{1993}]{padovani1993}
{Padovani} P.,  {Matteucci} F.,  1993, \mn@doi [\apj] {10.1086/173212}, \href
  {https://ui.adsabs.harvard.edu/#abs/1993ApJ...416...26P} {416, 26}

\bibitem[\protect\citeauthoryear{{Pipino} \& {Matteucci}}{{Pipino} \&
  {Matteucci}}{2004}]{P04}
{Pipino} A.,  {Matteucci} F.,  2004, \mn@doi [\mnras]
  {10.1111/j.1365-2966.2004.07268.x}, \href
  {http://adsabs.harvard.edu/abs/2004MNRAS.347..968P} {347, 968}

\bibitem[\protect\citeauthoryear{{Pipino} \& {Matteucci}}{{Pipino} \&
  {Matteucci}}{2008}]{pipino2008}
{Pipino} A.,  {Matteucci} F.,  2008, \mn@doi [\aap]
  {10.1051/0004-6361:200809395}, \href
  {http://adsabs.harvard.edu/abs/2008A%26A...486..763P} {486, 763}

\bibitem[\protect\citeauthoryear{{Pipino}, {Matteucci}, {Borgani}  \&
  {Biviano}}{{Pipino} et~al.}{2002}]{pipino2002}
{Pipino} A.,  {Matteucci} F.,  {Borgani} S.,   {Biviano} A.,  2002, \mn@doi
  [\na] {10.1016/S1384-1076(02)00136-7}, \href
  {http://adsabs.harvard.edu/abs/2002NewA....7..227P} {7, 227}

\bibitem[\protect\citeauthoryear{{Recchi}, {Matteucci}  \& {D'Ercole}}{{Recchi}
  et~al.}{2001}]{recchi2001}
{Recchi} S.,  {Matteucci} F.,   {D'Ercole} A.,  2001, \mn@doi [\mnras]
  {10.1046/j.1365-8711.2001.04189.x}, \href
  {http://adsabs.harvard.edu/abs/2001MNRAS.322..800R} {322, 800}

\bibitem[\protect\citeauthoryear{{Recchi}, {Calura}  \& {Kroupa}}{{Recchi}
  et~al.}{2009}]{recchi2009}
{Recchi} S.,  {Calura} F.,   {Kroupa} P.,  2009, \mn@doi [\aap]
  {10.1051/0004-6361/200811472}, \href
  {http://adsabs.harvard.edu/abs/2009A%26A...499..711R} {499, 711}

\bibitem[\protect\citeauthoryear{Renzini \& Greggio}{Renzini \&
  Greggio}{2012}]{renzini2012}
Renzini A.,  Greggio L.,  2012, Stellar populations: a guide from low to high
  redshift.
John Wiley \& Sons

\bibitem[\protect\citeauthoryear{Rosani, Pasquali, La~Barbera, Ferreras  \&
  Vazdekis}{Rosani et~al.}{2018}]{rosani2018}
Rosani G.,  Pasquali A.,  La~Barbera F.,  Ferreras I.,   Vazdekis A.,  2018,
  \mn@doi [Monthly Notices of the Royal Astronomical Society]
  {10.1093/mnras/sty528}, p. sty528

\bibitem[\protect\citeauthoryear{{Salpeter}}{{Salpeter}}{1955}]{salpeter1955}
{Salpeter} E.~E.,  1955, \mn@doi [\apj] {10.1086/145971}, \href
  {http://adsabs.harvard.edu/abs/1955ApJ...121..161S} {121, 161}

\bibitem[\protect\citeauthoryear{{Scalo}}{{Scalo}}{1986}]{scalo1986}
{Scalo} J.~M.,  1986, \fcp, \href
  {http://adsabs.harvard.edu/abs/1986FCPh...11....1S} {11, 1}

\bibitem[\protect\citeauthoryear{{Silk}}{{Silk}}{1995}]{silk1995}
{Silk} J.,  1995, \mn@doi [\apjl] {10.1086/187710}, \href
  {http://adsabs.harvard.edu/abs/1995ApJ...438L..41S} {438, L41}

\bibitem[\protect\citeauthoryear{{Smith} \& {Lucey}}{{Smith} \&
  {Lucey}}{2013}]{smith2013}
{Smith} R.~J.,  {Lucey} J.~R.,  2013, \mn@doi [\mnras] {10.1093/mnras/stt1141},
  \href {http://adsabs.harvard.edu/abs/2013MNRAS.434.1964S} {434, 1964}

\bibitem[\protect\citeauthoryear{{Smith}, {Lucey}  \& {Conroy}}{{Smith}
  et~al.}{2015}]{smith2015}
{Smith} R.~J.,  {Lucey} J.~R.,   {Conroy} C.,  2015, \mn@doi [\mnras]
  {10.1093/mnras/stv518}, \href
  {http://adsabs.harvard.edu/abs/2015MNRAS.449.3441S} {449, 3441}

\bibitem[\protect\citeauthoryear{{Spiniello}, {Trager}, {Koopmans}  \&
  {Chen}}{{Spiniello} et~al.}{2012}]{spiniello2012}
{Spiniello} C.,  {Trager} S.~C.,  {Koopmans} L.~V.~E.,   {Chen} Y.~P.,  2012,
  \mn@doi [\apjl] {10.1088/2041-8205/753/2/L32}, \href
  {http://adsabs.harvard.edu/abs/2012ApJ...753L..32S} {753, L32}

\bibitem[\protect\citeauthoryear{{Spiniello}, {Trager}, {Koopmans}  \&
  {Conroy}}{{Spiniello} et~al.}{2014}]{spiniello2014}
{Spiniello} C.,  {Trager} S.,  {Koopmans} L.~V.~E.,   {Conroy} C.,  2014,
  \mn@doi [\mnras] {10.1093/mnras/stt2282}, \href
  {http://adsabs.harvard.edu/abs/2014MNRAS.438.1483S} {438, 1483}

\bibitem[\protect\citeauthoryear{{Tantalo}, {Chiosi}  \& {Bressan}}{{Tantalo}
  et~al.}{1998}]{tantalo1998}
{Tantalo} R.,  {Chiosi} C.,   {Bressan} A.,  1998, \aap, \href
  {http://adsabs.harvard.edu/abs/1998A%26A...333..419T} {333, 419}

\bibitem[\protect\citeauthoryear{{Thomas}, {Greggio}  \& {Bender}}{{Thomas}
  et~al.}{1999}]{thomas1999}
{Thomas} D.,  {Greggio} L.,   {Bender} R.,  1999, \mn@doi [\mnras]
  {10.1046/j.1365-8711.1999.02138.x}, \href
  {http://adsabs.harvard.edu/abs/1999MNRAS.302..537T} {302, 537}

\bibitem[\protect\citeauthoryear{{Thomas}, {Maraston}  \& {Bender}}{{Thomas}
  et~al.}{2003}]{thomas2003stellar}
{Thomas} D.,  {Maraston} C.,   {Bender} R.,  2003, \mn@doi [\mnras]
  {10.1046/j.1365-8711.2003.06248.x}, \href
  {http://adsabs.harvard.edu/abs/2003MNRAS.339..897T} {339, 897}

\bibitem[\protect\citeauthoryear{{Thomas}, {Maraston}, {Schawinski}, {Sarzi}
  \& {Silk}}{{Thomas} et~al.}{2010}]{thomas2010}
{Thomas} D.,  {Maraston} C.,  {Schawinski} K.,  {Sarzi} M.,   {Silk} J.,  2010,
  \mn@doi [\mnras] {10.1111/j.1365-2966.2010.16427.x}, \href
  {http://adsabs.harvard.edu/abs/2010MNRAS.404.1775T} {404, 1775}

\bibitem[\protect\citeauthoryear{{Trager}, {Worthey}, {Faber}, {Burstein}  \&
  {Gonz{\'a}lez}}{{Trager} et~al.}{1998}]{trager1998}
{Trager} S.~C.,  {Worthey} G.,  {Faber} S.~M.,  {Burstein} D.,   {Gonz{\'a}lez}
  J.~J.,  1998, \mn@doi [\apjs] {10.1086/313099}, \href
  {http://adsabs.harvard.edu/abs/1998ApJS..116....1T} {116, 1}

\bibitem[\protect\citeauthoryear{{Treu}, {Auger}, {Koopmans}, {Gavazzi},
  {Marshall}  \& {Bolton}}{{Treu} et~al.}{2010}]{treu2010}
{Treu} T.,  {Auger} M.~W.,  {Koopmans} L.~V.~E.,  {Gavazzi} R.,  {Marshall}
  P.~J.,   {Bolton} A.~S.,  2010, \mn@doi [\apj]
  {10.1088/0004-637X/709/2/1195}, \href
  {http://adsabs.harvard.edu/abs/2010ApJ...709.1195T} {709, 1195}

\bibitem[\protect\citeauthoryear{{Vazdekis}, {Peletier}, {Beckman}  \&
  {Casuso}}{{Vazdekis} et~al.}{1997}]{vazdekis1997}
{Vazdekis} A.,  {Peletier} R.~F.,  {Beckman} J.~E.,   {Casuso} E.,  1997,
  \mn@doi [\apjs] {10.1086/313008}, \href
  {http://adsabs.harvard.edu/abs/1997ApJS..111..203V} {111, 203}

\bibitem[\protect\citeauthoryear{{Vazdekis}, {Cenarro}, {Gorgas}, {Cardiel}  \&
  {Peletier}}{{Vazdekis} et~al.}{2003}]{vazdekis2003}
{Vazdekis} A.,  {Cenarro} A.~J.,  {Gorgas} J.,  {Cardiel} N.,   {Peletier}
  R.~F.,  2003, \mn@doi [\mnras] {10.1046/j.1365-8711.2003.06413.x}, \href
  {http://adsabs.harvard.edu/abs/2003MNRAS.340.1317V} {340, 1317}

\bibitem[\protect\citeauthoryear{{Vazdekis} et~al.,}{{Vazdekis}
  et~al.}{2015}]{vazdekis2015}
{Vazdekis} A.,  et~al., 2015, \mn@doi [\mnras] {10.1093/mnras/stv151}, \href
  {http://adsabs.harvard.edu/abs/2015MNRAS.449.1177V} {449, 1177}

\bibitem[\protect\citeauthoryear{{Vazdekis}, {Koleva}, {Ricciardelli},
  {R{\"o}ck}  \& {Falc{\'o}n-Barroso}}{{Vazdekis} et~al.}{2016}]{vazdekis2016}
{Vazdekis} A.,  {Koleva} M.,  {Ricciardelli} E.,  {R{\"o}ck} B.,
  {Falc{\'o}n-Barroso} J.,  2016, \mn@doi [\mnras] {10.1093/mnras/stw2231},
  \href {http://adsabs.harvard.edu/abs/2016MNRAS.463.3409V} {463, 3409}

\bibitem[\protect\citeauthoryear{{Vincenzo}, {Matteucci}, {Vattakunnel}  \&
  {Lanfranchi}}{{Vincenzo} et~al.}{2014}]{vincenzo2014}
{Vincenzo} F.,  {Matteucci} F.,  {Vattakunnel} S.,   {Lanfranchi} G.~A.,  2014,
  \mn@doi [\mnras] {10.1093/mnras/stu710}, \href
  {http://adsabs.harvard.edu/abs/2014MNRAS.441.2815V} {441, 2815}

\bibitem[\protect\citeauthoryear{{Vincenzo}, {Matteucci}, {Recchi}, {Calura},
  {McWilliam}  \& {Lanfranchi}}{{Vincenzo} et~al.}{2015}]{vincenzo2015}
{Vincenzo} F.,  {Matteucci} F.,  {Recchi} S.,  {Calura} F.,  {McWilliam} A.,
  {Lanfranchi} G.~A.,  2015, \mn@doi [\mnras] {10.1093/mnras/stv357}, \href
  {http://adsabs.harvard.edu/abs/2015MNRAS.449.1327V} {449, 1327}

\bibitem[\protect\citeauthoryear{{Vincenzo}, {Matteucci}, {de Boer}, {Cignoni}
  \& {Tosi}}{{Vincenzo} et~al.}{2016}]{vincenzo2016}
{Vincenzo} F.,  {Matteucci} F.,  {de Boer} T.~J.~L.,  {Cignoni} M.,   {Tosi}
  M.,  2016, \mn@doi [\mnras] {10.1093/mnras/stw1145}, \href
  {http://adsabs.harvard.edu/abs/2016MNRAS.460.2238V} {460, 2238}

\bibitem[\protect\citeauthoryear{{Wang} et~al.,}{{Wang}
  et~al.}{2014}]{wang2014}
{Wang} L.,  et~al., 2014, \mn@doi [\mnras] {10.1093/mnras/stt2481}, \href
  {http://adsabs.harvard.edu/abs/2014MNRAS.439..611W} {439, 611}

\bibitem[\protect\citeauthoryear{{Weidner}, {Kroupa}  \& {Bonnell}}{{Weidner}
  et~al.}{2010}]{weidner2010}
{Weidner} C.,  {Kroupa} P.,   {Bonnell} I.~A.~D.,  2010, \mn@doi [\mnras]
  {10.1111/j.1365-2966.2009.15633.x}, \href
  {http://adsabs.harvard.edu/abs/2010MNRAS.401..275W} {401, 275}

\bibitem[\protect\citeauthoryear{{Weidner}, {Kroupa}  \&
  {Pflamm-Altenburg}}{{Weidner} et~al.}{2011}]{weidner2011}
{Weidner} C.,  {Kroupa} P.,   {Pflamm-Altenburg} J.,  2011, \mn@doi [\mnras]
  {10.1111/j.1365-2966.2010.17959.x}, \href
  {http://adsabs.harvard.edu/abs/2011MNRAS.412..979W} {412, 979}

\bibitem[\protect\citeauthoryear{{Weidner}, {Ferreras}, {Vazdekis}  \& {La
  Barbera}}{{Weidner} et~al.}{2013}]{weidner2013}
{Weidner} C.,  {Ferreras} I.,  {Vazdekis} A.,   {La Barbera} F.,  2013, \mn@doi
  [\mnras] {10.1093/mnras/stt1445}, \href
  {http://adsabs.harvard.edu/abs/2013MNRAS.435.2274W} {435, 2274}

\bibitem[\protect\citeauthoryear{{Whelan} \& {Iben}}{{Whelan} \&
  {Iben}}{1973}]{whelan1973}
{Whelan} J.,  {Iben} Jr. I.,  1973, \mn@doi [\apj] {10.1086/152565}, \href
  {http://adsabs.harvard.edu/abs/1973ApJ...186.1007W} {186, 1007}

\bibitem[\protect\citeauthoryear{{Wing} \& {Ford}}{{Wing} \&
  {Ford}}{1969}]{wing1969}
{Wing} R.~F.,  {Ford} Jr. W.~K.,  1969, \mn@doi [\pasp] {10.1086/128814}, \href
  {http://adsabs.harvard.edu/abs/1969PASP...81..527W} {81, 527}

\bibitem[\protect\citeauthoryear{{Yang}, {Mo}, {van den Bosch}, {Pasquali},
  {Li}  \& {Barden}}{{Yang} et~al.}{2007}]{yang2007}
{Yang} X.,  {Mo} H.~J.,  {van den Bosch} F.~C.,  {Pasquali} A.,  {Li} C.,
  {Barden} M.,  2007, \mn@doi [\apj] {10.1086/522027}, \href
  {http://adsabs.harvard.edu/abs/2007ApJ...671..153Y} {671, 153}

\bibitem[\protect\citeauthoryear{{Yoshii} \& {Arimoto}}{{Yoshii} \&
  {Arimoto}}{1987}]{yoshii1987}
{Yoshii} Y.,  {Arimoto} N.,  1987, \aap, \href
  {http://adsabs.harvard.edu/abs/1987A%26A...188...13Y} {188, 13}

\bibitem[\protect\citeauthoryear{{van Dokkum} \& {Conroy}}{{van Dokkum} \&
  {Conroy}}{2010}]{vandokkum2010}
{van Dokkum} P.~G.,  {Conroy} C.,  2010, \mn@doi [\nat] {10.1038/nature09578},
  \href {http://adsabs.harvard.edu/abs/2010Natur.468..940V} {468, 940}

\bibitem[\protect\citeauthoryear{{van Dokkum} \& {Conroy}}{{van Dokkum} \&
  {Conroy}}{2011}]{vandokkum2011}
{van Dokkum} P.~G.,  {Conroy} C.,  2011, \mn@doi [\apjl]
  {10.1088/2041-8205/735/1/L13}, \href
  {http://adsabs.harvard.edu/abs/2011ApJ...735L..13V} {735, L13}

\makeatother
\end{thebibliography}




\appendix

\section{Matches for central/satellites}
\begin{figure*}
	\includegraphics[width=.8\textwidth]{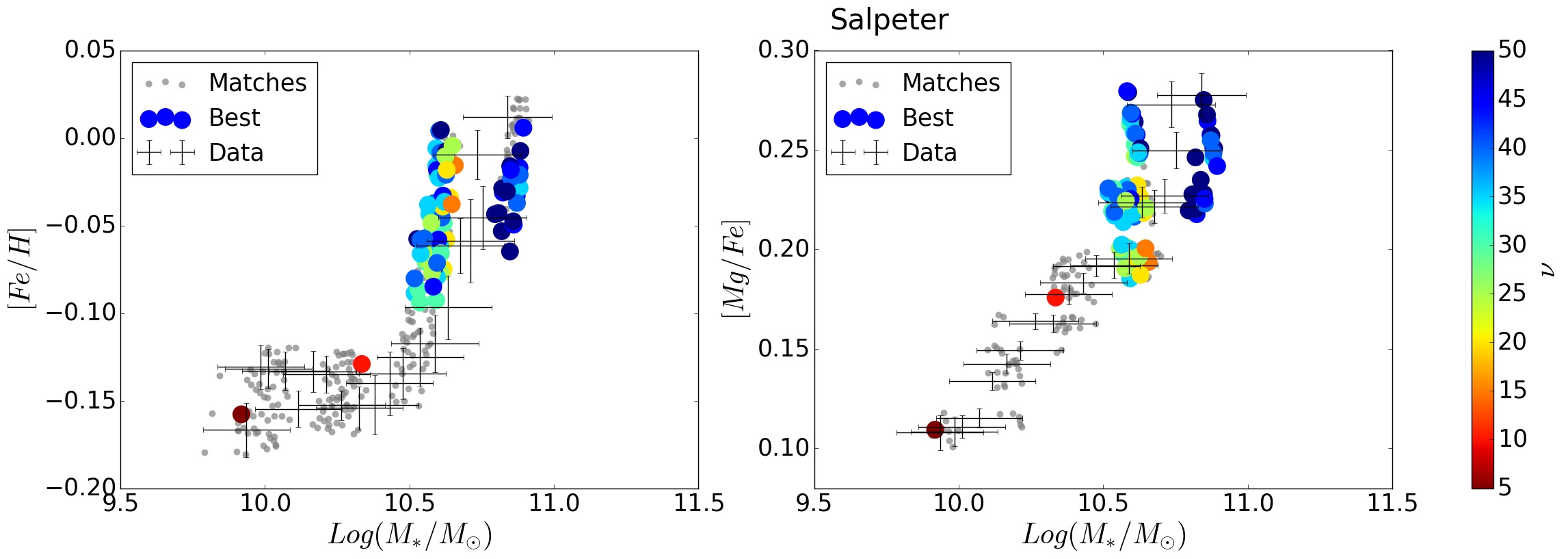}
	\includegraphics[width=.8\textwidth]{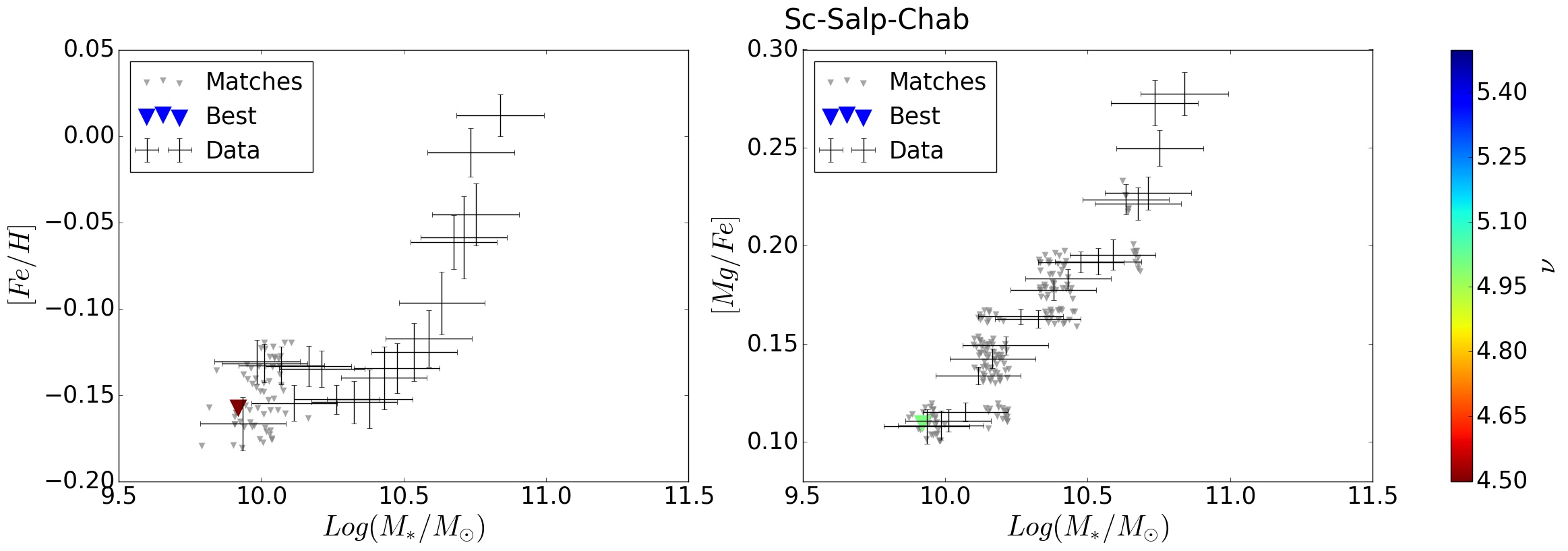}
	\includegraphics[width=.8\textwidth]{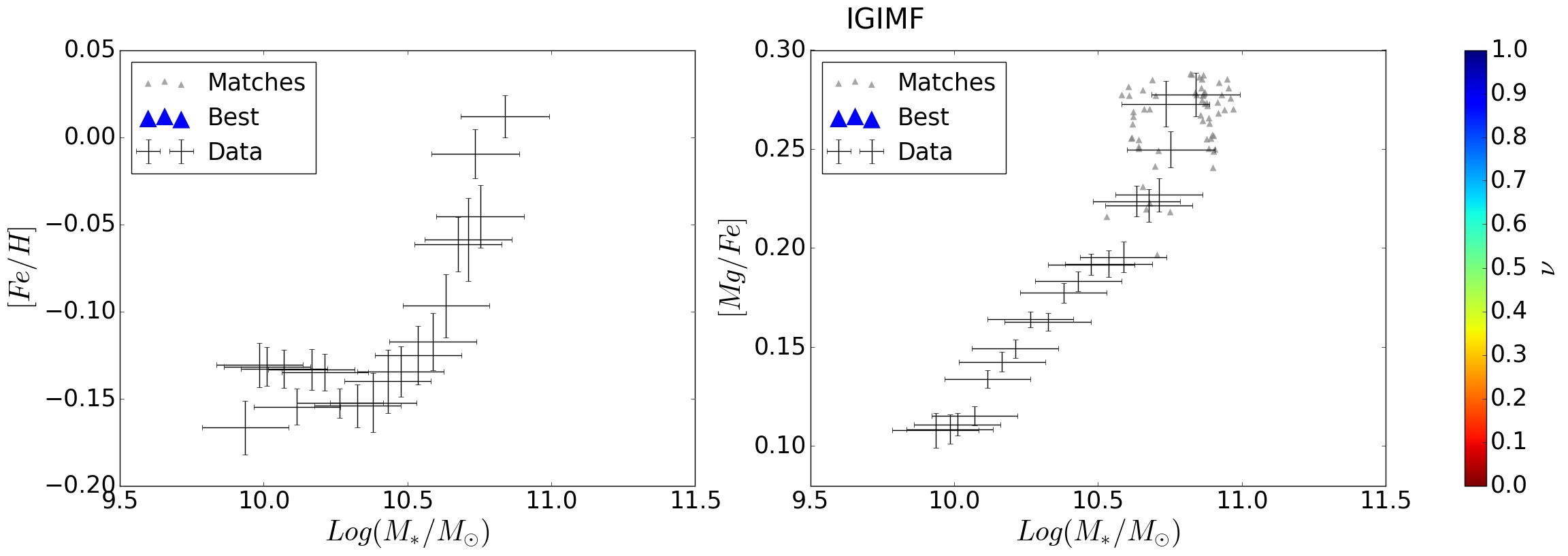}
	\caption{Comparison between central galaxies in the dataset data and Models01 (top row), 02 (central row) and 03 (bottom row) for the $[Fe/H]$ and $[Mg/Fe]$ abundance ratios. Matching models are color-coded according to their star formation efficiency, while not matching ones are shown with fading, smaller markers.}
	\label{fig:match_01_CN}
\end{figure*}

\begin{figure*}
	\centering
	\includegraphics[width=.8\textwidth]{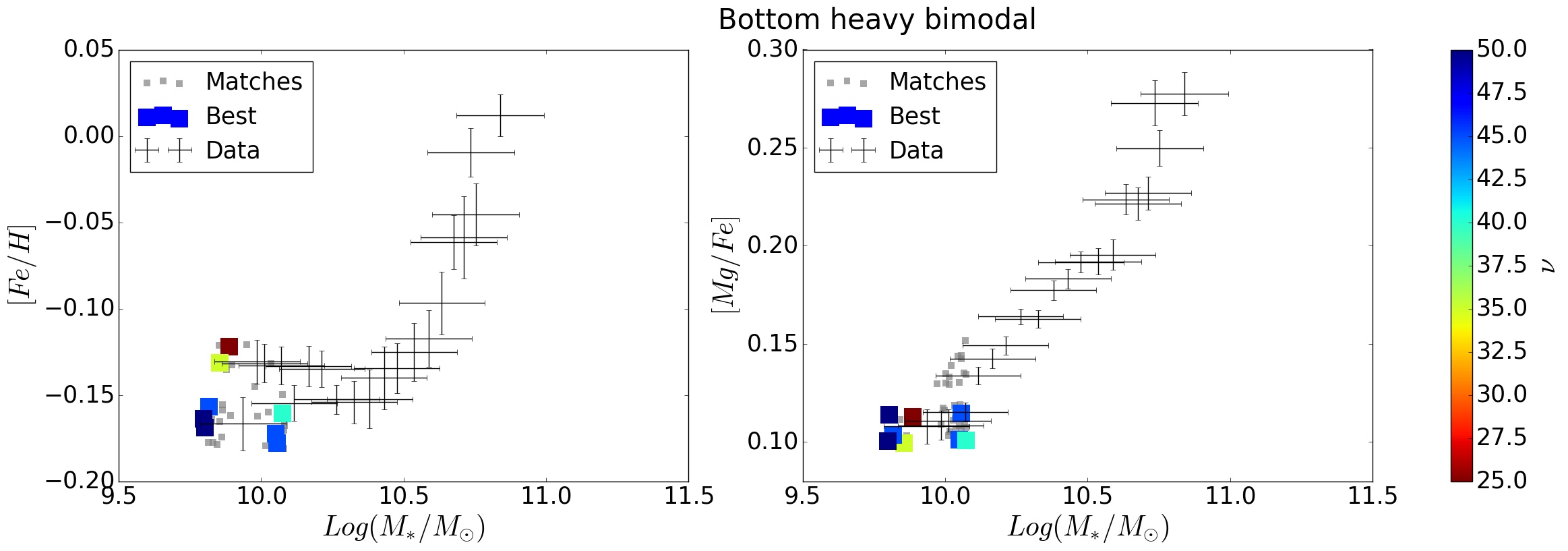}
	\includegraphics[width=.8\textwidth]{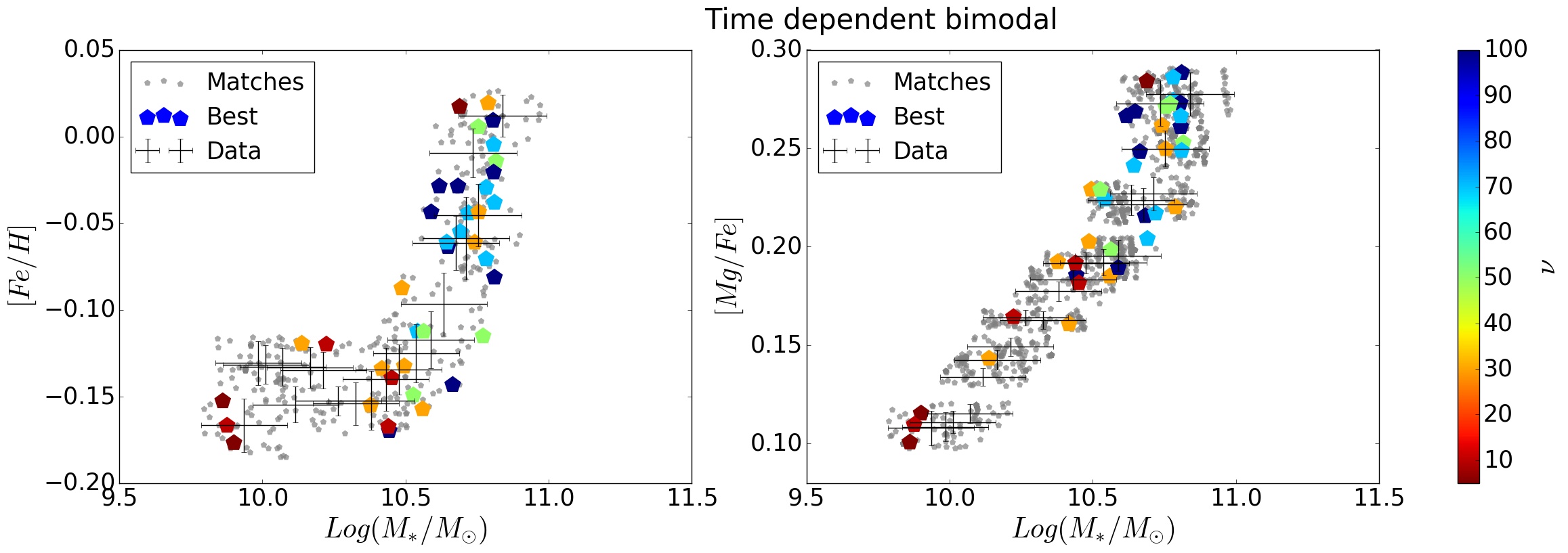}
	\caption{Comparison between central galaxies in the dataset and Models04 (top row) and 05 (bottom row) for the $[Fe/H]$ and $[Mg/Fe]$ abundance ratios. Matching models are color-coded according to their star formation efficiency, while not matching ones are shown with fading, smaller markers.}
	\label{fig:match_02_CN}
\end{figure*}

\begin{figure*}
	\includegraphics[width=.8\textwidth]{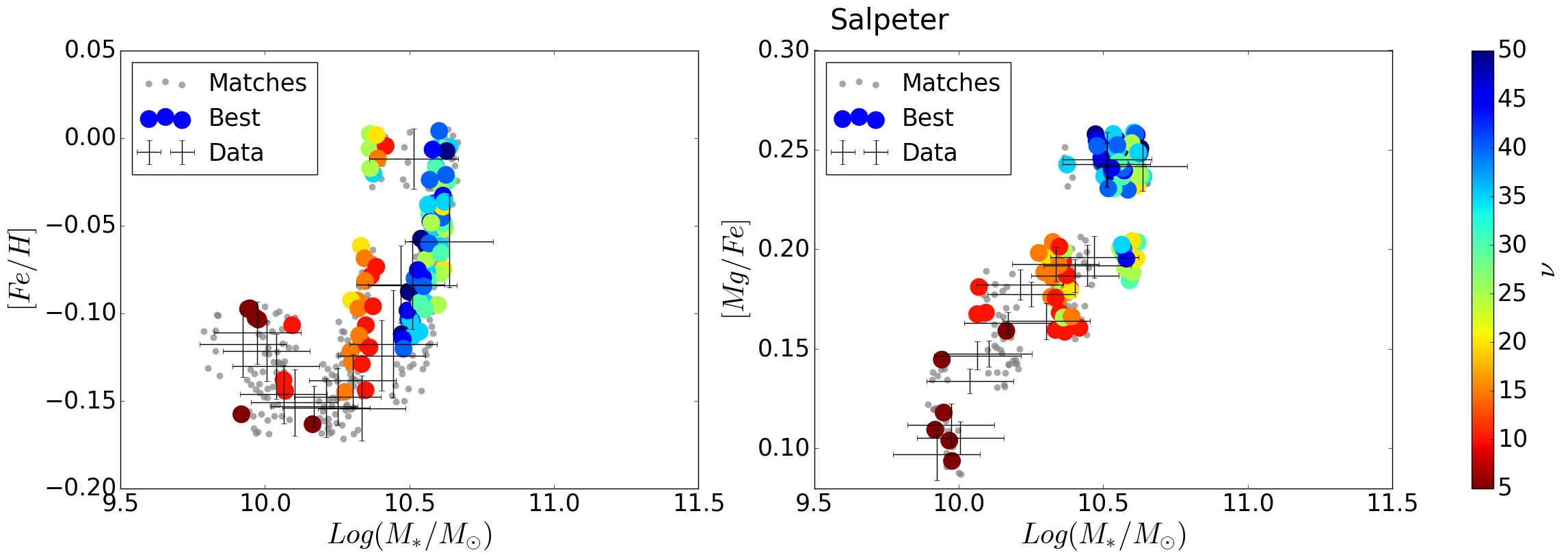}
	\includegraphics[width=.8\textwidth]{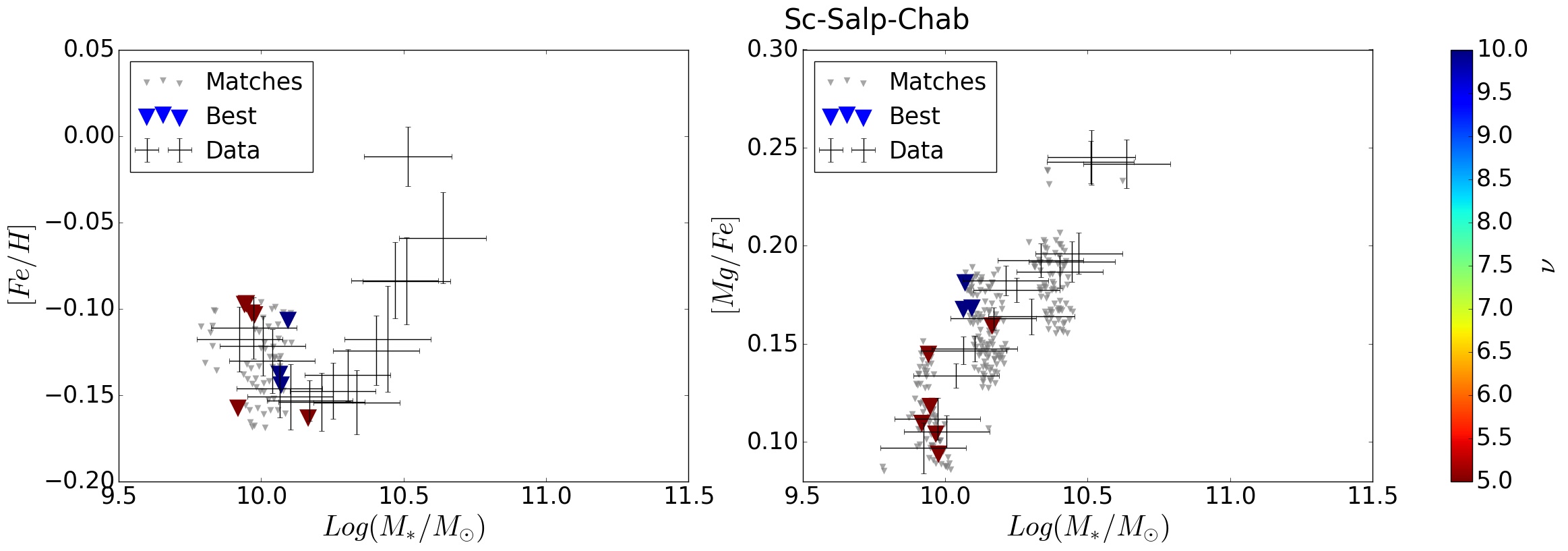}
	\includegraphics[width=.8\textwidth]{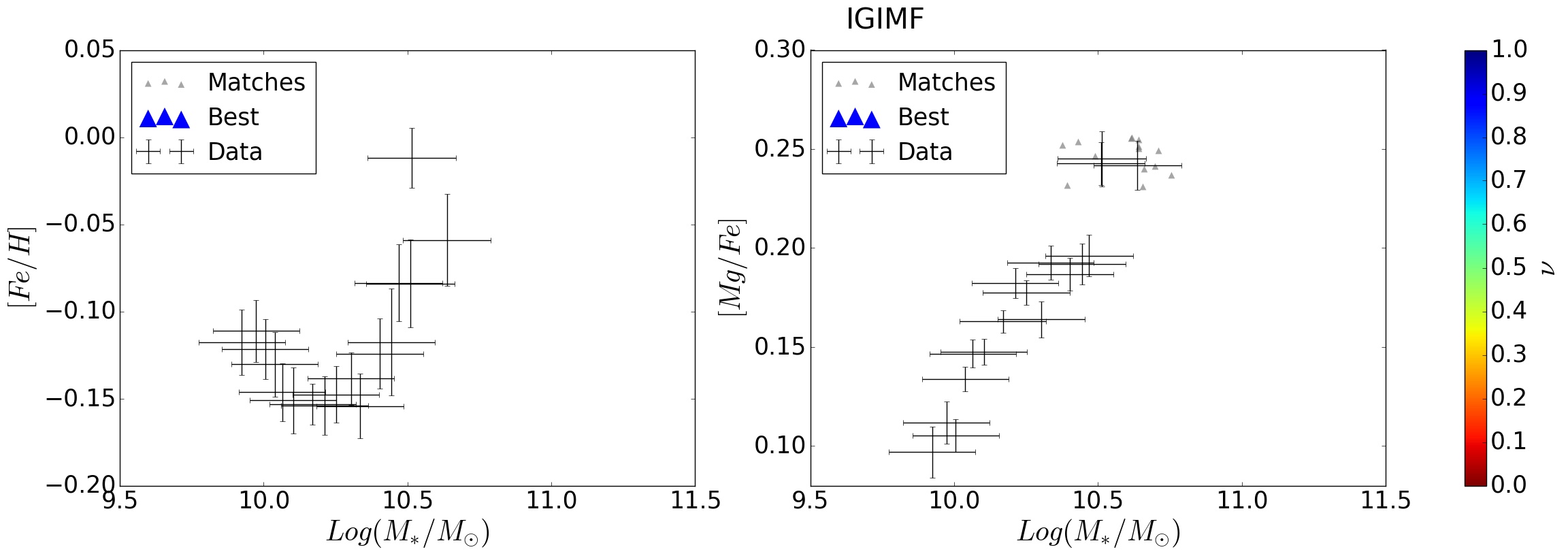}
	\caption{Comparison between satellite galaxies in the dataset data and Models01 (top row), 02 (central row) and 03 (bottom row) for the $[Fe/H]$ and $[Mg/Fe]$ abundance ratios. Matching models are color-coded according to their star formation efficiency, while not matching ones are shown with fading, smaller markers.}
	\label{fig:match_01_SAT}
\end{figure*}

\begin{figure*}
	\centering
	\includegraphics[width=.8\textwidth]{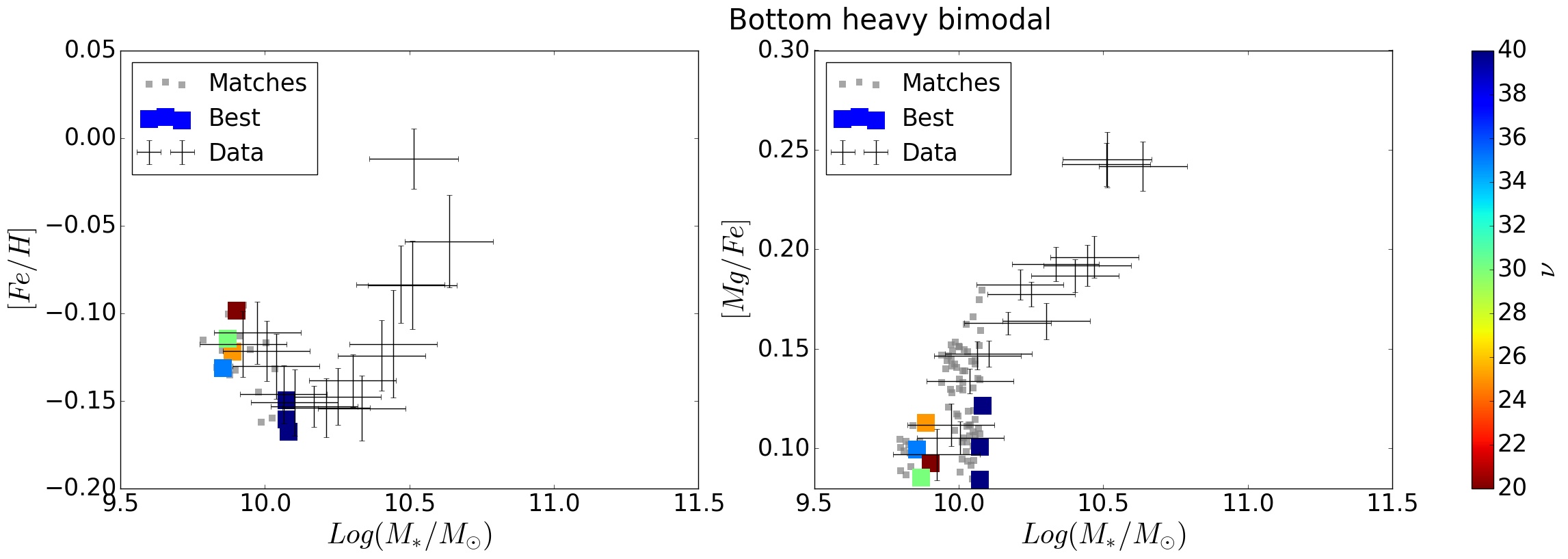}
	\includegraphics[width=.8\textwidth]{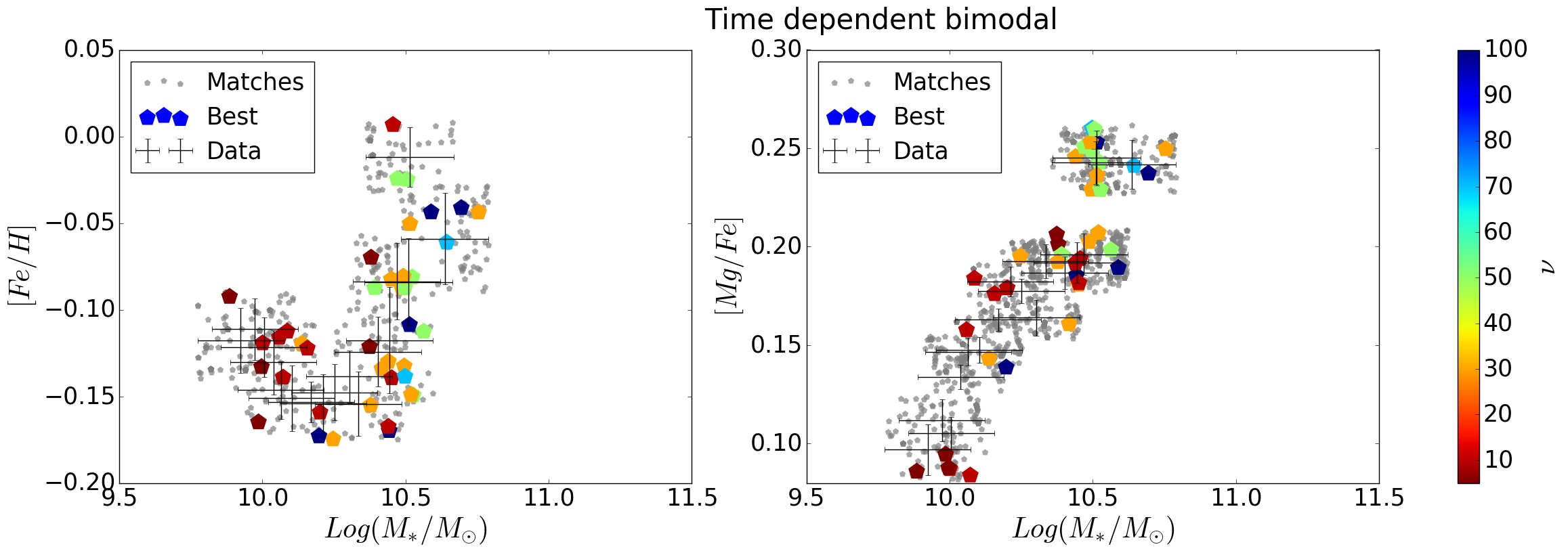}
	\caption{Comparison between satellite galaxies in the dataset and Models04 (top row) and 05 (bottom row) for the $[Fe/H]$ and $[Mg/Fe]$ abundance ratios. Matching models are color-coded according to their star formation efficiency, while not matching ones are shown with fading, smaller markers.}
	\label{fig:match_02_SAT}
\end{figure*}

As reported in section \ref{sec:environment}, we analyzed the impact of hierarchy on the matching between models and data. Figures \ref{fig:match_01_CN} to \ref{fig:match_02_SAT} are analogous to figures \ref{fig:match_01} and \ref{fig:match_02}, and show the results of the matching procedure with the subset of central/satellite galaxies in the dataset. As mentioned in the text, no significant difference with the general case (where we did not divide galaxies according to their hierarchy) is noticeable.



\section{Light averaged metallicities}\label{sec:mass_light_averages}

\begin{figure*}
	\includegraphics[width=.6\textwidth]{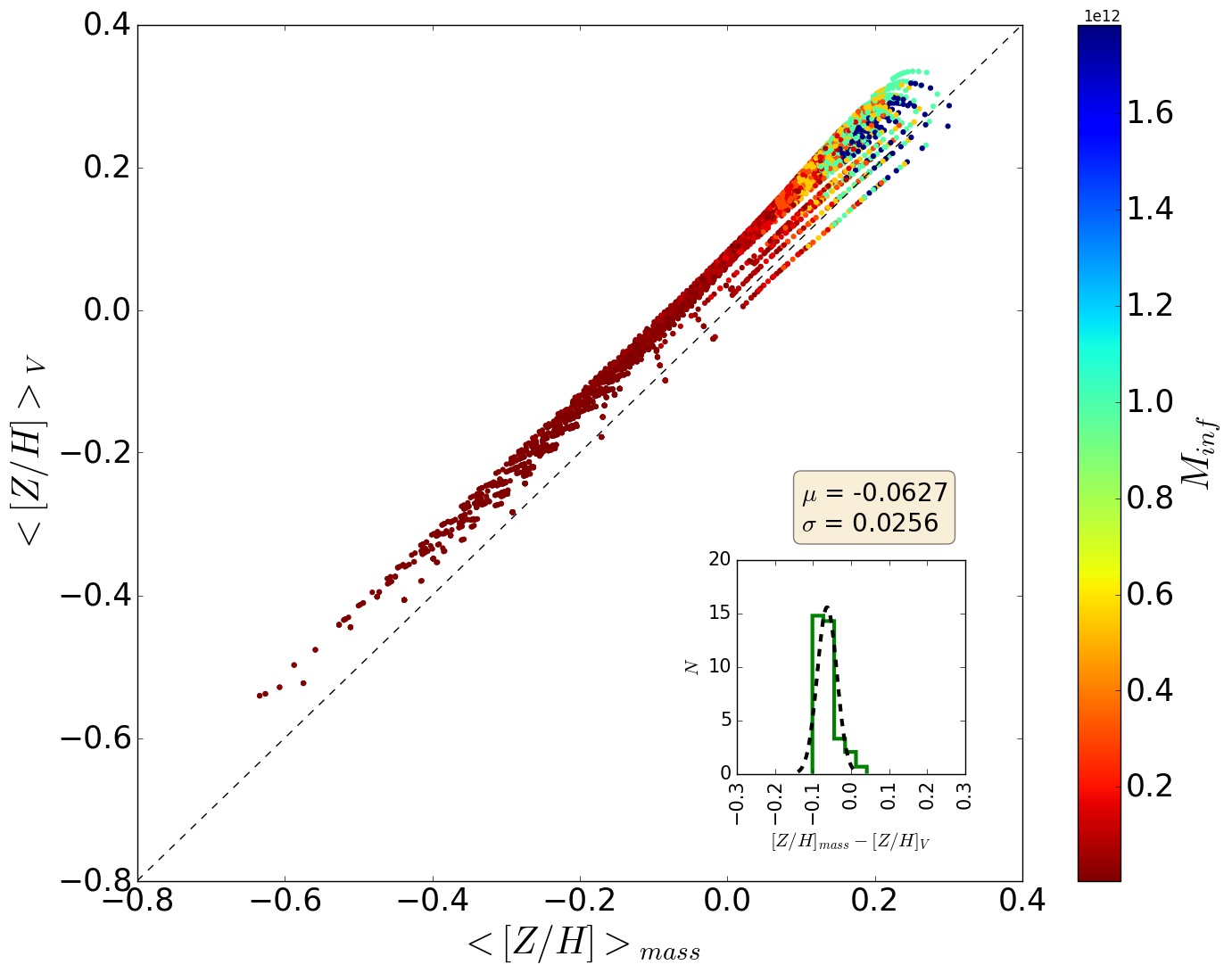}
	\caption{Scatter plot showing the comparison between the mass and light-weighted average $<[Z/H]>_{V}$ for the 4400 model galaxies obtained with the Salpeter IMF (Model 01). The points are color-coded according to the initial infall mass $M_{inf}$. The inserted histogram shows the distribution of the differences between the two averages for each model (solid line), while the dotted line shows a gaussian fit to the histogram, whose parameters are reported in the box above.}
	\label{fig:mass_light_comparison}
\end{figure*}

As described in Sec. \ref{sec:data_model_comparison}, in order to compare the results of our chemical evolution model with data it is first necessary to derive the chemical composition of the stellar population; this can be done by averaging the chemical abundances in ISM, either on mass or luminosity.\\
We always applied mass-weighted estimates, based on the results by \cite{yoshii1987}, \cite{gibson1997a} and \cite{matteucci1998}, showing that there is no significant difference for massive galaxies $(M>10^9\,M_{\odot})$ between light and mass weighted abundances.\\
We further tested this assumption by recomputing the light-averaged metallicity $[Z/H]$ for the sample of models obtained with the Salpeter IMF (Model 01), and we compared them with the corresponding mass-averaged ones, as shown in Figure \ref{fig:mass_light_comparison}.\\
As evident from the scatter plot, there is an average offset between the two quantities, with the light-weighted abundances being generally higher than the mass-averaged ones. The inserted histogram in Fig. \ref{fig:mass_light_comparison} shows the distribution of the differences between the two averages for each model (solid line), while the dotted line shows a gaussian fit to the histogram. The average difference is always less than 0,1 dex, within the observational error.  Our conclusions, however, were not significantly affected by this shift, since we are not interested in absolute abundances, but in abundance trends.


\bsp	
\label{lastpage}
\end{document}